\documentclass[longauth]{aa}  

\usepackage{graphicx}
\usepackage{txfonts}

\usepackage{txfonts}
\usepackage{lscape}
\usepackage{xcolor}
\usepackage{multirow}
\usepackage{graphicx}
\usepackage{orcidlink}
\usepackage{txfonts}
\usepackage{hyperref}  
\hypersetup{colorlinks=true,allcolors=blue}

\usepackage{xspace} 
\let\oldAA\AA
\renewcommand{\AA}{\text{\oldAA}\xspace}
\newcommand{\target}{GN-77652\xspace}

\newcommand{\nii}{[N{\sc{ii}}]}

\newcommand{\oiii}{[O\,{\sc{iii}}]}
\newcommand{\ha}{H$\alpha$\xspace}

\newcommand{\hg}{H$\gamma$\xspace}

\newcommand{\hb}{H$\beta$\xspace}

\newcommand{\sii}{[S\,{\sc{ii}}]\xspace}

\newcommand{\oi}{[O\,{\sc{i}}]$\lambda$6300\xspace}
\newcommand{\heii}{He\,{\sc{ii}}$\lambda$4686\xspace}

\newcommand{\hei}{He\,{\sc{i}}$\lambda$5876\xspace}

\newcommand{\sfr}{${M_\odot}$\,yr$^{-1}$\xspace}        
\newcommand{\msun}{${M_\odot}$\xspace}

\newcommand{\kms}{${\rm km~s^{-1}}$\xspace}

\newcommand{\ergscm}{${\rm erg~s^{-1}~cm^{-2}}$\xspace}

\begin{document} 





   \title{BlackTHUNDER Reveals a Massive Filament around\\a Compact AGN at $z$\,$\simeq$\,5.23}

   \authorrunning{G. Tozzi, H. Übler et al.}
   \titlerunning{BlackTHUNDER 77652}
   
   \author{Giulia~Tozzi\inst{\ref{MPE}}\thanks{gtozzi@mpe.mpg.de}, 
            Hannah~Übler\inst{\ref{MPE}}, 
            Eleonora~Parlanti \inst{\ref{iNorm}}, 
            Roberto~Maiolino \inst{\ref{kavli_cam},\ref{cavendish},\ref{UCL}}, 
            Claudia~Pulsoni \inst{\ref{MPE}}, 
            Rachel~Somerville\inst{\ref{CCA}},             Mirko~Curti\inst{\ref{inafbo}},
            Giovanni~Mazzolari \inst{\ref{MPE}}, 
            Capucine~Barfety\inst{\ref{MPE}}, 
            Elena~Bertola\inst{\ref{inafflo}}, 
            Andrew~J.~Bunker\inst{\ref{oxford}}, 
            Stefano~Carniani\inst{\ref{iNorm}},
            Giovanni~Cresci\inst{\ref{inafflo}}, 
            Richard~Davies\inst{\ref{MPE}}, 
            Francesco~D'Eugenio\inst{\ref{kavli_cam},\ref{cavendish}}, 
            Frank~Eisenhauer\inst{\ref{MPE},\ref{TUM}},
            Natascha~M.~F\"orster~Schreiber\inst{\ref{MPE}}, 
            Reinhard~Genzel \inst{\ref{MPE}, \ref{berkeley}},
            Lucy~R.~Ivey\inst{\ref{kavli_cam},\ref{cavendish}},
            Ignas~Juodžbalis\inst{\ref{kavli_cam},\ref{cavendish}},
            Dieter~Lutz\inst{\ref{MPE}}, 
            Cosimo~Marconcini\inst{\ref{inafflo}}, 
            Thorsten~Naab\inst{\ref{MPA}}, 
            Meghana~Pannikkote\inst{\ref{MPE}}, 
            Stavros~Pastras\inst{\ref{MPE},\ref{MPA}}, 
            Michele~Perna\inst{\ref{CAB}}, 
            Letizia~Scaloni \inst{\ref{unibo},\ref{inafbo}}, 
            Raffaella~Schneider \inst{\ref{uniro_sap},\ref{inafro},\ref{infnro},\ref{ssas}},
            Linda~J.~Tacconi \inst{\ref{MPE}}, 
            Giacomo~Venturi\inst{\ref{inafflo},\ref{iNorm}}
          }
   \institute{
        Max-Planck-Institut f\"ur extraterrestrische Physik (MPE), Gie{\ss}enbachstra{\ss}e 1, 85748 Garching, Germany\label{MPE}
   \and
        Scuola Normale Superiore, Piazza dei Cavalieri 7, I-56126 Pisa, Italy\label{iNorm}
    \and 
        Kavli Institute for Cosmology, University of Cambridge, Madingley Road, Cambridge, CB3 0HA, UK\label{kavli_cam} 
    \and 
        Cavendish Laboratory, University of Cambridge, 19 JJ Thomson Avenue, Cambridge CB3 0HE, UK\label{cavendish} 
    \and 
        Department of Physics and Astronomy, University College London, Gower Street, London WC1E 6BT, UK\label{UCL}
    \and
        Center for Computational Astrophysics, Flatiron Institute, 162 5th Ave, New York, NY 10010, USA\label{CCA}
    \and
        INAF – Astrophysics and Space Science Observatory of Bologna, Via Piero Gobetti 93/3, I-40129 Bologna, Italy \label{inafbo}
    \and 
        INAF - Osservatorio Astrofisico di Arcetri, Largo E. Fermi 5, 50125, Florence, Italy\label{inafflo}
    \and
        Department of Physics, University of Oxford, Denys Wilkinson Building, Keble Road, Oxford OX13RH, UK\label{oxford}
    \and
        Department of Physics, Technical University of Munich, 85748 Garching, Germany\label{TUM}
    \and 
        Max-Planck-Institut für Astrophysik (MPA), Karl-SchwarzschildStr. 1, D-85748 Garching, Germany \label{MPA}
    \and
    Departments of Physics and Astronomy, University of California,
Berkeley, CA 94720, USA \label{berkeley}
    \and
        Centro de Astrobiología (CAB), CSIC-INTA, Ctra. de Ajalvir km 4, Torrejón de Ardoz, E-28850, Madrid, Spain\label{CAB}
    \and
        Department of Physics and Astronomy ‘Augusto Righi’, University of Bologna, Via Piero Gobetti 93/2, I-40129 Bologna, Italy\label{unibo}
    \and
        Dipartimento di Fisica, ‘Sapienza’ Università di Roma, Piazzale Aldo Moro 2, I-00185 Roma, Italy\label{uniro_sap}
    \and
        INAF/Osservatorio Astronomico di Roma, Via di Frascati 33, I-00040 Monte Porzio Catone, Italy\label{inafro}
    \and
        INFN, Sezione Roma1, Dipartimento di Fisica, ‘Sapienza’ Università di Roma, Piazzale Aldo Moro 2, I-00185 Roma, Italy\label{infnro}
    \and
        Sapienza School for Advanced Studies, Viale Regina Elena 291, I-00161 Roma, Italy\label{ssas}\\
             }

   \date{}

 
  \abstract
   {Despite the growing number of compact broad-line (BL) active galactic nuclei (AGN) at $z$\,$>$\,4 discovered by \textit{James Webb Space Telescope} ({\it JWST}), their formation and evolution remain poorly understood. In this paper, we investigate the large-scale environment of \target, a compact BL AGN at $z$\,=\,5.229 observed as part of the {\it JWST} NIRSpec IFU Large Program BlackTHUNDER and complemented by deep multi-band NIRCam imaging. \target lies in close proximity to a 12~kpc-long filament composed of multiple sources at $z$\,$\simeq$\,5.23, spanning a remarkable range in stellar masses ($M_{\star}$\,$=$\,0.7\,--\,13 $\times$\,10$^8$~\msun), gas phase metallicities (12\,$+$\,log\,(O/H)\,$=$\,7.6\,--\,8.5) and star formation rates (SFR\,$=$\,0.4\,--\,6~\sfr). The \oiii$\lambda$5007 kinematics reveals a smooth large-scale velocity gradient centred on the massive ($M_{\star}\simeq1.1\times10^9$~\msun) and metal rich ($Z$\,$\sim$\,0.6~$Z_{\odot}$) system residing at the centre of the complex. In this source, only 2.4~kpc (projected) from \target, \oiii$\lambda$4363 auroral-line diagnostics provide possible evidence for a second AGN. \target exhibits a shallow ($-30$ to $+20$~\kms) velocity gradient that is consistent with disk rotation according to dynamical modelling. The Lyman-Werner radiation field produced by the filament is too weak for the black hole (BH) in \target to have formed recently via direct collapse. However, the required conditions may have existed at earlier epochs, or alternative scenarios, like a recoiling BH ejected from the filament, could also be plausible. Owing to their small separations, the multiple sources are expected to coalesce in $150-440$~Myr, also motivating an exploration of the system's future evolution through toy-model extrapolations and numerical simulations. Although robust constraints cannot be placed, our analysis suggests that the compact BL AGN appearance of \target represents a transient evolutionary phase, consistent with the apparent decline with redshift in number density of compact, red AGN identified with {\it JWST}.}


   \maketitle
%
 
\section{Introduction}\label{sec:intro}

The {\it James Webb Space Telescope} ({\it JWST}) has discovered a novel population of broad-line (BL) active galactic nuclei (AGN) at $z$\,$>$\,4 \citep{Harikane:2023,Greene:2024,Maiolino:2024,Matthee:2024,Hviding:2025,Taylor:2025,Juodzbalis:2026_jades} of moderate luminosity ($L_{\rm bol}$\,$\sim$\,10$^{44}-10^{46}$ erg s$^{-1}$), powered by a central accreting black hole (BH). These {\it JWST}-discovered AGN feature peculiar spectral properties, such as X-ray \citep{Ananna:2024,Yue:2024,Maiolino:2025_xray,Mazzolari:2025} and radio weakness \citep{Gloudemans:2025,Mazzolari:2026}.
For typical host galaxies with stellar mass of $M_\star$\,$\sim$\,10$^8$\,--\,10$^{10}$ \msun, single-epoch scaling relations imply BH masses of 
$M_{\rm BH}$\,$\sim$\,10$^6$\,--\,10$^8$ \msun\ for these AGN, thus BHs 10\,--\,100 times more massive \citep{Kocevski:2023,Kokorev:2023,Pacucci:2023,Ubler:2023,Furtak:2024,Juodzbalis:2024, Maiolino:2024,Napolitano:2025,Tripodi:2025} than expectations from local $M_{\rm BH} - M_{\star}$ scaling relations (e.g. \citealt{Reines:2015,Kormendy:2013}).
Assuming that the BH masses are accurately measured, this significant offset could be partly a consequence of: selection biases (i.e. finite detection limit and required detection of broad lines), which would favour the detection of more massive, highly accreting BHs \citep{Volonteri:2023,Li:2025,Ziparo:2026}; and/or underestimated host stellar mass, due to difficulties in constraining via spectral energy distribution (SED) modelling the relative AGN and stellar contribution to the total observed emission \citep{Perez-Gonzalez:2024,Ma:2026}. Yet, even attributing all continuum emission to the stars in the SED modelling, \citet{Maiolino:2024} found that the {\it JWST} BL AGN still remain well above the local expectations.

Early overmassive BHs are predicted to form by several theoretical models via a combination of seeding and growth mechanisms (e.g. see \citealt{Volonteri:2021,Schneider:2023}): light mass seeds ($\sim$\,$10^2$~\msun) as remnants of metal-free stars (e.g. \citealt{Bromm:2002,Madau:2014_seeds,Pezzulli:2016}), or intermediate mass seeds ($\sim$\,$10^{3-4}$~\msun) from runaway stellar mergers or via gas accretion in dense star clusters (e.g. \citealt{Devecchi:2009,Davies:2011}), accreting at super- or nearly-Eddington rate; or heavy seeds ($\sim$\,$10^{4-6}$~\msun) formed as direct collapse black holes (DCBHs; e.g. \citealt{Loeb:1994,Begelman:2006,Habouzit:2016}), also allowing for sub-Eddington accretion.

Another striking hallmark of many {\it JWST} BL AGN is their ‘little dot' appearance in rest-frame optical imaging (i.e. sizes $\lesssim$\,0.15$^{\prime\prime}$, corresponding to $<$\,1~kpc at $z$\,$>$\,4), with most of them being unresolved or marginally resolved by {\it JWST}. Recently, \citet{Brazzini:2026} proposed a unified picture that divides compact {\it JWST} BL AGN into two categories based on their SED shape (see also \citealt{Madau:2026,Matthee:2026,Perez-Gonzalez:2026}): ‘Little Red Dots' (LRDs, as originally named by \citealt{Matthee:2024}), a subset of {\it JWST} BL AGN ($\sim$\,10\%\,--\,30\%; \citealt{Hainline:2025}) that features a V-shaped continuum sharply turning from a negative rest-frame UV to a positive rest-frame optical slope \citep{Kocevski:2023,Greene:2024,Setton:2025,Ji:2025}; and ‘Little Blue Dots' (LBDs), the remaining ones exhibiting bluer continua (i.e.~negative UV and optical slopes; see also \citealp{Geris:2026}).

{\it JWST} BL AGN are increasingly found to feature off-centre, kiloparsec-extended rest-frame UV emission ($M_{\rm UV}$\,$<$\,$-18$; e.g. \citealt{Matthee:2024,Chen:2025,Rinaldi:2025,Baggen:2026,DEugenio:2026,Golubchik:2026,Torralba:2026}). \citet{Baggen:2026} detected such blue knots in 43\% of their examined LRD sample, and interpreted them as young, star-forming companions of modest stellar mass ($M_{\star}$\,$\sim$\,10$^8$\,--\,10$^9$~\msun), potentially producing Lyman-Werner (LW) radiation intense enough to suppress H$_2$ gas cooling and trigger DCBHs.
These results point towards a possible heavy-seed channel for at least some {\it JWST} BL AGN.

\begin{figure}
    \centering
    \includegraphics[width=0.8\linewidth]{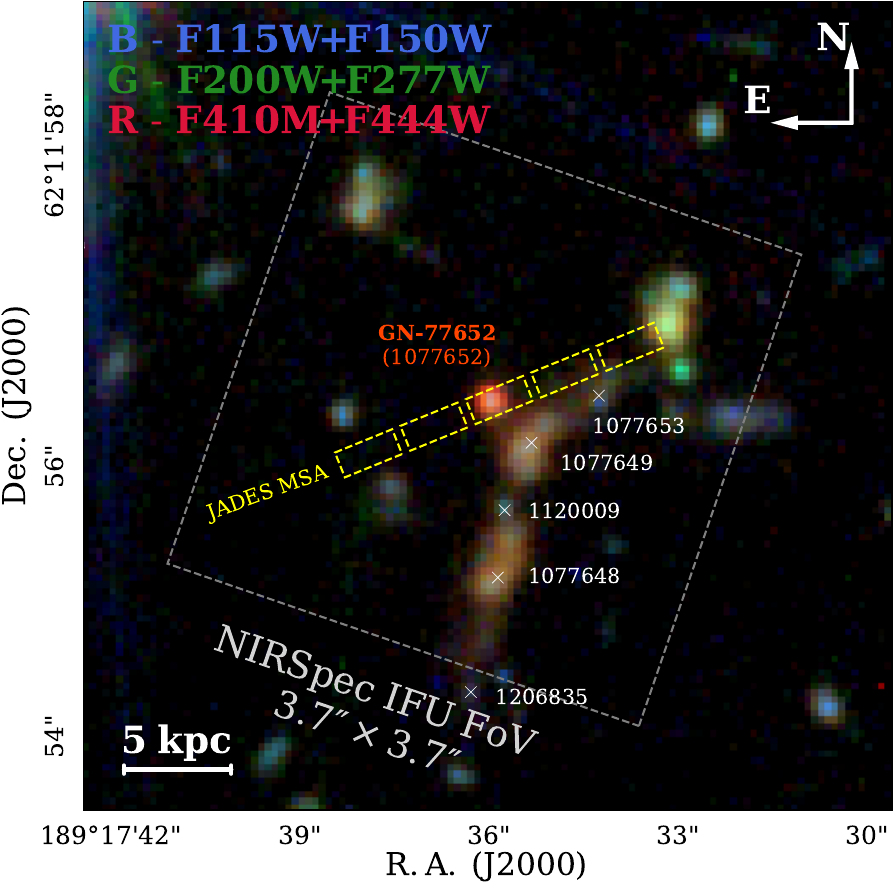}
    \caption{{\it JWST} data of \target at $z$\,=\,5.229. We show a 6$''$\,$\times$\,6$''$ colour-composite NIRCam image centred on \target (red label), where the location of the JADES NIRSpec-MSA shutters is indicated in yellow, and the BlackTHUNDER 3.6$''$\,$\times$\,3.7$''$ NIRSpec IFU FoV in grey. Some sources in the FoV are foreground galaxies, while the NW-to-SE filamentary structure consists of multiple sources at the redshift of \target. Sources identified in JADES at $z$\,$\sim$\,5 (i.e. photometric redshift) are marked by white crosses and labels.}        
    \label{fig:rgb}
\end{figure}

In this paper, we present a detailed study of the large-scale environment of JADES-GN+189.29323+62.199 (hereafter \target), a BL AGN at $z$\,$=$\,5.229 that resides close to a 12~kpc-extended filament of massive sources at the same redshift (Fig.~\ref{fig:rgb}).
Located in the GOODS-N field \citep{Giavalisco:2004}, \target was first identified as a compact source in deep multi-band {\it JWST}/NIRCam imaging, obtained as part of the {\it JWST} Advanced Deep Extragalactic Survey (JADES; \citealt{Eisenstein:2023a}), and then spectroscopically followed up in JADES \citep{Bunker:2024,CurtisLake:2026,DEugenio:2025} with the {\it JWST}/NIRSpec Micro Shutter Array (MSA; \citealt{Ferruit:2022}). The NIRSpec MSA medium-resolution spectra revealed a broad \ha component 
consistent with the presence of a moderate-luminosity ($L_{\rm bol}$\,$\sim$\,10$^{44}$~erg~s$^{-1}$) BL AGN, powered by an accreting BH of mass $M_{\rm BH}$\,$\simeq$\,4.2\,$\times$\,10$^{6}$~\msun (\citealt{Juodzbalis:2026_jades}; see also \citealt{Maiolino:2024}). This BH appears overmassive by more than two orders of magnitude relative to the local $M_{\rm BH}-M_\star$ relation of \citet{Reines:2015}, with a host stellar mass of ($M_{\star}$\,$\simeq$\,2\,$\times$\,10$^{8}$~\msun; \citealt{Juodzbalis:2026_jades}). Moreover, as is the case for the vast majority of high-$z$ {\it JWST} AGN, \target is undetected at X-ray \citep{Maiolino:2025_xray} and radio frequencies \citep{Mazzolari:2026}, and features a monotonically decreasing SED shape at increasing wavelengths ($\beta_{\rm UV}=-2.10$, $\beta_{\rm opt}=-1.56$; Mazzolari et al., in prep.), which classifies it as a LBD.

\target and its immediate ($<$\,15~kpc) environment were next observed in {\it JWST}/NIRSpec IFU mode \citep{Jakobsen:2022,Boker:2022} by the BlackTHUNDER program (PID: 5015; PIs: H. Übler, R. Maiolino), which acquired the spatially resolved high- ($R$\,$\sim$\,2700) and low-resolution ($R$\,$\sim$\,100) spectroscopy presented in this work. Exploiting all available {\it JWST} data, in this paper we investigate gas kinematics, interstellar medium (ISM) and stellar properties across the entire $z$\,$\simeq$\,5.23 complex, finding differences by up to an order of magnitude in $M_\star$, star formation rate (SFR) and gas phase metallicity for individual sources. The close proximity of the compact BL AGN to the filament also allows us to explore early BH formation scenarios for \target, and the future coalescence of the whole complex into a single massive source.

This paper is structured as follows. In Sect.~\ref{sec:data} we present all {\it JWST} data used in this work, including NIRSpec spectroscopy and NIRCam imaging. In Sect.~\ref{sec:sec3}, we investigate the morphology, gas kinematics and stellar properties of the system, and in Sect.~\ref{sec:diag} the dominant ionisation mechanisms via optical emission-line diagrams. In Sect.~\ref{sec:discussion}, we discuss the properties of the various sources in relation with main galaxy scaling relations, possible early BH formation scenarios for \target, and the possible future evolution of the complex, including comparison with simulations. In Sect.~\ref{sec:conclusions} we draw our conclusions. 

In this work, we adopt the cosmology from \citet{Planck:2015} ($H_0
= 67.7$ \kms\,Mpc$^{-1}$, $\Omega_{\rm m}$ = 0.307, and $\Omega_\Lambda$= 0.691), and a \citet{Chabrier:2003} initial mass function (IMF).

\section{Data}\label{sec:data}

Figure~\ref{fig:rgb} shows all {\it JWST} data of \target: multi-band NIRCam imaging and NIRSpec MSA spectroscopy of \target from JADES (JADES ID: 1077652); and the NIRSpec IFU observations from BlackTHUNDER, covering the crowded $\sim$\,3.7$''$\,$\times$\,3.7$''$ field-of-view (FoV) around \target. In the following, we describe the various data and reduction procedure.

\subsection{JADES JWST/NIRCam imaging}\label{sec:nircam}
\target benefits from a rich dataset including deep NIRCam multi-band imaging obtained as part of the JADES program 1181 in GOODS-N (PI: Eisenstein; \citealt{Eisenstein:2023a}), which consists of seven broad-band filters (F090W, F115W, F150W,
F200W, F277W, F356W, F444W) and two medium-band filters (F335M, F410M). Additional NIRCam data from general observer (GO) programs were included in the JADES Fifth Data Release (DR5; \citealt{Johnson:2026}), which released total multi-band mosaics of the GOODS-N and GOODS-S fields. At the location of \target, supplementary coverage is provided by the {\it JWST} GO programs 2674 (PI: Arrabal Haro) and 6434 (PI: Eiichi Egami), leading to total exposure times of $3.1-6.6$~hr and $2.0-4.5$~hr for the short- and long-wavelength filters, respectively. F182M imaging is also available for \target from the program 2674, but too shallow ($\sim$\,0.9 hr) to robustly detect emission across the whole structure (see Fig.~\ref{fig:nircam}). We retrieved the latest DR5 reduced multi-band NIRCam imaging (\citealt{Johnson:2026}; see also \citealt{Rieke:2023} and \citealt{Eisenstein:2023} for more details on JADES data reduction) from the JADES webpage\footnote{\url{https://slate.ucsc.edu/~brant/jades-dr5/}}. We extracted 6$''$\,$\times$\,6$''$ cutouts from the original mosaics (Figs.~\ref{fig:rgb} and \ref{fig:nircam}), and used the F444W PSF-matched mosaics \citep{Robertson:2026} for resolved SED fitting (Sect.~\ref{sec:sed_maps}).

\subsection{JADES JWST/NIRSpec MSA spectroscopy}\label{sec:nirspec_msa}
JADES also obtained NIRSpec MSA spectroscopy of \target, as part of program ID 1181 in the Medium {\it JWST} GOODS-N tier of JADES \citep{Eisenstein:2023a}. The exposure times of the MSA observations of \target are about 3060 sec with PRISM-CLEAR, G140M-F070LP, G235M-F170LP, G395M-F290LP and G395H-F290LP. The total exposures were split into two repeated 3-position nodding sequences, using the NRSIRS2 readout pattern. In this paper, we use the 1D spectra of \target released as part of the JADES DR4 \citep{CurtisLake:2026,Scholtz:2026_jades}, extracted from the central 3 pixels of each 2D spectrum (i.e. the central 0.3$''$ region), which maximises the signal-to-noise (S/N) ratio for compact sources. Low- and medium-resolution spectra of \target from previous JADES reductions were analysed by \citet{Maiolino:2024} and more recent ones by \citet{Juodzbalis:2026_jades}, establishing the BL AGN nature of \target and providing a first characterisation of its properties. Here, we perform an independent spectral fitting of the available MSA data, using more recent reductions and exploiting high-resolution (G395H-F290LP) data for accurate line modelling, in particular of broad \ha emission. Unlike in the NIRSpec high-resolution IFU observations from BlackTHUNDER, \ha does not fall into the detector gap in the NIRSpec MSA data, since in this mode the wavelength cutoffs depend on the shutter position, in addition to the physical edges of the two detectors and the separation between them.

\subsection{BlackTHUNDER NIRSpec IFU observations}\label{sec:blackthunder_data}
\target was next observed in NIRSpec IFU mode by the BlackTHUNDER survey (PID: 5015; PIs: Übler, Maiolino), on March 10-11, 2025 with a medium cycling 8-point dithering pattern and a total integration time of 5.1~hr and 1~hr with the G395H-F290LP ($\Delta\lambda$\,=\,2.87\,--5.14~$\mu$m; $R$\,$\sim$\,2000\,--\,3500) and PRISM-CLEAR ($\Delta\lambda$\,=\,0.6\,--\,5.3~$\mu$m; $R$\,$\sim$\,30\,--\,300) configurations, respectively. We retrieved the raw data files from the Barbara A. Mikulski
Archive for Space Telescopes (MAST) portal\footnote{\url{https://archive.stsci.edu/}}, and processed them with the {\it JWST} Science Calibration pipeline\footnote{\url{https://jwst-pipeline.readthedocs.io/en/stable/jwst/introduction.html}} version
1.15.0 under the Calibration Reference Data System (CRDS)
context jwst\_1293.pmap. To improve the data reduction, we applied some modifications to the standard reduction steps (see \citealt{Perna:2023} for details). We corrected the count-rate frames for 1/$f$ noise via a polynomial fit, and masked in the calibration stage 2 regions in each frame affected by failed open MSA shutters, cosmic rays and bad pixels. Following \citet{DEugenio:2024}, we thus flagged the outliers identified in individual exposures, and rejected the 98th (99.5th) percentile of the resulting distribution in the grating (prism) data. The final cubes were combined using the ‘drizzle’ method with a pixel scale of 0.05$''$, 
and cover a total 3.7$''$\,$\times$\,3.7$''$ FoV.

No dedicated off-source sky exposures were acquired in the BlackTHUNDER observations, the background was therefore subtracted using spaxels away from the central source and free of source emission. To obtain reliable flux uncertainties, we scaled, spaxel-by spaxel, the formal error extension of the two final reduced datacubes to match the rms in regions free of spectral features \citep[see][]{Ubler:2023,Ubler:2025}. We measured average scaling factors of about 1.3 and 2.0 in the PRISM-CLEAR and G395H-F290LP datacubes, respectively. From the \oiii$\lambda$5007 (\oiii\ hereafter) line fitting in the high-resolution data, we measure a spectroscopic redshift of $z$\,=\,5.2294\,$\pm$\,0.0007 for \target, consistent with the previous measurement in JADES MSA data ($z$\,=\,5.22943; \citealt{Maiolino:2024}).

\section{\target and its complex environment}\label{sec:sec3}

\subsection{A overdense filament of multiple sources at $z$\,$\simeq$\,5.23}\label{sec:sources}

\begin{figure*}
    \centering
    \includegraphics[width=0.95\linewidth]{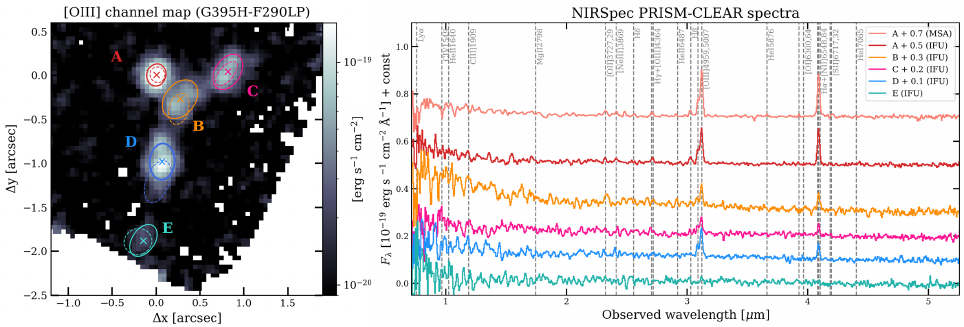}
    \caption{The whole source complex at $z$\,$\simeq$\,5.23 consisting of \target\ and a nearby filament of multiple sources. \textit{Left.} \oiii\ flux map, obtained by collapsing the \oiii\ channels at $|\Delta v|$\,$<$\,200 km s$^{-1}$ in the NIRSpec high-resolution datacube, revealing five distinct sources: \target (A) and another four extended sources (B, C, D, E). Solid (dashed) elliptical apertures are based on the 2D Gaussian fitting to the \oiii\ (F444W) light distribution (see Sect.~\ref{sec:sources}). \textit{Right.} Integrated spectra of the various sources, extracted from the NIRSpec IFU low-resolution cube with the solid apertures shown in the left panel. The JADES NIRSpec MSA spectrum of \target is also shown at the top, after rescaling its \oiii\ peak to that of the NIRSpec IFU spectrum (red). The spectra feature differences in the continuum shape and in the relative emission-line intensity from rest-frame UV to optical wavelengths, indicating different physical properties for the various sources. For visualisation purposes, an arbitrary constant has been added to each spectrum as indicated in the legend.}
    \label{fig:fig2}
\end{figure*}

\renewcommand{\arraystretch}{1.2}  
\begin{table*}
    \centering
    \footnotesize
\caption{Characterisation of the source morphology based on their \oiii\ line and F444W continuum light distributions.}
\label{tab:morphology}
    \resizebox{0.94\linewidth}{!}
{\begin{tabular}{l|cccc|cccc|ccc|c}
    \hline
    \hline
     \multirow{2}{*}{ } & \multicolumn{4}{c|}{\oiii} & \multicolumn{4}{c|}{F444W} & \multicolumn{3}{c|}{JADES DR5} &\multirow{2}{*}{$\delta_{\rm c}$} \\
     \cline{2-12}
     & $d_{\rm A} $ & FWHM$_{1,2}$ & FWHM & $R_{\rm e}$ & $d_{\rm A}$ & FWHM$_{1,2}$ & FWHM & $R_{\rm e}$ & $d_{\rm A}$ & FWHM$_{1,2}$ & FWHM & \\
     \hline\hline
    A & - & 0.25, 0.22 & 0.23 & 0.098 [0.60] & - & 0.18, 0.17 & 0.17 & 0.041 [0.25] & - & 0.18, 0.17 & 0.18 & - \\
    B & 0.39 [2.38] & 0.49, 0.34 & 0.41 & 0.19 [1.19] & 0.46 [2.81] & 0.36, 0.22 & 0.28 & 0.12 [0.72] & 0.44 [2.7] & 0.38, 0.25 & 0.31 & 0.11 [0.7] \\
    C & 0.82 [5.01] & 0.44, 0.29 & 0.36 & 0.17 [1.02] & 0.79 [4.82] & 0.33, 0.16 & 0.23 & 0.085 [0.52] & 0.79 [4.9] & 0.34, 0.24 & 0.29 & - \\
    D & 0.99 [6.07] & 0.41, 0.28 & 0.34 & 0.16 [0.96] & 1.22 [7.48] & 0.50, 0.27 & 0.37 & 0.17 [1.02] & - & - & - & 0.23 [1.4] \\
    E & 1.89 [11.6] & 0.40, 0.26 & 0.32 & 0.15 [0.91] & 1.91 [11.7] & 0.32, 0.30 & 0.31 & 0.13 [0.82] & - & - & - & 0.06 [0.4]\\
    
    \hline

    \end{tabular}
    }
    \tablefoot{All quantities are in units of arcsecond, except for values in squared brackets in units of kpc (6.141 kpc/$''$ at $z$\,$=$\,5.229). The three four-column groups list the main parameters resulting from our 2D Gaussian fit to the \oiii\ and F444W light distributions, and based on the JADES DR5 Gaussian regression modelling \citep{Robertson:2026}. Each group of columns reports: the projected separation of each source from A ($d_{\rm A}$, i.e. from \target); the major and minor axis (FWHM$_1$, FWHM$_2$) of the 2D Gaussian model; the circularised FWHM\,$=$\,$\sqrt{\rm FWHM_1 \times FWHM_2}$; for our measurements, also the effective radius $R_{\rm e}$ corrected for the instrumental PSF. The last column displays the distance $\delta_{\rm c}$ between the \oiii\ and F444W Gaussian centroids within the same source, when $\delta_{\rm c}$\,$>$\,0.05$''$. JADES DR5 measurements refer to the sources 1077652, 1077649, and 1077653, roughly corresponding to A, B and C in our source identification, respectively.}
    
\end{table*}

As shown by Fig.~\ref{fig:rgb}, \target is surrounded by several sources and extended structures within the NIRSpec IFU FoV, for which spectroscopic redshifts can now be measured robustly.
The main filamentary structure, extending from NW to SE, is composed of multiple sources (see also Fig.~\ref{fig:nircam}) at the same redshift of \target ($z$\,$\simeq$\,5.23), as shown by the \oiii\ channel map in the left panel of Fig.~\ref{fig:fig2}.
Five main sources can be visually identified in the \oiii\ channel map: \target (labelled as A), and four sub-sources along the filament (B, C, D, E). Interestingly, the compact BL AGN is offset with respect to the overdense filament. In addition, seven foreground objects are detected in the NIRSpec IFU low-resolution data (Appendix~\ref{apx:specz}).

To characterise the five main sources at $z$\,$\simeq$\,5.23, we fit five 2D (spatial) Gaussian components (one for each source) to the \oiii\ and F444W continuum images, tracing gas and stars, respectively.
The main best-fit parameters are listed in Table~\ref{tab:morphology},
and the resulting \oiii\ (F444W) 2D Gaussian models are plotted in the left panel of Fig.~\ref{fig:fig2} as solid (dashed) elliptical apertures with the major and minor axes equal to the best-fit FWHMs. The average projected separations between closest pairs of sources is 5~kpc. The integrated NIRSpec IFU low-resolution spectra of the various sources (right panel), extracted using the \oiii-based apertures, exhibit differences in both the continuum shape and the relative emission-line intensity, which point to intrinsically different properties between the multiple sources.

\begin{figure*}
    \centering
    \includegraphics[width=0.95\linewidth]{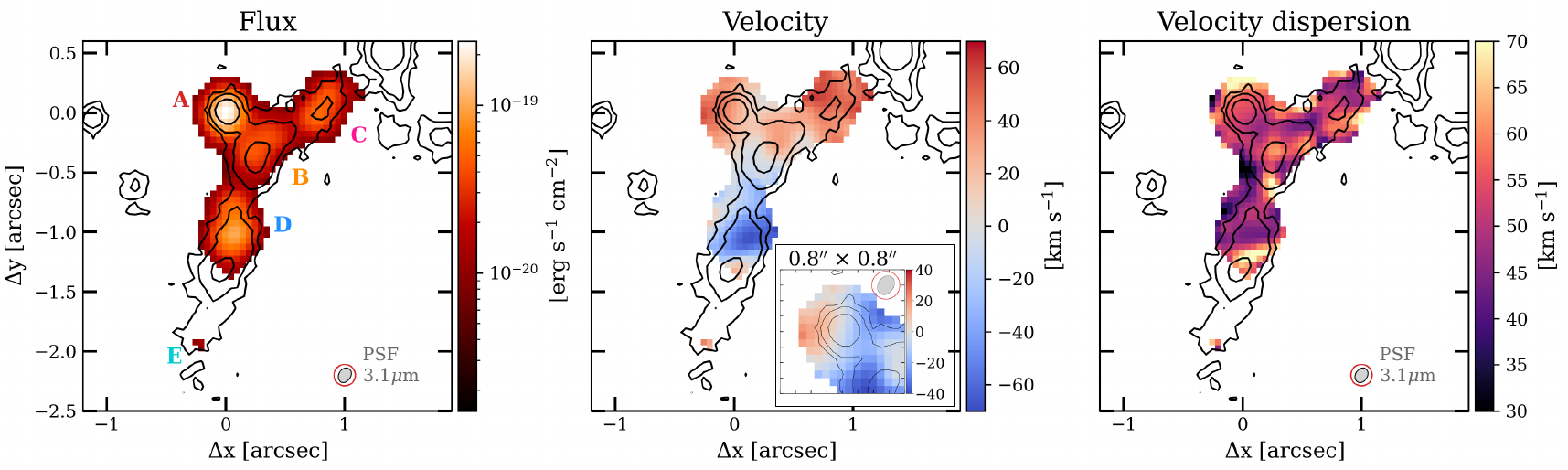}
    \caption{Maps of \oiii\ flux, velocity and velocity dispersion corrected for instrumental broadening, as resulting from the spectral fitting of the NIRSpec high-resolution datacube. The maps only display pixels with S/N(\oiii)\,$\geq$\,3. Black contours correspond to 20$\sigma$, 7$\sigma$, and 4$\sigma$ levels of F444W continuum emission. 
    The velocity field in the main middle panel is centred on the systemic velocity of B, and reveals offsets $|v|$\,$<$\,70 km~s$^{-1}$ across the various sources, suggesting converging gas flows in B along the filament as a possible interpretation. The small inset displays the zoomed-in velocity field centred on the systemic velocity of \target, which features a shallow velocity gradient across this compact source of $\Delta v$\,$\simeq$\,50~km~s$^{-1}$ magnitude. Enhanced velocity dispersions are observed corresponding to the F444W peak in D, and of the common F444W and \oiii\ peak in B. The grey filled and red empty ellipses at the bottom of the panels represent the NIRSpec PSF at the 3.1~$\mu$m \oiii\ observed wavelength inferred by \citet{DEugenio:2024} and more recently \citet{Jones:2026}, respectively.}
    \label{fig:oiii_map_R2700}
\end{figure*}

Most of the sources feature a relative offset between the \oiii\ gas distribution and the F444W stellar continuum emission, and/or a more extended \oiii\ morphology (see Table~\ref{tab:morphology}). Such differences could be partially due to the gravitational interaction between the sources, during which gas and stars are affected differently because of their dissipative and collisionless nature, respectively. The largest relative offset is observed in D ($\delta_{\rm c}$\,$=$\,0.23$''$), where continuum emission is more elongated and becomes brighter towards the South than \oiii. Such difference can also be appreciated in the left panel of Fig.~\ref{fig:oiii_map_R2700}, which displays the \oiii\ flux map obtained from the spectral fitting of the NIRSpec IFU high-resolution datacube, with the F444W continuum emission contours drawn on the top. Moreover, the PSF-corrected size of \target in \oiii\ line emission ($R_{\rm e}$\,$\sim$\,0.1$''$) is larger by more than a factor of 2 than continuum emission ($R_{\rm e}$\,$\sim$\,0.04$''$), suggesting a more extended gas distribution. 

All sources within the NIRSpec IFU FoV correspond to entries in the DR5 JADES 
photometric catalogue \citep{Robertson:2026}\footnote{Source identified in the JADES DR5 can be quickly visualised online using the interactive web interface at \url{https://jades-survey.github.io/viewer/}, developed by the JADES team and based on \texttt{FitsMap} \citep{Hausen:2022}.}. In particular, five sources (JADES IDs in Fig.~\ref{fig:rgb}) at $z_{\rm phot}$\,$=$\,5.17\,--\,5.25 ($z_{\rm spec}$\,$=$\,5.227\,--\,5.230 from NIRSpec IFU data; see Table~\ref{tab:results}) are identified along the filament, three of which well match our sources A, B and C. For them, we infer the same structural parameters as for our \oiii\ and F444W Gaussian modellings (listed in Table~\ref{tab:morphology}), using the JADES DR5 results from a regression Gaussian fit to a detection image created by combining individual NIRCam long-wavelength images \citep{Robertson:2026}. Conversely, our source D corresponds to two distinct sources in JADES, whereas the source E to the northernmost emission of the JADES source 120683, whose centre falls outside the NIRSpec IFU FoV.

\subsection{Large-scale gas kinematics of the complex}\label{sec:gas_kinematics}

We use the fitting code presented in \citet{Marasco:2020} (see also \citealt{Cresci:2023}) for the spectral modelling of the NIRSpec IFU datacubes (details in Appendix~\ref{apx:specfit}). The same procedure is applied to the NIRSpec MSA spectra of \target, and the integrated NIRSpec IFU spectra of the various sources, extracted with the \oiii-based apertures in Sect.~\ref{sec:sources}, assuming uncorrelated errors between spaxels. 
Figure~\ref{fig:oiii_map_R2700} shows \oiii\ maps of flux, velocity and velocity dispersion corrected for the instrumental broadening\footnote{We adopt $\sigma_{\rm inst}$\,$=$\,44~km~s$^{-1}$ at 3.12~$\mu$m, applying a factor 0.7 following \citet{Shajib:2025}, as compared to the nominal values by \citet{Jakobsen:2022}.}, as obtained from the spectral fitting of the NIRSpec high-resolution datacube. \oiii\ is detected at S/N\,$\gtrsim$\,3 in only very few pixels of E, which is insufficient for the scope of the following spatially resolved analysis. Resolved flux maps of additional emission lines are provided in Appendix~\ref{apx:specfit}, namely: \ha and \hb, showing an extended flux distribution similar to \oiii; and He\,{\sc i}$\lambda$5876, which is only detected in \target.

The velocity field (Fig.~\ref{fig:oiii_map_R2700}, middle panel) has the zero velocity centred on the systemic velocity of B (corresponding to $z$\,$=$\,5.230), which appears roughly at the centre of the complex in the projected plane of the sky. We observe velocity offsets $|v|$\,$<$\,70 km~s$^{-1}$ across the whole complex, precisely a smooth gradient along the filament, from most positive values in C to most negative ones in D. This suggests that the northern sources A and C could be moving back along the line of sight, whereas D could be moving forward. The mass-weighted mean velocity of only $-1$~\kms inferred in the inner region of B, obtained by weighting each pixel velocity by the pixel stellar mass $M_\star$ from resolved SED fitting (presented in Sect.~\ref{sec:sed_maps}) also supports the central position of B in the 3D space too. Considering the continuous filamentary appearance of the complex, another possible explanation to such large-scale kinematic pattern is the existence of converging gas flows in B, along the main NW-SE filament and also from \target.

The velocity dispersion mostly varies in the range $\sigma$\,$=$\,40\,--70~\kms (Fig.~\ref{fig:oiii_map_R2700}, third panel), with a relatively flat distribution of values ($\sigma$\,$\simeq$\,55~\kms) across \target. Enhanced velocity dispersions are observed corresponding to the F444W emission peak of D ($\sigma$\,$\simeq$\,70~\kms), while lower values ($\sigma$\,$\simeq$\,40\,--\,50~\kms) are found in the \oiii\ bright region. This points to the F444W emission peak as the likely kinematic centre of D, hosting the bulk of the stars and exhibiting a peak in velocity dispersion as expected for gravitationally bound systems (e.g. \citealt{Genzel:2014a,Genzel:2023, Wisnioski:2015,Ubler:2019}). 
Higher velocity dispersions are also found in the inner region of B ($\sigma$\,$\simeq$\,60~\kms), compared to the outskirts ($\sigma$\,$<$\,50~\kms). Finally, in C the largest line widths are found to gather south of the common \oiii\ and F444W emission peak.



\subsection{Resolved rotation in \target}
\label{sec:vel_grad}

The small inset in the middle panel of Fig.~\ref{fig:oiii_map_R2700} displays a zoomed-in view of the velocity field of \target, centred on its systemic velocity. The map reveals a shallow velocity gradient ($-30$~km~s$^{-1}$ to 20~km~s$^{-1}$), roughly along the WE direction and encompassing at least three resolution elements of the NIRSpec PSF derived by \citet{Jones:2026} (FWHM\,$=$\,0.18$^{\prime\prime}$). This kinematic pattern indicates possible resolved disk rotation, so far detected in only a handful of {\it JWST} BL AGN at $z$\,$>$\,4 \citep{Ubler:2023,Ubler:2025,Parlanti:2024,Juodzbalis:2026_moka}, as a consequence of their compact morphology (barely resolved by {\it JWST}) and the elusive emission of their host galaxy (e.g. \citealt{Xiao:2025,Akins:2026,Mazzolari:2026}).

To test whether the velocity gradient observed in \target is associated with disk rotation, we carry out a dynamical modelling of \target using \texttt{DysmalPy} \citep{Davies:2004,Cresci:2009,Genzel:2023,Ubler:2018,Price:2021,Lee:2025_dysmalpy,Jolly:2026}, a 3D forward-modelling code that accounts for beam smearing, line broadening and finite spatial resolution. As detailed in the following, we only fit the (high-resolution) \oiii\ kinematics (i.e. velocity and velocity dispersion corrected for instrumental broadening), while fixing the morphological parameters derived in Sect.~\ref{sec:sources} and the centre of the model to the central flux-peak pixel. 

Our model only includes a baryonic disk, since the dark matter (DM) component is expected to be negligible at such small ($<$\,1~kpc) radii. Moreover, as will be discussed in Sect.~\ref{sec:merger_tscale}, the whole source complex is likely to be embedded in a common DM halo, which makes uncertain any constraint on the DM halo properties derived from the stellar component (e.g. the stellar-to-halo mass relation; \citealt{Girelli:2020}). For the disk, we fix the following structural parameters: the Sérsic index $n=1$ (i.e. we assume an exponential disk); the effective radius $R_{\rm e,d} = 0.6$~
kpc (inferred from \oiii\ in Sect.~\ref{sec:sources}); and the minor-to-major axis
ratio $b/a = 0.95$, as measured via 2D Gaussian fit to the sharp F115W imaging. The scale height of the disk is tied to $R_{\rm e,d}$ (see Eq. 2 in \citealt{Lee:2025_dysmalpy}). Assuming a thick disk with a scale height-to-length ratio of $q_0 = 0.2$ (e.g. \citealt{Wuyts:2016}), $b/a = 0.95$ corresponds to an inclination $i=18^\circ$. We adopt a flat prior on the intrinsic velocity dispersion $\sigma_0$, with bounds of 5~\kms and 100~\kms and assumed to be constant and isotropic throughout the disk, and allow the position angle (PA) to vary in the range PA\,$=$\,[250$^\circ$, 290$^\circ$] (measured with respect to the blueshifted part, from North to East). Finally, we model the PSF as a Gaussian profile with FWHM\,$=$\,0.18$''$, which approximately corresponds to the NIRSpec PSF derived by \citet{Jones:2026} at the \oiii\ wavelength.

\begin{figure}
    \centering    \includegraphics[width=0.98\linewidth]{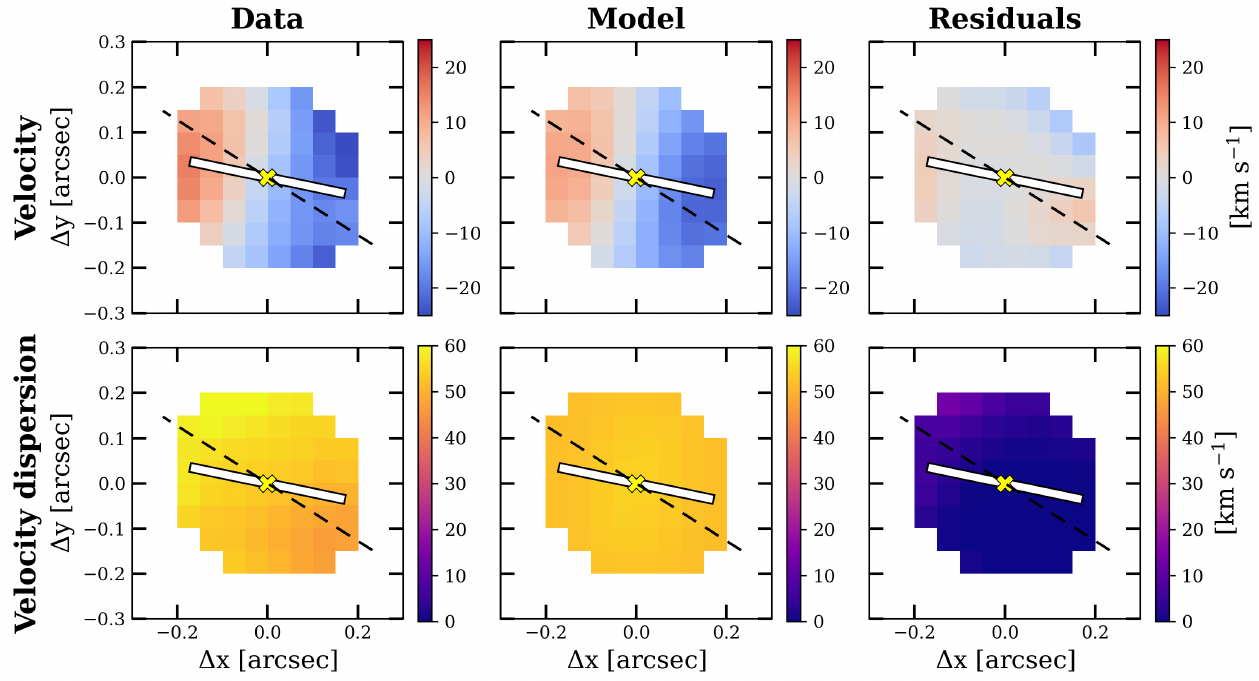}
    \caption{Best-fit results for \oiii\ velocity (top) and velocity dispersion (bottom) from our dynamical modelling of \target. From the left, the maps show the observed data, the model field and the residuals (data -- model). The white segment and the black dashed line indicate the PA\,$=$\,$258.3^\circ$ and the cross slice direction ($\theta$\,$=$\,$237.5^\circ$), respectively, revealing a relative offset of $21^\circ$, which supports the truthfulness of the detected disk rotation against artificial instrumental effects.}
    \label{fig:oiii_kinmod}
\end{figure}

We infer our best-fit model through a Markov chain Monte Carlo (MCMC) approach, using 300 walkers and 500 steps, of which 200 in burn-in phase. This delivers as best-fit parameters total dynamical mass log($M_{\rm dyn}$/\msun)\,$=$\,$9.72^{+0.02}_{-0.02}$, an intrinsic velocity dispersion $\sigma_0=50.24^{+0.87}_{-0.57}$ and a $V_{\rm rot}/\sigma_0=2.0$, consistent with dominant gravity driving as measured in 4\,$<$\,$z$\,$<$\,6 star-forming galaxies \citep{Danhaive:2025,Lee:2025_cristal}, and largely observed in $z$\,$\sim$\,2 galaxies (e.g. \citealt{Forster:2009,Forster:2018,Wisnioski:2015,Ubler:2019,Birkin:2024}). We notice that the uncertainty on $M_{\rm dyn}$ is significantly underestimated, as the inclination in our fitting is fixed, which avoids additional uncertainty due the well known degeneracy between galaxy mass and inclination (e.g. \citealt{Pensabene:2020,Parlanti:2023}). If we leave the inclination free to vary, we obtain an error of about 0.2~dex on $M_{\rm dyn}$, which we assume as our fiducial uncertainty. From the integrated \oiii\ line width, as described in Sect.~\ref{sec:mdyn}, we independently infer a range $M_{\rm dyn}=(1-9) \times 10^9$~\msun, consistent with the results from our \texttt{DysmalPy} fit ($M_{\rm dyn}=5.2^{+3.1}_{-1.9} \times 10^9$~\msun).

The best-fit velocity and velocity dispersion fields are shown in Fig.~\ref{fig:oiii_kinmod}, along with the data and the residual maps. The best-fit model well reproduces the velocity gradient and the relatively flat velocity dispersion distribution observed in the data. Instrumental effects have been noted to potentially introduce artificial velocity gradients in NIRSpec IFU observations of compact sources in the cross slice direction\footnote{See {\it JWST} NIRSpec technical documents \url{https://www.stsci.edu/files/live/sites/www/files/home/jwst/documentation/technical-documents/_documents/JWST-STScI-009029.pdf}.}. For \target, we find a difference of $21^\circ$ between the PA\,$=$\,$258.3^{+4.0}_{-3.8}$ (white segment) and the across slit direction (dashed black line), which supports the truthfulness of the detected disk rotation.

\subsection{Dynamical mass of the sources along the filament}\label{sec:mdyn}

For the sources along the filament, less isolated than \target, we only provide a range of $M_{\rm dyn}$ using the \oiii\ line width measured in the integrated NIRSpec high-resolution spectra ($\sigma_{\rm [OIII]}$ in Table~\ref{tab:specfit_R2700}). According to the virial mass calibration by \citet{vanderWel:2022}, the dynamical mass can be computed as:

\begin{equation}
    M_{\rm dyn}=K(n)~K(q)~\frac{\sigma_\star^2~R_{\rm e}}{G},
\label{eq:2}
\end{equation}

where $K(n)=8.87-0.831n+0.0241n^2$ and $K(q)=(0.87+0.38e^{-3.7(1-q)})^2$, with $n$ and $q$ being the Sérsic index \citep{Cappellari:2006} and the projected axis ratio \citep{vanderWel:2022}, respectively; $\sigma_\star$ is the integrated stellar velocity dispersion, $R_{\rm e}$ is the effective radius, and $G$ is the gravitational constant. All quantities are in SI units. Due to the complex source morphologies and to the differences in their inferred size based on \oiii\ or F444W continuum emission (see Sect.~\ref{sec:sources}), we infer a minimum-to-maximum range of $M_{\rm dyn}$ values by considering $n$\,$=$\,[0.5, 4] and $q$\,$=$\,[0.2, 1] as possible ranges, and both (PSF-corrected) \oiii- and F444W-based $R_{\rm e}$ values inferred for each source (see Table~\ref{tab:morphology}). A $+0.1$~dex logarithmic correction \citep{Bezanson:2018} is then applied to $\sigma_{\rm [OIII]}$ (see Table~\ref{tab:specfit_R2700}) to convert to $\sigma_\star$. With these ingredients, we find approximate ranges of $M_{\rm dyn}\sim10^9-10^{10}$~\msun (see Table~\ref{tab:results} for individual ranges). 

\subsection{Massive sources along the filament}\label{sec:sed_maps}

\begin{figure*}
    \centering
    \includegraphics[width=0.95\linewidth]{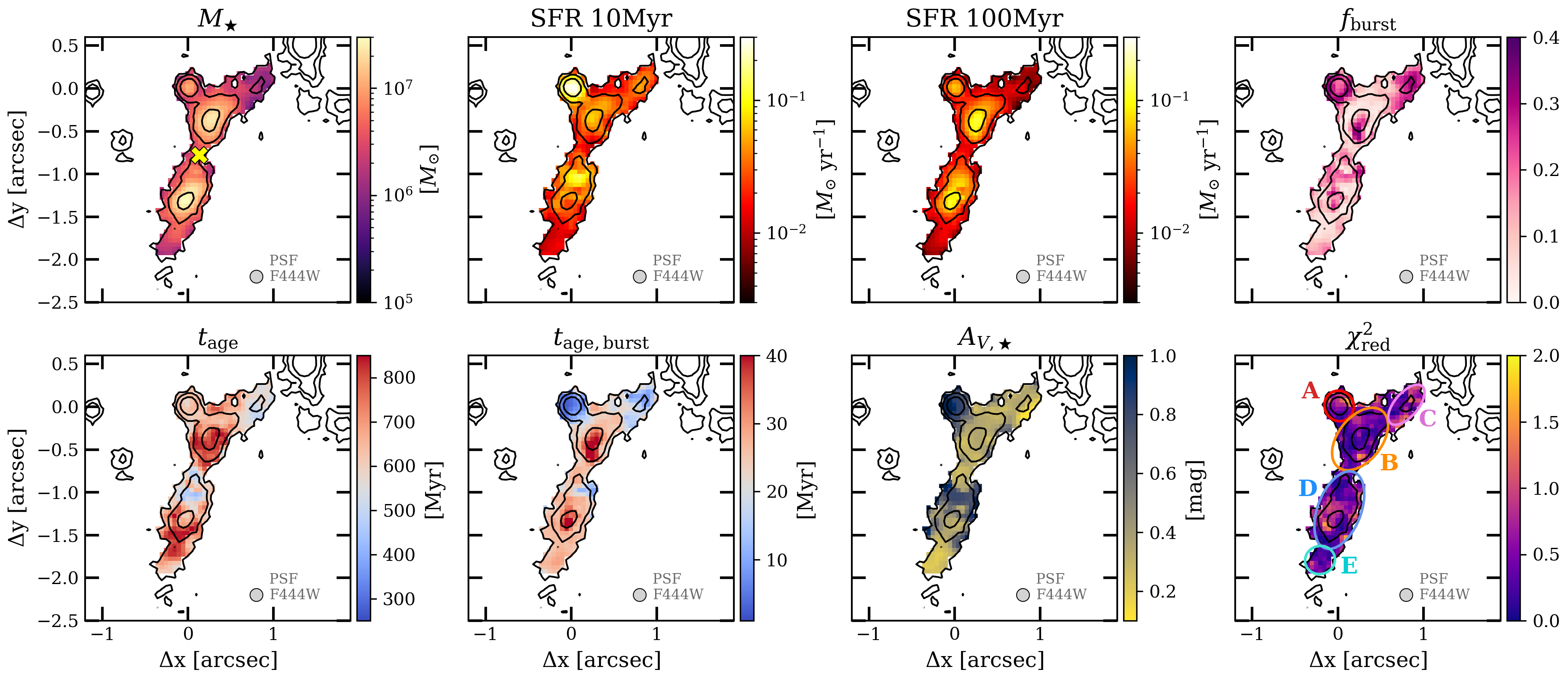}
    \caption{2D maps (0.05$''$ pixel scale) of the main physical properties of \target and its extended environment, as inferred from resolved SED fitting of {\it JWST}/NIRCam imaging with \texttt{CIGALE}. From left, stellar mass, SFRs averaged over the last 10 Myr and 100 Myr, and mass fraction produced during the late burst are shown in the upper row; age of the main and late-burst populations, stellar continuum dust attenuation, and $\chi^2_{\rm red}$ in the lower row. The yellow cross in the $M_\star$ map marks the position of the mass centre of the whole source complex. The black contours are the same as in Fig.~\ref{fig:oiii_map_R2700}. Altogether the maps reveal substantial variations between the various sources and across the complex, in particular with the sources B and D standing out for the massive nature. The inferred $\chi^2_{\rm red}$\,$\approx$\,1 values demonstrate the goodness of our resolved SED modelling.}
    \label{fig:sed_maps}
\end{figure*}

To investigate stellar populations across the source complex, we perform spatially resolved SED fitting of {\it JWST}/NIRCam imaging using the code \texttt{CIGALE} \citep{Boquien:2019,Yang:2020}. The SED of each pixel is modeled independently under the same set of assumptions (details in Appendix~\ref{apx:sed_analysis}), including a delayed exponential star formation history (SFH) with a late (i.e. younger) burst of star formation (SF). We include an AGN component only for those pixels in the region of \target (red aperture in the bottom right map of Fig.~\ref{fig:sed_maps}), although its contribution turns out to be less important than stellar emission (see Fig.~\ref{fig:sed_plots}). Conversely, adding an AGN component to the fit in the other regions results in a negligible (unconstrained) AGN contribution, potentially consistent with either a real lack of AGN emission or an obscured AGN that is overwhelmed by the stellar component at rest-frame UV/optical wavelengths.

Figure~\ref{fig:sed_maps} displays the full 2D maps inferred for the following main physical parameters: stellar mass ($M_{\star}$), SFRs averaged over the last 10~Myr and 100~Myr (SFR10, SFR100), and the fraction of total mass produced during the late starburst episode ($f_{\rm burst}$) in the top panels; the age of the main and late-burst populations ($t_{\rm age}$, $t_{\rm age,burst}$), stellar continuum dust attenuation ($A_{V,\rm \star}$) and the reduced $\chi^2$ ($\chi^2_{\rm red}$) in the bottom panels. The $\chi^2_{\rm red}$ map shows the goodness of our resolved SED fitting ($\chi^2_{\rm red}$\,$\approx$\,1), which reproduces sufficiently well the diverse SED shapes of individual pixels (see Fig.~\ref{fig:sed_plots} for representative best-fit SEDs), by accounting for intrinsic physical differences among distinct regions.

The maps shown in Fig.~\ref{fig:sed_maps} display clear spatial variations in stellar properties across the entire source complex. The 2D mass distribution (top left panel) well follows the intensity of F444W continuum emission (black contours), and points to the sources B and D as the most prominent structures of the source complex. Both sources feature a peak in $M_{\star}$ that coincides with that of enhanced \oiii\ velocity dispersion observed too (see Fig.~\ref{fig:oiii_map_R2700} and related text), thus strengthening the interpretation of such central (high-mass, high-$\sigma$) regions as the kinematic centre of two gravitationally bounded systems. The inner regions of B and D also feature the most intense SFR100 of the entire complex, whereas differences on shorter timescales are observed in the SFR10 map. In particular, D exhibits enhanced recent SF in its most gaseous region, north of its $M_\star$ peak, which could also explain the higher local dust attenuation ($A_{V, \star}$\,$\simeq$\,0.8) compared to the more massive, less star forming southern region ($A_{V, \star}$\,$\simeq$\,0.4). Conversely, B features a flat distribution of relatively low values of recent SFR10 activity. Among all derived physical parameters, $t_{\rm age}$ and $t_{\rm age,burst}$ are the worst constrained, with uncertainties of 30\%\,--\,60\% and 60\%\,--\,80\%, respectively. The stellar population ages are consistent within such uncertainties between the various regions of the complex, with approximate ages $t_{\rm age}$\,$\simeq$\,650~Myr and $t_{\rm age,burst}$\,$\simeq$\,20~Myr, by averaging over the whole source complex.

\renewcommand{\arraystretch}{1.2}
\begin{table*}[]
\caption{Main properties of \target and of the four sources identified along the nearby filament.}
\centering
\footnotesize
\label{tab:results}
\resizebox{0.75\linewidth}{!}{
\begin{tabular}{l|c|ccccc}
\hline\hline

 & Method & A & B & C & D & E \\
\hline\hline

$z_{\rm spec}$ $^{\rm (a)}$ & \oiii\ & 5.229 & 5.230 & 5.230 & 5.227 & 5.229\\
$M_{\rm dyn}$ [10$^9$\,$\times$\,M$_\odot$] $^{\rm (b)}$ & \oiii\ & $5.2^{+3.1}_{-1.9}$ & $4-17$ & $2-14$ & $4-12$ & $7-20$\\
$M_{\star}$ [10$^8$\,$M_\odot$] & SED & $1.7 \pm 0.8$ & $11 \pm 2$ & $0.7 \pm 0.3$ & $13 \pm 3$ & $0.8 \pm 0.3$\\
$M_{\rm gas}$ [10$^8$\,$\times$\,M$_\odot$] $^{\rm (c)}$ & SED & 35 & 27 & 11 & 42 & 2.7\\
$f_{\rm gas}=M_{\rm gas}/(M_{\rm gas}+M_{\star})$ & SED & 0.95 & 0.71 & 0.94 & 0.76 & 0.77 \\
SFR10 [M$_\odot$~yr$^{-1}$] & SED & $5.0 \pm 1.0$ & $3.9 \pm 1.3$ & $1.5 \pm 0.4$ & $6 \pm 2$ & $0.39 \pm 0.14$\\
SFR100 [M$_\odot$~yr$^{-1}$] & SED & $1.0 \pm 0.4$ & $4.4 \pm 1.3$ & $0.4 \pm 0.2$ & $5.0 \pm 1.8$ & $0.33 \pm 0.10$\\
$f_{\rm burst}$ & SED & $0.28 \pm 0.14$ & $<0.20$ & $0.25 \pm 0.17$ & $<0.28$ & $<0.25$\\
$t_{\rm age}$ [Myr] & SED & $630 \pm 320$ & $740 \pm 230$ & $550 \pm 330$ & $650 \pm 250$ & $690 \pm 280$\\
$t_{\rm age,burst}$ [Myr] & SED & $6 \pm 5$ & $26 \pm 17$ & $16 \pm 13$ & $26 \pm 17$ & $27 \pm 17$\\
$A_{V,\star}$ [mag] & SED & $0.83 \pm 0.20$ & $0.34 \pm 0.12$ & $0.33 \pm 0.10$ & $0.52 \pm 0.19$ & $0.23 \pm 0.13$\\
$A_{V, \rm neb}$ [mag] $^{\rm (d)}$ & \ha, \hb & $0.69_{-0.60}^{+0.62}$ & $1.45_{-0.51}^{+0.52}$ & $0.49_{-0.33}^{+0.33}$ & $1.36_{-0.84}^{+0.52}$ & - \\
12\,+\,log(O/H) $^{\rm (e)}$ & [O\,{\sc ii}], \oiii, \hb $^{(\dagger)}$ & $8.26\pm0.15$ & $8.49\pm0.15$ & $7.56_{-0.20}^{+0.26}$ & $7.75_{-0.16}^{+0.12}$ & $>7.0$\\
$M_{\rm BH}$ [10$^7$\,$\times$\,$M_\odot$] $^{\rm (f)}$ & \ha $^{(\dagger)}$ & $1.1^{+0.4}_{-0.3}$ & - & - & - & - \\
$L_{\rm bol}$ [10$^{44}$\,erg s$^{-1}$] $^{\rm (f)}$ & \ha $^{(\dagger)}$ & $2.35^{+0.18}_{-0.13}$ & - & - & - & -\\
$\lambda_{\rm Edd}=L_{\rm bol}/L_{\rm Edd}$ $^{\rm (f)}$ & \ha $^{(\dagger)}$ & $0.18^{+0.10}_{-0.05}$ & - & - & - & -\\
\hline
\end{tabular}
}

\tablefoot{As specified in the column ‘Method', the various properties have been derived from: either emission lines using the \oiii-based apertures inferred in Sect.~\ref{sec:sources}; or SED fitting with \texttt{CIGALE} using the apertures defined in Sect~\ref{sec:sed_maps}, where details on our SED modelling can also be found.\\
$^{\rm (a)}$ Spectroscopic redshift measured from \oiii\ in the integrated NIRSpec IFU high-resolution spectra (\oiii-based apertures in Sect.~\ref{sec:sources}).\\
$^{\rm (b)}$ For A, best-fit value resulting from dynamical modelling of the observed \oiii\ kinematics using \texttt{DysmalPy} (Sect.~\ref{sec:vel_grad}); for the other sources, approximate ranges inferred through the virial calibration by \citet{vanderWel:2022} (Sect.~\ref{sec:mdyn}).\\
$^{\rm (c)}$ Computed as $M_{\rm gas}$\,$=$\,$t_{\rm depl}\times$\,SFR10, assuming $t_{\rm depl}=0.7$~Gyr (Sect.~\ref{sec:sed_maps}).\\
$^{\rm (d)}$ Nebular dust attenuation inferred from narrow \ha/\hb ratios using the \citet{Cardelli:1989} extinction law (Appendix~\ref{apx:ism}).\\
$^{\rm (e)}$ Gas phase metallicities derived from various narrow emission-line ratios using the \citet{Dors:2021} calibration for the sources A and B, which show AGN signatures, and the recent calibrations by \citet{Cataldi:2025} for the other sources (Sect.~\ref{sec:metallicity}).\\
$^{\rm (f)}$ Properties of the AGN in \target (Appendix~\ref{apx:ism}) based on the broad \ha\ best fit in high-resolution NIRSpec MSA data, namely, BH mass ($M_{\rm BH}$; \citealt{Reines:2013}), bolometric luminosity ($L_{\rm bol}$; \citealt{Stern:2012}) and Eddington ratio ($\lambda_{\rm Edd}$).\\
$^{(\star)}$ Based on the \ha narrow flux, inferred by correcting the total one for 0.345 (Appendix~\ref{apx:specfit_msa}).\\
$^{(\dagger)}$ Dust-corrected fluxes using $A_{V, \rm neb}$ (Appendix~\ref{apx:ism}).} 
\end{table*}

To facilitate direct and quantitative comparison between the various regions, we use apertures maximising the source coverage (shown in the $\chi^2_{\rm red}$ map of Fig.~\ref{fig:sed_maps}) to compute representative global properties of each region, by summing (for $M_{\star}$ and SFRs) or averaging (for all other properties) the pixel values within each aperture (all listed in Table~\ref{tab:results}). Consistent results are also obtained by fitting the integrated photometry extracted from the various apertures. The first noteworthy aspect is the difference by more than one order of magnitude between the stellar mass of the various sources. As highlighted by the $M_\star$ map, interestingly, the BL AGN is not hosted by the most massive system of the complex, unlike what found for standard AGN (e.g. \citealt{Allevato:2012,Haines:2012,Powell:2018}). In fact, B and D are the most massive sources ($M_{\star}$\,$\sim$\,10$^9$~\msun), about one order of magnitude more massive than \target ($M_{\star}$\,$=$\,(1.7\,$\pm$\,0.8)\,$\times$\,10$^8$~\msun; see also \citealt{Juodzbalis:2026_jades}); whereas C and E are the least massive systems ($M_{\star}$\,$\simeq$\,$7-8$\,$\times$\,10$^7$~\msun). As a consequence, the centre of mass of this 2D (projected) mass distribution of the complex lies between the massive sources B and D (yellow cross in the $M_\star$ map). Despite the large uncertainties, \target and C might have formed a significant fraction of their stellar mass ($f_{\rm burst}$\,$\sim$\,0.25) during a late burst ($t_{\rm age,burst}$\,$<$\,20~Myr). In \target, such starburst episode has produced a 5~$\times$ higher SFR in the last 10~Myr (SFR10\,$=$\,5.0\,$\pm$\,1.0~\sfr) than in the past (SFR100\,$=$\,1.0\,$\pm$\,0.4~\sfr), which is also likely the cause of the relatively high continuum attenuation in this region ($A_{V, \star}$\,$=$\,0.83\,$\pm$\,0.20).
Unlike \target, the other sources do not exhibit significantly enhanced recent SF.

Because of lack of millimeter observations, we can only indirectly infer an approximate estimate of the gas mass, $M_{\rm gas}$, from the inferred SFR, assuming a certain gas depletion timescale ($t_{\rm depl}$\,$=$\,$M_{\rm gas}$\,/\,SFR), that is the time by when the gas reservoir is depleted by the current SF activity. To a first approximation, $t_{\rm depl}$ depends on $z$, $M_\star$ and SFR \citep{Tacconi:2020}. \citet{Accard:2025} found depletion times of $\sim$\,0.5\,--\,1~Gyr for the 4\,$<$\,$z$\,$<$\,6 star-forming galaxies observed by the CRISTAL survey \citep{Herrera-Camus:2025}. Assuming $t_{\rm depl}$\,$=$\,0.7~Gyr and using the SED-based SFR10 values, we find $M_{\rm gas}$\,$\sim$\,$0.3-4$\,$\times$\,10$^9$~\msun for the various sources, yielding gas fractions $f_{\rm gas}$\,$=$\,$M_{\rm gas}$\,/\,($M_{\rm gas}+M_\star$)\,$\sim$\,0.70\,--\,0.95 (see Table~\ref{tab:results}). These should only be considered as rough estimates, with additional uncertainty introduced by the presence of AGN in \target and possibly in B (illustrated in the next Sect.~\ref{sec:diag}), which could alter gas depletion,
as a consequence of gas heating or ejection due to AGN feedback. 
The resulting total baryonic mass ($M_{\rm bar}$\,$=$\,$M_{\star}$\,$+$\,$M_{\rm gas}$) is consistent or nearly consistent with the inferred $M_{\rm dyn}$ value or range of values for all sources but E, for which $M_{\rm bar}$ is about one order of magnitude lower. This could indicate a dominant DM component or an overestimate of $M_{\rm dyn}$ for E, likely due to non-gravitational contributions to the measured velocity dispersion ($\sigma$\,$\simeq$\,70~\kms against $\sigma$\,$\simeq$\,50~\kms in the other sources; see Table~\ref{tab:specfit_R2700}), for instance, shocks and turbulence associated with accreting gas flows along the filament.

\section{Supporting evidence for a dual AGN}\label{sec:diag}

\begin{figure*}
    \centering
    \includegraphics[width=0.9\linewidth]{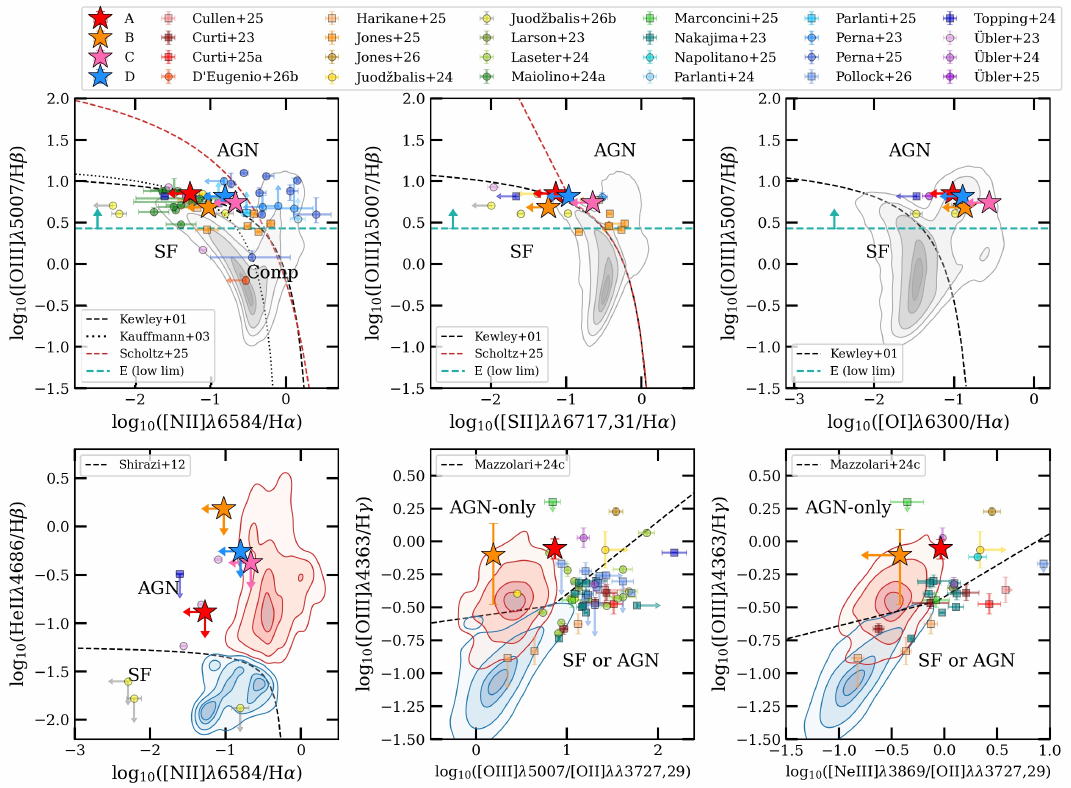}
    \caption{Optical emission-line diagrams displaying \target and its companions (coloured stars), along with other {\it JWST}-based measurements of AGN (circles) and non-AGN galaxies (squares) at $z$\,$>$\,3 from the literature. \textit{Top.} Standard diagnostic diagrams \citep{Baldwin:1981,Veilleux:1987}, separating between SF and AGN ionisation through the demarcation lines by \citet{Kauffmann:2003} (dotted black) and \cite{Kewley:2001} (dashed black). The gray shading indicates low-$z$ galaxies from the SDSS DR7 \citep{Abazajian:2009}, with the contours corresponding to 90, 70, 30, 10 and 1 percent of the sample. The dashed red demarcation curve identifies to its right a more conservative region for high-$z$ objects to be classified as AGN \citep{Scholtz:2025_diag}. For source E, we only plot the lower limit to \oiii/\hb (horizontal dashed turquoise line). \textit{Bottom.} Alternative diagnostic diagrams employing \heii/\hb (left, \citealt{Shirazi:2012}) and \oiii$\lambda$4363/\hg (middle and right, \citealt{Mazzolari:2024}), respectively. Red and blue shaded areas indicate the distribution of local AGN and SF galaxies, respectively, with the contours tracing 90, 70, 30 and 10 percent of the sample: He\,{\sc ii}-classified spatially resolved line ratios (\citealt{Tozzi:2023}; left panel), and BPT-selected AGN and star-forming galaxies from the SDSS DR7 (middle and right panels), used to define the AGN-only region of the two \oiii$\lambda$4363-based diagrams \citep{Mazzolari:2024}. While the classical diagnostic diagram do not provide any conclusive classification, the \oiii$\lambda$4363 auroral diagrams points to AGN ionisation not only in \target, but also in the massive close by source B.}
    \label{fig:diagrams}
\end{figure*}

\subsection{Emission-line diagrams}\label{sec:diagrams}

In this section, we examine the dominant ionisation mechanism of the various sources through multiple narrow emission-line diagrams (Fig.~\ref{fig:diagrams}), comparing them with other {\it JWST} $z$\,$>$\,3 measurements from the literature, including known BL AGN \citep{Larson:2023,Ubler:2023,Ubler:2024,Ubler:2025,Juodzbalis:2024,Juodzbalis:2026_jades,Laseter:2024,Maiolino:2024,Parlanti:2024,Perna:2025,DEugenio:2026_abel,Jones:2026}, confirmed or possible obscured (type 2) AGN \citep{Perna:2023,Perna:2025,Ubler:2024,Napolitano:2025}, and galaxies with no clear (or at least prominent) AGN signature \citep{Curti:2023,Curti:2025,Nakajima:2023,Topping:2024,Cullen:2025,Harikane:2025,Jones:2025,Marconcini:2025,Parlanti:2025a,Pollock:2026}. For our sources, we infer all line ratios from NIRSpec IFU low-resolution fluxes except for \oiii$\lambda$5007/\hb, for which we adopt the more accurate high-resolution ones. For \target and the source B, we also infer \oiii$\lambda$4363 and \hg fluxes from the total measured \oiii$\lambda$4363\,$+$\,\hg flux\footnote{The \oiii$\lambda$4363 and \hg are blended at the spectral resolution of PRISM-CLEAR data ($R$\,$\sim$\,100).} (Appendix~\ref{apx:ism}), after correcting for extinction and assuming a theoretical \hg/\hb\,$=$\,0.466 (recombination case B, low-density limit, $T$\,$=$\,10$^4$ K; \citealt{Osterbrock:2006}). Details on the adopted line ratios and dust corrections are provided in Appendix~\ref{apx:ism}. For \target, the line ratios measured in the NIRSpec IFU data are consistent with the NIRSpec MSA-inferred ones at various spectral resolutions.

The top panels of Fig.~\ref{fig:diagrams} display the classical line ratio diagnostics used to differentiate between SF and AGN activity \citep{Baldwin:1981,Veilleux:1987}. While they have proven to be informative up to $z$\,$\simeq$\,3 for roughly solar-metallicity systems, they are less effective at low metallicity, as pointed out by several models and {\it JWST} observations at $z$\,$>$\,4 (e.g. \citealt{Kewley:2001,Feltre:2016,Hirschmann:2019,Nakajima:2022,Ubler:2023,Maiolino:2024,Scholtz:2025_diag}). In fact, our sources (coloured stars) ‘ambiguously' lie close to the SF/AGN demarcation line (dashed black line; \citealt{Kewley:2001}), along with most of the other high-$z$ {\it JWST} systems, with $x$-axis ratio upper limits. The \nii, \sii and [O\,{\sc i}] non-detections at high redshift can be ascribed to low metallicity and/or a combination of different physical conditions (e.g. stellar and gas phase metallicity, ionisation parameter, electron density; e.g. \citealt{Allen:2008,Feltre:2016,Reddy:2023}). For the source E, only a lower limit to the \oiii/\hb ratio can be inferred (turquoise dashed line).

As a consequence of the high-$z$ low metallicity and diverse physical conditions, effort has been put to develop novel emission-line diagnostics capable of identifying obscured (type 2) and/or faint AGN signatures in high-redshift galaxies. Three of such novel diagrams are shown in the bottom panels of Fig.~\ref{fig:diagrams}: the He\,{\sc ii} diagram (left panel; \citealt{Shirazi:2012}), and two distinct \oiii$\lambda$4363/H$\gamma$-based diagrams (middle and right panels; \citealt{Mazzolari:2024}). The integrated NIRSpec spectra yield He\,{\sc ii}/\hb upper limits for all the four brightest sources of the complex, similar to most of {\it JWST}-discovered AGN (e.g. \citealt{Topping:2024,Juodzbalis:2026_jades}), apart from few individual detections (e.g. \citealt{Perna:2023,Ubler:2023,Scholtz:2025_diag}). 
Being \heii a high-ionisation line insensitive to metallicity, its frequent non-detection could be due to dense gas obscuring AGN inner regions and/or intrinsic differences in the AGN accretion disk at high redshift (e.g. \citealt{Madau:2024,Maiolino:2025_xray,Lambrides:2026}), which prevent efficient production of enough hard ionising photons.

Unlike He\,{\sc ii}, the \oiii$\lambda$4363 auroral line is frequently detected at high redshift (e.g. \citealt{Laseter:2024,Cataldi:2025,Juodzbalis:2026_jades,Sanders:2026}), and acts as an effective, temperature-sensitive diagnostic to disentangle the main excitation mechanism, when combined with emission lines directly related with the ISM ionising conditions (e.g. ionisation parameter, hardness of ionising photons).  
The bottom middle and right panels of Fig.~\ref{fig:diagrams} display two distinct versions of the novel \oiii\ auroral diagram proposed by \citet{Mazzolari:2024}, which combine the \oiii$\lambda$4363/H$\gamma$ ratio with the \oiii$\lambda$5007/[O\,{\sc ii}]$\lambda\lambda$3727,29 and [Ne\,{\sc iii}]$\lambda$3869/[O\,{\sc ii}]$\lambda\lambda$3727,29, respectively. At a fixed ionisation parameter (i.e. roughly at fixed $x$-axis ratio), AGN photoionisation typically leads to much higher \oiii$\lambda$4363/H$\gamma$ ratios, which imply more effective gas heating than photoionisation due to hot stars. The demarcation line (dashed) inferred by \citet{Mazzolari:2024}, based on both photoionisation models \citep{Feltre:2016,Gutkin:2016,Nakajima:2022} and observations at low and high redshift, separate the upper ‘AGN-only' region, where the fraction of star-forming contaminants is expected to be less than 1\,--\,2~\% \citep{Mazzolari:2024}. Due to faintness of the emission lines involved, A and B are the only sources of the complex for which we can infer the needed line ratios (or upper limits). Interestingly, we find both sources to lie in the AGN-only region of the diagrams, where other high-$z$ AGN are found too (e.g. \citealt{Kokorev:2023,Ubler:2024,Jones:2026}).

Whereas the position of \target in the AGN-only region of the \oiii\ auroral diagrams is consistent with the confirmed presence of an AGN in this source, the detection of AGN ionisation also in B could be a consequence of either ‘external' ionisation from \target, or ‘intrinsic' ionisation from a low luminosity and/or highly obscured AGN residing in B. 
The second hypothesis appears particularly plausible considering the massive nature of B ($M_{\star}$\,$\sim$\,10$^9$\,$M_\odot$ from Sect.~\ref{sec:sed_maps}) compared to the rest of the complex, along with its more mature chemical enrichment ($\sim$\,0.6\,$Z_\odot$), as will be inferred in  Sect.~\ref{sec:metallicity}.

\subsection{External or intrinsic AGN ionising source in B?}\label{sec:agn_ion}

\begin{figure*}
    \centering
    \includegraphics[width=0.95\linewidth]{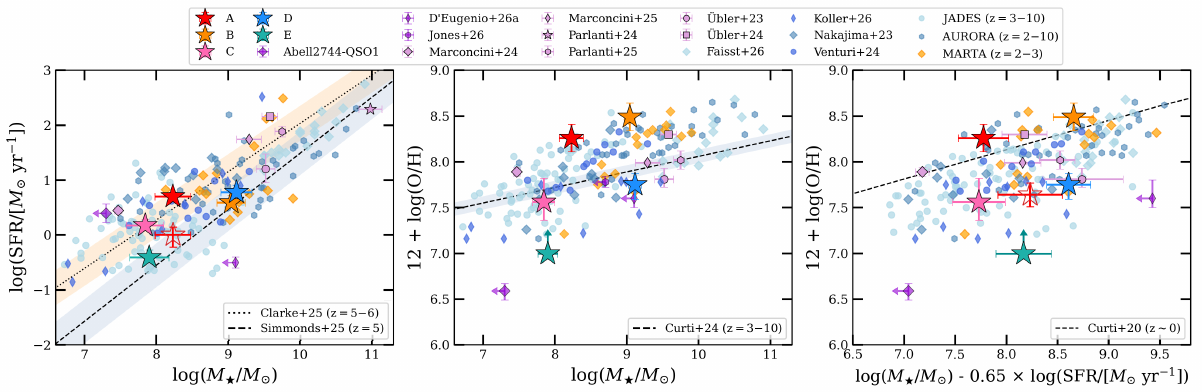}
    \caption{Global scaling relations between stellar mass, SFR (i.e. SFR10) and gas phase metallicity for the multiple sources of the complex (coloured stars), along with other JWST-based high-$z$ measurements from BlackTHUNDER (violet markers), GA-NIFS (pink), and other programs (see Sect.~\ref{sec:metallicity} for the full list of references). The empty red star shows the location of \target based on its (lower) inferred SFR100. Specifically, the left panel shows the $M_\star$\,--\,SFR relation, with two distinct best fits of the star-forming main sequence at $z$\,$\sim$\,5 (dotted and dashed lines; \citealt{Clarke:2025,Simmonds:2025}). The middle panel displays the distribution of measurements with respect to the mass\,--\,metallicity relation at 3\,$<$\,$z$\,$<$\,10 from JADES (dashed line; \citealt{Curti:2024}), while the right panel shows a 2D projected plane of the local FMR (dashed line; \citealt{Curti:2020}). Shaded areas indicate the intrinsic 1$\sigma$ scatter of corresponding best-fit relations. In spite of their diverse properties (about one order of magnitude in $M_\star$, SFR and metallicity), the multiple sources of the complex follow fundamental galaxy scaling relations.}
    \label{fig:mzsfr}
\end{figure*}

In this section we test whether the AGN ionisation revealed in B by the \oiii\ auroral diagrams is due to the AGN in \target (i.e. external source) or to a local ionising source in B (i.e. a second distinct AGN). As recently done by \citet{Perna:2025} to confirm a sample of dual AGN at $z$\,$\sim$\,3, we follow the prescriptions by \citet{Keel:2012} to estimate a lower limit to the incident ionising luminosity from \target (i.e. the primary AGN) required for ionising the gas in B, given the \hb luminosity, $L({\rm H}\beta)$, of B and its distance from \target. This calculation adopts the observed projected separation (smaller to equal than the actual one),
which would imply an even higher incident flux from \target, assuming that all observed \hb emission in B is due to AGN photoionisation.
Assuming a circular geometry for the \hb emitting source in B, its subtended solid angle can be computed as $\theta=2\,{\rm arctan}(r/d)$, where $r$ is its radius and $d$ is its angular distance from the primary AGN, that is \target. For $r$ and $d$, we respectively adopt the \oiii-based $R_{\rm e}$\,$=$\,0.19$''$ and $d_{\rm A}$\,$=$\,2.4~kpc values inferred for B in Sect.~\ref{sec:sources}, relying on the same shared \oiii\ and \hb morphology (see Figs.~\ref{fig:oiii_map_R2700} and \ref{fig:flux_maps}). With these observed quantities, the lower limit to the ionising luminosity required for B can be computed as:



\begin{equation}
    L^{\rm B}_{\rm ion}>L({\rm H}\beta) \times k_{\rm bol} \times \gamma \times \left( \frac{4 \pi}{\theta^2} \right),
    \label{eq:1}
\end{equation}

where $k_{\rm bol}$ is the \hb bolometric correction from \citet{Netzer:2019}; and $\gamma$ is the ionising-to-bolometric luminosity fraction based on the mean radio-quiet SED ($\gamma$\,$=$\,0.14, \citealt{Keel:2019}). With this formula, we find a lower limit of $L^{\rm B}_{\rm ion}$\,$>$\,8.1\,$\times$\,10$^{45}$ erg s$^{-1}$, to be compared with the incident ionising luminosity originating from \target, calculated as $L_{\rm inc}=L_{\rm bol} \times \gamma$ \citep{Keel:2019}. Using our fiducial value of $L_{\rm bol}\simeq2.4\times10^{44}$ erg~s$^{-1}$ inferred for \target from \ha BL emission (\citealt{Stern:2012}; Appendix~\ref{apx:ism}), we obtain $L_{\rm inc}$\,$=$\,3.3\,$\times$\,10$^{43}$ erg s$^{-1}$, which is more than two orders of magnitude lower than the inferred lower limit to $L^{\rm B}_{\rm ion}$. If we consistently use the \citet{Netzer:2019} \hb bolometric correction for \target too, we find $L_{\rm bol}\simeq1.6\times10^{45}$ erg~s$^{-1}$, which yields $L_{\rm inc}$\,$=$\,2.2\,$\times$\,10$^{44}$ erg s$^{-1}$, still lower by one order of magnitude than $L^{\rm B}_{\rm ion}$. Therefore, the AGN-like ionisation observed for B cannot stem from the AGN in \target, thus supporting the presence of a second AGN in the complex, hosted by the massive source B, at just $\sim$\,2.4~kpc (projected) separation from \target. If confirmed, this would be one of the very few dual (or multiple) AGN at a few kiloparsec separations discovered at $z$\,$>$\,3 \citep{Perna:2023,Perna:2025,Ubler:2024,Ubler:2025}.

\section{Heterogeneous galaxy assembly at $z$\,$=$\,5.23}\label{sec:discussion}

The results presented in the previous sections have revealed the diverse nature of the various sources composing the complex, pointing to their mutual influence and possible presence of a second AGN in B. In this section, we discuss this diversity in the framework of possible formation and evolution scenarios of both the individual sources and the complex as a whole.  

\subsection{Diverse chemical enrichment of the multiple sources}\label{sec:metallicity}

Since \oiii$\lambda$4363 flux can only be indirectly inferred for \target and the source B (Sect.~\ref{sec:diagrams}, details in Appendix~\ref{apx:ism}), we opt for the strong-ling method to derive the gas phase metallicity for the various sources. For \target and B, which show AGN signatures, we adopt the metallicity calibration by \citet{Dors:2021}, which combines [O\,{\sc ii}]$\lambda\lambda$3727,29, \oiii$\lambda\lambda$4959,5007 and \hb line ratios; for the others we use the recent calibration by \citet{Cataldi:2025} based on the R3\,$=$\,log$_{10}$(\oiii$\lambda$5007/\hb) diagnostic. We exploit the NIRSpec IFU high-resolution fluxes to calculate R3, checking consistency with the low-resolution R3 values. All derived gas-phase metallicities are reported in Table~\ref{tab:results}.

For \target, we infer independent metallicity values separately from NIRSpec IFU low-resolution data and NIRSpec MSA low- and medium-resolution spectra, finding consistent values within the uncertainties. We adopt as fiducial value the average of the three distinct metallicity values, that is 12\,+\,log(O/H)\,$=$\,$8.26\pm0.15$. For B, the NIRSpec IFU low-resolution fluxes yield a metallicity of $8.49\pm0.15$ (i.e. ($Z$\,$\sim$\,0.6\,$Z_\odot$). For the sources with no AGN signatures, we obtain a lower chemical abundance for C and D, namely, 12\,+\,log(O/H) of $7.56^{+0.26}_{-0.20}$ and $7.75^{+0.12}_{-0.16}$, respectively, and a lower limit to the metallicity of E, where \hb is undetected.
Altogether the multiple sources of the complex reveal diverse chemical enrichment, spanning about one order of magnitude in gas phase metallicity (Fig.~\ref{fig:mzsfr}, middle and right panels), from the subsolar metallicities (12\,+\,log(O/H)\,$\simeq$\,$7.6-7.8$, corresponding to $Z$\,$\sim$\,0.1\,$Z_\odot$) found in C and D at the two opposite edges of the filament, to the higher metallicity ($Z$\,$\sim$\,$0.4-0.6$\,$Z_\odot$) measured for \target and the massive source B at the centre of the filament, where emission-line diagnostics support the presence of an AGN and gas flows from the other sources are seen to converge.

In addition to the differences in the global chemical abundance of the various sources, local metallicity gradients could also be present within each source and across the filament. The 2D map of the relatively dust-insensitive (high-resolution) \oiii/\hb ratio indeed features spatial variations across the multiple sources (Fig.~\ref{fig:flux_maps}), potentially due to local differences in the chemical enrichment and/or ionisation conditions. To break degeneracies between distinct physical properties, we would need spatially resolved maps of multiple emission-line diagnostics, which we cannot infer from the currently available data.


Considering also the properties derived in Sect.~\ref{sec:sed_maps}, the various sources overall differ by stellar mass, SFR and metallicity, spanning about one order of magnitude in all three properties ($M_{\star}\simeq0.7-13\times10^8$~$M_{\odot}$, SFR10\,$\simeq$\,0.4\,--\,6~\sfr, 12\,+\,log(O/H)\,$\simeq$\,$7.6-8.5$). However, all sources individually follow global scaling relations between these galaxy properties (Fig.~\ref{fig:mzsfr}), namely, the star-forming main sequence (left panel), the mass\,--\,metallicity relation (middle), and the fundamental metallicity relation (FMR; a 2D projection of the relation is shown in the right panel). In fact, Fig.~\ref{fig:mzsfr} displays the various sources occupying distinct regions of the various planes, yet still being consistent with the scatter of other {\it JWST}-based measurements at high redshift from the literature, including results from BlackTHUNDER \citep{Ji:2025,DEugenio:2026,DEugenio:2026_abel,Jones:2026,Maiolino:2026_abel}, the NIRSpec Guaranteed Time Observations (GTO) program GA-NIFS (PIs: S. Arribas, R. Maiolino; \citealt{Ubler:2023,Ubler:2024,Marconcini:2024,Marconcini:2025,Parlanti:2024,Parlanti:2025a}), and other high-$z$ observations \citep{Nakajima:2023,Curti:2024,Venturi:2024,Cataldi:2025,Faisst:2026,Koller:2026,Sanders:2026}. Among all sources, B and D are the most massive systems of the complex ($M_\star$\,$\sim$\,$10^9$~\msun) and feature comparable SFRs of a few \sfr. Yet, their different (about one order of magnitude) chemical abundance  and the possible existence of an AGN in B suggest that these two sources are experiencing different evolutionary stages. Despite showing no AGN signature, the massive source D could still host an inactive BH or a low-luminosity AGN that eludes observations.


\subsection{Possible BH formation scenarios for \target}\label{sec:dcbh}

From our broad \ha best fit in the NIRSpec MSA high-resolution data of \target, we infer $M_{\rm BH}$\,$=$\,$1.1^{+0.4}_{-0.3}$\,$\times$\,10$^7$~\msun (Appendix~\ref{apx:ism}). Thus, for its host stellar mass of $M_{\star}=1.7\pm0.8\times10^8$~$M_{\odot}$ (inferred in Sect.~\ref{sec:sed_maps}), \target falls a factor $\sim$\,300 above the local $M_{\rm BH}-M_\star$ relation by \citet{Reines:2015}. According to theoretical models (e.g. \citealt{Volonteri:2021,Schneider:2023}), early overmassive BHs can form from light ($\sim$\,$10^2$~\msun; e.g. \citealt{Bromm:2002,Madau:2014_seeds,Pezzulli:2016}) or intermediate seeds ($\sim$\,$10^{3-4}$~\msun; e.g. \citealt{Devecchi:2009,Davies:2011,Rantala:2025,Vergara:2025}), followed by super- or nearly-Eddington growth; or, alternatively, heavy seeds ($\sim$\,$10^{4-6}$~\msun; e.g. DCBHs, \citealt{Loeb:1994,Begelman:2006,Habouzit:2016}) accreting at sub-Eddington rate.

The frequent detection of UV-bright extended sources at a few kiloparsec from {\it JWST} BL AGN (i.e in $\sim$\,43\% of LRDs; \citealt{Baggen:2026}) has motivated several studies to explore DCBHs as a possible BH formation mechanism for these objects (e.g.~\citealt{Cenci:2025,Baggen:2026,Fei:2026,Pacucci:2026}). In this picture, the nearby extended UV-bright emission has been interpreted as young star-forming companions, which could produce a LW radiation field intense enough to suppress H$_2$ gas cooling \citep{Draine:1996} and trigger rapid direct collapse, with no intermediate gas fragmentation \citep{Loeb:1994,Bromm:2003,Volonteri:2005}. In order to occur, this process requires nearly metal-free gas ($Z$\,$<$\,$10^{-5}$~$Z_{\odot}$) and an incident LW field in the dimensionless form $J_{\rm 21}$ (i.e. specific intensity divided by $10^{21}$ erg~s$^{-1}$~cm$^{-2}$~Hz$^{-1}$~sr$^{-1}$) that exceeds a critical threshold $J_{\rm crit}$\,$\approx$\,$10^3$ \citep{Sugimura:2014,Wolcott-Green:2017}.

The close proximity of the massive filament of sources gives us the opportunity to explore whether this could have enabled BH formation in \target via DCBH, despite the inferred $Z$\,$\sim$\,0.4~$Z_{\odot}$ metallicity (Sect.~\ref{sec:metallicity}), far from metal-free conditions. As observed in many other {\it JWST} BL AGN (e.g. \citealt{Ubler:2023,Ubler:2024,Labbe:2024,DEugenio:2025_iron,Jones:2026}), the enriched chemical abundance of \target could indeed be a consequence of either enriched gas from close companions, or SF taking place during or after the initial collapse (e.g. \citealt{Natarajan:2017}). 
Following \citet{Baggen:2026}, we compute the mean flux density in the rest-frame LW band (91.2\,--\,111~nm) from the best-fit SED model of each pixel outside the aperture associated with \target (i.e. red aperture in the $\chi^2_{\rm red}$ map of Fig.~\ref{fig:sed_maps}). By then converting to intrinsic LW luminosity $L_{\nu,i}^{\rm LW}$ and assuming isotropic emission from each emitting pixel $i$ at projected distance $d_i$, the total LW radiation field at the position of \target can be calculated as:

\begin{equation}
    J_{\nu} = \frac{1}{4\pi} \sum_i \frac{L_{\nu,i}^{\rm LW}}{4\pi d^2_i}.
\end{equation}

Dividing by $10^{21}$ erg~s$^{-1}$~cm$^{-2}$~Hz$^{-1}$~sr$^{-1}$, we find the total LW intensity $J_{\rm 21,LW}$ of about 500 incident on \target, which is lower than the critical threshold predicted for DCBH. Among all sources, B provides the most effective LW radiation field ($J_{\rm 21,LW}$\,$\sim$\,350), this being also the closest source to \target.

This result therefore disfavours BH formation via direct collapse, based on the currently measured total LW radiation field incident on \target from the massive companions along the filament. However, a more intense LW radiation field might have been emitted in the past (for instance, from B), possibly matching the required conditions for direct collapse to take place.

It is worthwhile to reiterate that \target is spatially offset with respect to the filament of sources and, interestingly, does not reside in the most massive galaxy of the complex, unlike standard AGN (e.g. \citealt{Allevato:2012,Haines:2012,Powell:2018}). This invites speculation on alternative scenarios, such as: a primordial black hole \citep{Hawking:1971,Carr:1974,Carr:2024,Dayal:2026}, which plunged into the overdensity and only became visible when it gathered enough gas around it to accrete \citep[see][]{Maiolino:2026_abel}; or a BH that was expelled by one of the massive galaxies in the filament, as a consequence of a gravitational wave kick off or multi-body interactions \citep{Peres:1962,Bekenstein:1973,Madau:2004}, as recently proposed for two other {\it JWST} BL AGN \citep{Juodzbalis:2026_moka,Ubler:2025}.

\begin{figure*}
    \centering
    \includegraphics[width=0.93\linewidth]{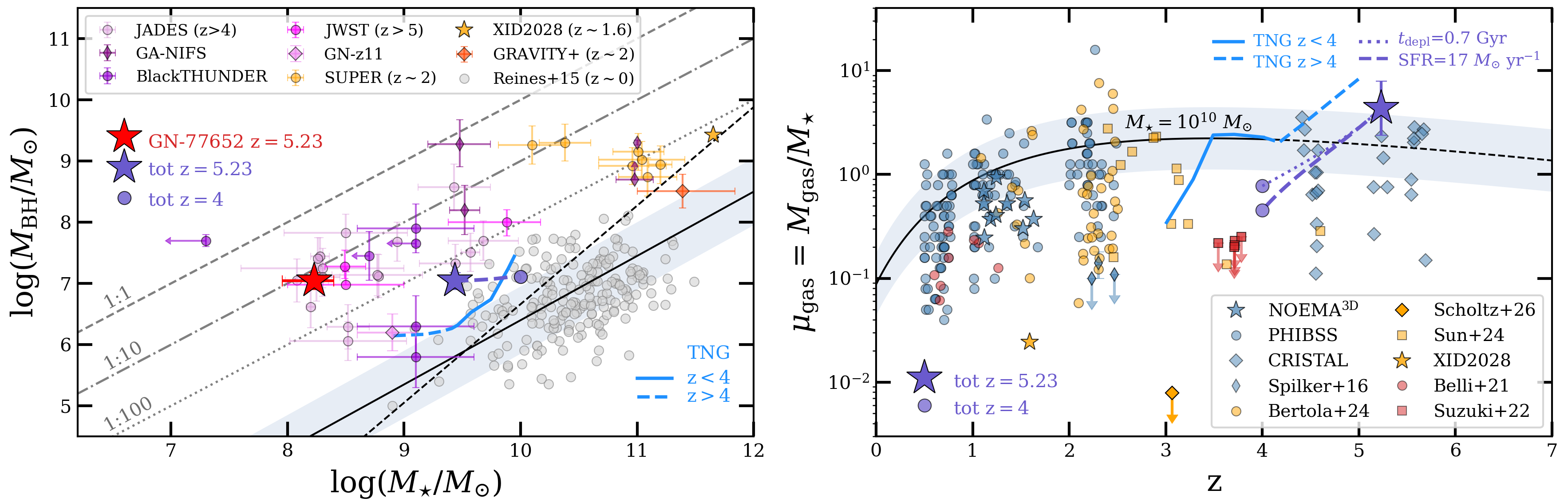}
    \caption{$M_{\rm BH}-M_\star$ (left) and $\mu_{\rm gas}-z$ (right) planes showing the evolution of our source complex at $z$\,$=$\,5.23, as sum of the multiple components (violet star), extrapolated to $z$\,$=$\,4 (violet circle), under our toy-model assumptions of host galaxy growth at either constant SFR\,$=$\,17~\sfr (violet dashed line; the only one shown in the left panel) or constant $t_{\rm depl}$\,$=$\,0.7~Gyr (violet dotted line), and BH growth at constant $\lambda_{\rm Edd}$\,$=$\,0.18 and with an AGN duty cycle of 10\% (Appendix~\ref{apx:toy_model}). The red star (left panel only) shows \target as individually observed at $z$\,$=$\,5.23. In addition, we plot measurements from the literature (full list of references in Sect.~\ref{sec:discussion}): {\it JWST}-discovered AGN (purple/pink), AGN at cosmic noon (yellow/orange) and local AGN (grey) in the left panel; star-forming galaxies (light blue), AGN hosts (yellow) and quiescent systems with no clear AGN signature (red) in the right panel. Black solid and dashed lines in the left panel represent local relations derived by \citet{Reines:2015} and \citet{Greene:2020}, whereas the black solid one in the right panel indicates the $\mu_{\rm gas}$ evolution of a $M_\star$\,$=$\,10$^{10}$~\msun main-sequence galaxy \citep{Tacconi:2020}, extrapolated at $z$\,$>$\,4 (dashed), with its 0.3~dex scatter shown as shaded area. The light blue curve shows the evolution from $z$\,$=$\,5 to $z$\,$=$\,3 of the TNG100 galaxy group analogue, as sum of its multiple components, featuring similar properties to our observed system (Appendix~\ref{apx:simulations}).}
    \label{fig:evolution}
\end{figure*}

\subsection{Rapid evolution to final coalescence}\label{sec:merger_tscale}

The compact configuration of the (12~kpc) extended complex, consisting of multiple massive, kiloparsec-size sources with gas velocity dispersion of about 50~\kms, suggests that all sources are gravitationally bound systems embedded in a common DM halo. In fact, the stellar-to-halo mass relation at $z$\,$=$\,5 by \citet{Shuntov:2025} predicts a halo mass of about $2 \times 10^{11}$~\msun for the most massive sources B and D, yielding an expected virial radius $R_{\rm vir} \sim 30$~kpc and virial velocity $V_{\rm vir} \sim 170$~\kms, which correspond to a dynamical time $\tau_{{\rm dyn}}=R_{\rm vir}/V_{\rm vir}\sim170$~Myr. This strongly supports that the whole source complex is within a common host halo, and will ultimately merge into a more massive galaxy, as predicted by the $\Lambda$CDM hierarchical merging scenario \citep{Lacey:1993,Cole:2000}.

As they in-spiral inward in the halo potential well, galaxies lose energy and angular momentum via dynamical friction \citep{Chandrasekhar:1943}. At kiloparsec separations, strong gravitational torques and multi-body interactions accelerate the merging process (e.g. \citealt{Athanassoula:2000}), with the final time to coalescence depending on the local dynamical timescale (e.g. \citealt{Lacey:1993,Boylan-Kolchin:2008,Solanes:2018}). At high redshift, dynamical friction is even more efficient thanks to the high gas densities and high gas fractions \citep{Tacconi:2020}, with remaining merger times $\tau_{{\rm merger}}$ of few hundreds of Myr \citep{Boylan-Kolchin:2008,Jiang:2008,McCavana:2012,Keitaanranta:2026}. To quantify the expected $\tau_{{\rm merger}}$ for our $z$\,$\simeq$\,5.23 source complex, we use Eq.~9 of \citet{Jiang:2008}, assuming a minimum and a maximum primary-to-satellite mass ratio of 0.1 and 1, based on the inferred $M_{\rm dyn}$ ranges (see Table~\ref{tab:results}), an average separation of 5~kpc between closest pairs of sources, as measured, and a circular velocity $V_{\rm vir} \sim 170$~\kms. We thus find a range of $\tau_{{\rm merger}}=150-440$~Myr, consistent with N-body/hydrodynamical simulations (e.g. \citealt{Lotz:2008}), implying rapid coalescence by $z$\,$\sim$\,4.

Such short merger timescales suggest that these structures could be relatively short lived and rare in the early Universe. To better assess how common groups of multiple massive components are at $z$\,$\sim$\,5, we rely on the Santa Cruz semi-analytic models (SC-SAMs; \citealt{Somerville:1999,Somerville:2015_sam,Somerville:2021,Yung:2019a,Yung:2019b,Yung:2023}). In each of the five SC-SAM lightcone realizations, we search for systems at 4.5\,$<$\,$z$\,$<$\,6 composed of at least three components of $M_\star$\,$>$\,10$^{7}$~\msun, within the same DM halo and relative projected separations $<$\,15~kpc, similarly to our observations. The inferred comoving number density ranges from about $8.0\times10^{-7}$~Mpc$^{-3}$ to about $2.9\times10^{-8}$~Mpc$^{-3}$ for systems consisting as three and seven galaxies, respectively, and goes down to $4.4\times10^{-9}$~Mpc$^{-3}$ for systems with at least eight components. The abundance of three-to-seven component groups implies characteristic separations of $\sim100-330$ comoving Mpc, indicating that such structures are uncommon but not exceedingly rare at $z$\,$\sim$\,5, roughly as abundant as local massive galaxy clusters (typical number densities of $10^{-7}-10^{-6}$~Mpc$^{-3}$ for halo masses larger than $10^{14.5-15}$~\msun (e.g. \citealt{Bahcall:1993,Tinker:2008}). 

\subsection{How could \target and its companions evolve?}\label{sec:evolution}

In this section, we explore whether the compact appearance and overmassive BH of \target just represent a temporary phase, before evolving into a more common system by merging with the filament. In particular, we speculate on the possible evolution of the sum of the multiple components at $z$\,$=$\,5.23 in $M_{\rm BH}$, $M_{\star}$ and $\mu_{\rm gas}$\,$=$\,$M_{\rm gas}$\,/\,$M_\star$ (Fig.~\ref{fig:evolution}), relying on: a toy-model extrapolation until $z$\,$=$\,4 (i.e. over $\sim$\,400~Myr, by the expected final coalescence; Appendix~\ref{apx:toy_model}); and numerical predictions for a similar system (Appendix~\ref{apx:simulations}) identified in the 
100 Mpc cosmological volume of the hydrodynamic simulation IllustrisTNG (i.e. TNG100; \citealt{2018MNRAS.475..676S, 2018MNRAS.475..624N, 2018MNRAS.475..648P,2018MNRAS.477.1206N, 2018MNRAS.480.5113M}). 
As detailed in Appendix~\ref{apx:simulations}, in TNG100 we only identify one isolated group at $z$\,$=$\,5 (group ID 1010 at the snapshot 17), containing at least three sufficiently massive components ($0.5 \times 10^8$~\msun$<$\,$M_{\star}$\,$<$\,$ 20 \times 10^8$~\msun) at small ($<$\,15~kpc) galaxy-to-galaxy separations, as observed for our system. 
The most massive galaxy of the group hosts a BH of $M_{\rm BH}$\,$=$\,$1.4 \times 10^{6}$~\msun, and the entire group coalesces into a single galaxy by $z$\,$=$\,4.4 (i.e. in less than 200~Myr), and later, at $z$\,$=$\,2.9, undergoes a major merger with a much larger system (group ID 98 at snapshot 26), which dramatically affects its subsequent evolution and properties. Therefore, we only follow the evolution of the $z$\,$=$\,5 group 1010 until $z$\,$=$\,3.

Figure~\ref{fig:evolution} displays our toy-model evolution of the source complex from $z$\,$=$\,5.23 to $z$\,$=$\,4 in the $M_{\rm BH}$\,--\,$M_{\star}$ (left) and $\mu_{\rm gas}$\,--\,$z$ (right) planes, along with the TNG100 group analogue and other systems at $z$\,$\sim$\,1 from the literature: star-forming galaxies from \citet{Spilker:2016}, PHIBSS \citep{Tacconi:2018}, CRISTAL \citep{Lee:2025_cristal}, and NOEMA$^{\rm 3D}$ \citep{Jolly:2026}; {\it JWST}-discovered AGN at $z$\,$>$\,3 from JADES \citep{Maiolino:2024,Juodzbalis:2026_jades}, GA-NIFS \citep{Perna:2023,Ubler:2023,Parlanti:2024,Marshall:2025}, BlackTHUNDER \citep{Ubler:2025,Juodzbalis:2026_moka,DEugenio:2026,Jones:2026} and others at $z$\,$>$\,5 \citep{Tripodi:2025,Maiolino:2024_GNz11,Akins:2025,Rinaldi:2025b}; other AGN at $z$\,$\sim$\,1\,--\,4 \citep{Brusa:2018,Kakkad:2020,Cresci:2023,Bertola:2024,Sun:2024,Scholtz:2026}, including a bright AGN at $z$\,$\sim$\,2 with dynamical measurement of $M_{\rm BH}$ obtained with GRAVITY+ \citep{Abuter:2024}; finally, some quiescent systems with no clear AGN signature \citep{Belli:2021,Suzuki:2022}.

As shown in the left panel of Fig.~\ref{fig:evolution}, the sum of the multiple components of the $z$\,$=$\,5.23 complex corresponds to a total $M_{\rm BH}$/$M_{\star}$\,$=$\,0.004 (violet star), compatible with the scatter of the local AGN distribution \citep{Reines:2015}, as opposed to \target if considered alone ($M_{\rm BH}$/$M_{\star}$\,$=$\,0.06; red star). If we let the system to grow at constant SFR\,$=$\,17~\sfr and $\lambda_{\rm Edd}$\,$=$\,0.18, assuming a 10\% AGN duty cycle (details in Appendix~\ref{apx:toy_model}), we obtain $M_{\star}$\,$=$\,$1.0$\,$\times$\,10$^{10}$~\msun and $M_{\rm BH}$\,$=$\,$1.3$\,$\times$\,10$^7$~\msun at $z$\,$=$\,4, indicating a growth by only 17\% and by a factor of 2 for the BH and the host galaxy, respectively. This corresponds to a $M_{\rm BH}$/$M_{\star}$\,$=$\,0.001 ratio at $z$\,$=$\,4 (violet circle), which makes the system compatible with the local scaling relations \citep{Reines:2015,Greene:2020}. Assuming galaxy growth at constant $t_{\rm depl}$\,$=$\,0.7~Gyr, we find a comparable stellar mass at $z$\,$=$\,4 ($M_{\star}$\,$=$\,$8.2$\,$\times$\,10$^9$~\msun, not shown in this plot). The TNG100 group undergoes a $M_{\rm BH}$ and $M_{\star}$ growth by 50\% and about a factor of 3 between $z$\,$=$\,5 and 4 (light blue dashed line), respectively, similar to our toy-model extrapolation. At 3\,$<$\,$z$\,$<$\,4 (light blue solid line), the group next experiences a much more intense BH growth (by a factor of 12), while the host galaxy mass still grows by a factor of 2 only. Although we do not follow its evolution at $z$\,$<$\,3, we notice that at $z$\,$=$\,0 the simulated system features $M_{\rm BH}=3.0\times10^9$~\msun and $M_{\star}=8.1\times10^{11}$~\msun, corresponding to $M_{\rm BH}/M_\star \sim 0.004$, thus lying close to the $M_{\rm BH}-M_\star$ relation derived for local massive systems by \citet{Kormendy:2013}.

As discussed in Sects.~\ref{sec:diag} and \ref{sec:metallicity}, additional BHs could reside in the massive source B and possibly, in the form of an inactive BH, in the other massive source D too. Given the small kiloparsec separations, multiple massive BHs would be expected to merge rapidly, within a few tens to hundreds of Myr \citep{Keitaanranta:2026}. If B and D hosted standard BHs following the local $M_{\rm BH}$\,--\,$M_{\star}$ relation by \citet{Reines:2015}, these would have masses of only $M_{\rm BH}$\,$\sim$\,$10^5$~\msun, much smaller than the BH hosted by \target ($M_{\rm BH}$\,$\simeq$\,$1.1\times10^7$~\msun). Therefore, they would play just a minor role in the subsequent BH growth of the global merged system. Conversely, assuming for B and D the same $M_{\rm BH}$/$M_{\star}$\,$=$\,0.06 and $\lambda_{\rm Edd}$\,$=$\,0.18 as measured for \target, our toy model would deliver a BH of $M_{\rm BH}$\,$\sim$\,$1.8\times10^8$~\msun at $z$\,$=$\,4, well above the local $M_{\rm BH}$\,--\,$M_{\star}$ relation by more than one order of magnitude, corresponding to $M_{\rm BH}$/$M_{\star}$\,$=$\,0.02.

In terms of relative gas to stellar content (Fig.~\ref{fig:evolution}, right panel), the whole complex at $z$\,$=$\,5.23 corresponds to $\mu_{\rm gas}$\,$=$\,4.3, consistent with CRISTAL star-forming galaxies at 4\,$<$\,$z$\,$<$\,6 \citep{Lee:2025_cristal}. According to our toy model at constant $t_{\rm depl}$\,$=$\,0.7~Gyr (violet dotted line) or constant SFR\,$=$\,17~\sfr (violet dashed line), the gas reservoir of the overall system, within $z$\,$=$\,4, will get depleted by about a factor of 6 and 10, respectively, that is well below the expectations for a $M_{\star}$\,$=$\,$10^{10}$~\msun main-sequence galaxy, yet comparable with other AGN at $z$\,$\sim$\,3. In comparison, the TNG100 group features a similar decreasing trend until $z$\,$=$\,4.2, although with absolute $\mu_{\rm gas}$ values larger by a $2-3$ factor. From $z$\,$=$\,4.2 to $z$\,$=$\,3.5, the trend in $\mu_{\rm gas}$ flattens around a value of 2, as the system starts interacting with a more massive group, with which it will merge after $z$\,$=$\,3. Such interaction funnels gas into the group while increasing the SFR and thus maintaining $\mu_{\rm gas}$ constant. The intense BH growth starting at $z$\,$=$\,4 (Fig.~\ref{fig:evolution}, left panel) is likely triggered by such gas accretion, which also leads to reduced gas reservoir and SF after $z$\,$=$\,3.5 ($\mu_{\rm gas}$\,$=$\,0.34 at $z$\,$=$\,3), possibly due to AGN feedback at play. 

Although we cannot robustly constrain the future evolution of the system, the merger with the nearby (diverse and extended) sources is likely to drive the transition of \target from its current compact, red appearance to a more typical AGN. A similar environment-driven evolutionary scenario has recently been proposed by \citet{Merida:2026} for a possible emerging or fading LRD, as part of a galaxy group at $z=5.12$. Such a transient evolutionary phase would be consistent with the apparent decline in the number density of compact, red AGN identified with {\it JWST} from the first few billion years to the present day \citep{Kocevski:2025,Inayoshi:2025,Ma:2026_ndensity}.

\section{Conclusions}\label{sec:conclusions}
In this paper, we studied the large-scale environment of \target (Fig.~\ref{fig:rgb}, dubbed A), a compact BL AGN at $z$\,$=$5.229 with a nearby (a few kpc apart) extended ($\sim$\,12~kpc) filament of four main kiloparsec-size sources (B, C, D and E) at the same redshift of \target. By combining {\it JWST}/NIRSpec IFU spectroscopy and multi-band {\it JWST}/NIRCam imaging, we investigated the morphology, gas kinematics and the physical properties of the individual sources, and their mutual connection. The main results are summarised in the following:

\begin{itemize}
    \item The \oiii\ and F444W continuum emission of the various sources exhibit different morphologies and/or spatial offsets (Fig.~\ref{fig:fig2}), partly as a possible consequence of the dissipative and collisionless behaviour of gas and stars during gravitational interactions, respectively.

    \item The large-scale \oiii\ kinematics reveals a smooth South-to-North velocity gradient ($\Delta v<70$~\kms; Fig.~\ref{fig:oiii_map_R2700}) across the entire complex, centred on the source B, which is one of the two most massive sources (along with D, $M_\star\sim10^9$~\msun) and likely residing at the centre of the complex in the 3D space. The continuous filamentary structure also suggests possible gas flows converging into B, both along the main NW-SE filament and from \target, as an alternative explanation to the observed kinematic pattern. 

    \item We detect a shallow velocity gradient ($-30$ to $+20$~\kms) in \target (Fig.~\ref{fig:oiii_kinmod}), indicative of resolved disk rotation. Our \oiii\ dynamical modelling yields $V_{\rm rot}/\sigma_0=2.0$ and $M_{\rm dyn}=5.2^{+3.1}_{-1.9}\times10^9$~\msun, consistent with $M_{\star}+M_{\rm gas}$.

    \item The various sources feature diverse properties (Figs.~\ref{fig:sed_maps} and \ref{fig:mzsfr}), despite individually following main galaxy scaling relations. In particular, they span about one order of magnitude in stellar mass ($M_{\star}=0.7-13 \times 10^8$~\msun), SFR\,$=$\,0.4\,--\,6~\sfr and gas phase metallicity (12\,+\,log(O/H)\,$=$\,7.6\,--\,8.5), suggesting they are in distinct evolutionary stages.
  
    \item The source B is one of the most massive systems ($M_{\star}=(11\pm2) \times 10^8$~$M_{\odot}$) and the metal-richest  one of the complex (12\,+\,log(O/H)\,$=$\,8.49\,$\pm$\,0.15), only 2.4~kpc apart (in projection) from \target. The \oiii$\lambda$4363 auroral diagrams also show evidence for AGN ionisation in this source (Fig.~\ref{fig:diagrams}), likely due to another AGN residing in B. If confirmed, this would be one of the few dual (or multiple) AGN at a few kiloparsec separations discovered at $z$\,$>$\,3 so far.

    \item The LW field emitted by the filament is too weak for the BH in \target to have formed as a DCBH. More favourable conditions for direct collapse could have taken place in the past, or other mechanisms, such as a PBH or a BH ejected from the filament via a gravitational wave recoil or multi-body interactions, could also be possible.

    \item The multiple sources should coalesce in 150\,--\,440~Myr (i.e. by $z$\,$\sim$\,4). Up to that point, according to our toy model, they may evolve towards the local $M_{\rm BH}-M_\star$ relation and rapidly deplete gas (Fig.~\ref{fig:evolution}), whereas numerical predictions for a similar galaxy group hint at a more complex evolution, where later major mergers with other groups could play a role too. Overall, the compact BL AGN appearance of \target most likely represents a transient evolutionary phase.
        
\end{itemize} 

Future {\it JWST}/NIRSpec IFU follow-ups targeting multiple emission lines at medium or high spectral resolution will be fundamental to obtain accurate spatially resolved emission-line diagnostics across the whole source complex. These will allow us to robustly constrain metallicity gradients and the presence of additional AGN along the filament, providing crucial insights into the heterogeneous assembly history of this early system and into the possible evolution of high-$z$ {\it JWST}-discovered AGN.

\begin{acknowledgements}
G.T., N.M.F.S. and C.B. acknowledge funding by the European Union (ERC Advanced Grant GALPHYS, 101055023).
H.\"U. acknowledges support by the Max Planck Society through the Lise Meitner Excellence Program. H.\"U. and G.M. acknowledge funding by the European Union (ERC APEX, 101164796). A.J.B. acknowledges funding from the “FirstGalaxies” Advanced Grant from the European Research Council (ERC) under the European Union’s Horizon 2020 research and innovation program (Grant agreement No. 789056). Views and opinions expressed are, however, those of the authors only and do not necessarily reflect those of the European Union or the European Research Council. Neither the European Union nor the granting authority can be held responsible for them. E.B. acknowledges the support of Ricerca Fondamentale 2024” INAF program (INAF GO grant ``A JWST/MIRI MIRACLE: Mid-IR Activity of Circumnuclear Line Emission'' and mini-grant 1.05.24.07.01). T.N. acknowledges
support from the Deutsche Forschungsgemeinschaft (DFG, German
Research Foundation) under Germany’s Excellence Strategy -
EXC-2094 - 390783311 from the DFG Cluster of Excellence "ORIGINS".
M.Pa. acknowledges funding through the German Aerospace Center (DLR) under grant number 50OR2514. M.Pe. acknowledges support through the grants PID2021-127718NB-I00, PID2024-159902NA-I00, and RYC2023-044853-I, funded by the Spain Ministry of Science and Innovation/State Agency of Research MCIN/AEI/10.13039/501100011033 and El Fondo Social Europeo Plus FSE+. L.S. acknowledges the financial support from the PhD grant funded on PNRR Funds Notice No. 3264 28-12-2021 PNRR M4C2 Reference IR0000034 STILES Investment 3.1 CUP C33C22000640006. G.V. acknowledges support from the Italian National Institute for Astrophysics (INAF) under the IAF - Astrophysics Fellowships in Italy grant CUP C59J21034720001 - ``AD MAJORA'' and European Union's HE ERC Starting Grant No. 101040227 - WINGS.

\end{acknowledgements}

\bibliographystyle{aa}
\bibliography{aa}

@ARTICLE{Geris:2026,
       author = {{Geris}, Sophia and {Maiolino}, Roberto and {Ji}, Xihan and {Risaliti}, Guido and {Lanzuisi}, Giorgio and {D'Eugenio}, Francesco and {Isobe}, Yuki and {Jones}, Gareth and {Harshan}, Anishya and {Brazzini}, Matilde and {Juod{\v{z}}balis}, Ignas and {Scholtz}, Jan and {Rinaldi}, Pierluigi and {{\"U}bler}, Hannah and {Baker}, William and {Bunker}, Andrew J. and {Brusa}, Marcella and {Carniani}, Stefano and {Charlot}, Stephane and {Curti}, Mirko and {Comastri}, Andrea and {Lake}, Emma Curtis and {Gilli}, Roberto and {Hainline}, Kevin and {Madau}, Piero and {Marchesi}, Stefano and {Mazzolari}, Giovanni and {Napolitano}, Lorenzo and {Parlanti}, Eleonora and {Pentericci}, Laura and {Ramos Almeida}, Cristina and {Robertson}, Brant and {Silcock}, Maddie S. and {Tripodi}, Roberta and {Venturi}, Giacomo and {Vignali}, Cristian and {Vito}, Fabio and {Zhu}, Yongda},
        title = "{Little Red and Blue Dots: AGN-excited narrow lines, Lyman-$α$ emission, and resemblance to standard quasars}",
      journal = {arXiv e-prints},
         year = 2026,
        month = jun,
          eid = {arXiv:2606.21614},
        pages = {arXiv:2606.21614},
          doi = {10.48550/arXiv.2606.21614},
archivePrefix = {arXiv},
       eprint = {2606.21614},
 primaryClass = {astro-ph.GA},
       adsurl = {https://ui.adsabs.harvard.edu/abs/2026arXiv260621614G}
}

@ARTICLE{2015MNRAS.449...49R,
       author = {{Rodriguez-Gomez}, Vicente and {Genel}, Shy and {Vogelsberger}, Mark and {Sijacki}, Debora and {Pillepich}, Annalisa and {Sales}, Laura V. and {Torrey}, Paul and {Snyder}, Greg and {Nelson}, Dylan and {Springel}, Volker and {Ma}, Chung-Pei and {Hernquist}, Lars},
        title = "{The merger rate of galaxies in the Illustris simulation: a comparison with observations and semi-empirical models}",
      journal = {\mnras},
         year = 2015,
        month = may,
       volume = {449},
       number = {1},
        pages = {49-64},
          doi = {10.1093/mnras/stv264},
archivePrefix = {arXiv},
       eprint = {1502.01339},
 primaryClass = {astro-ph.GA},
       adsurl = {https://ui.adsabs.harvard.edu/abs/2015MNRAS.449...49R}
}

@ARTICLE{DEugenio:2025,
       author = {{D'Eugenio}, Francesco and {Cameron}, Alex J. and {Scholtz}, Jan and {Carniani}, Stefano and {Willott}, Chris J. and {Curtis-Lake}, Emma and {Bunker}, Andrew J. and {Parlanti}, Eleonora and {Maiolino}, Roberto and {Willmer}, Christopher N.~A. and {Jakobsen}, Peter and {Robertson}, Brant E. and {Johnson}, Benjamin D. and {Tacchella}, Sandro and {Cargile}, Phillip A. and {Rawle}, Tim and {Arribas}, Santiago and {Chevallard}, Jacopo and {Curti}, Mirko and {Egami}, Eiichi and {Eisenstein}, Daniel J. and {Kumari}, Nimisha and {Looser}, Tobias J. and {Rieke}, Marcia J. and {Rodr{\'\i}guez Del Pino}, Bruno and {Saxena}, Aayush and {{\"U}bler}, Hannah and {Venturi}, Giacomo and {Witstok}, Joris and {Baker}, William M. and {Bhatawdekar}, Rachana and {Bonaventura}, Nina and {Boyett}, Kristan and {Charlot}, Stephane and {Danhaive}, A. Lola and {Hainline}, Kevin N. and {Hausen}, Ryan and {Helton}, Jakob M. and {Ji}, Xihan and {Ji}, Zhiyuan and {Jones}, Gareth C. and {Juod{\v{z}}balis}, Ignas and {Maseda}, Michael V. and {P{\'e}rez-Gonz{\'a}lez}, Pablo G. and {Perna}, Michele and {Pusk{\'a}s}, D{\'a}vid and {Shivaei}, Irene and {Silcock}, Maddie S. and {Simmonds}, Charlotte and {Smit}, Renske and {Sun}, Fengwu and {Villanueva}, Natalia C. and {Williams}, Christina C. and {Zhu}, Yongda},
        title = "{JADES Data Release 3: NIRSpec/Microshutter Assembly Spectroscopy for 4000 Galaxies in the GOODS Fields}",
      journal = {\apjs},
         year = 2025,
        month = mar,
       volume = {277},
       number = {1},
          eid = {4},
        pages = {4},
          doi = {10.3847/1538-4365/ada148},
archivePrefix = {arXiv},
       eprint = {2404.06531},
 primaryClass = {astro-ph.GA},
       adsurl = {https://ui.adsabs.harvard.edu/abs/2025ApJS..277....4D}
}

@ARTICLE{DEugenio:2025_iron,
       author = {{D'Eugenio}, Francesco and {Nelson}, Erica and {Ji}, Xihan and {Baggen}, Josephine and {Greene}, Jenny and {Labb{\'e}}, Ivo and {Pezzulli}, Gabriele and {Brown}, Vanessa and {Maiolino}, Roberto and {Matthee}, Jorryt and {Terlevich}, Elena and {Terlevich}, Roberto and {Torralba}, Alberto and {Carniani}, Stefano},
        title = "{Irony at z=6.68: a bright AGN with forbidden Fe emission and multi-component Balmer absorption}",
      journal = {arXiv e-prints},
         year = 2025,
        month = sep,
          eid = {arXiv:2510.00101},
        pages = {arXiv:2510.00101},
          doi = {10.48550/arXiv.2510.00101},
archivePrefix = {arXiv},
       eprint = {2510.00101},
 primaryClass = {astro-ph.GA},
       adsurl = {https://ui.adsabs.harvard.edu/abs/2025arXiv251000101D}
}

@ARTICLE{Schneider:2023,
       author = {{Schneider}, Raffaella and {Valiante}, Rosa and {Trinca}, Alessandro and {Graziani}, Luca and {Volonteri}, Marta and {Maiolino}, Roberto},
        title = "{Are we surprised to find SMBHs with JWST at z {\ensuremath{\geq}} 9?}",
      journal = {\mnras},
         year = 2023,
        month = dec,
       volume = {526},
       number = {3},
        pages = {3250-3261},
          doi = {10.1093/mnras/stad2503},
archivePrefix = {arXiv},
       eprint = {2305.12504},
 primaryClass = {astro-ph.GA},
       adsurl = {https://ui.adsabs.harvard.edu/abs/2023MNRAS.526.3250S}
}

@ARTICLE{Scholtz:2026,
       author = {{Scholtz}, Jan and {D'Eugenio}, Francesco and {Maiolino}, Roberto and {P{\'e}rez-Gonz{\'a}lez}, Pablo G. and {Circosta}, Chiara and {Tacchella}, Sandro and {Williams}, Christina C. and {Alberts}, Stacey and {Arribas}, Santiago and {Baker}, William M. and {Bertola}, Elena and {Bunker}, Andrew J. and {Carniani}, Stefano and {Charlot}, Stephane and {Cresci}, Giovanni and {Jones}, Gareth C. and {Kumari}, Nimisha and {Lamperti}, Isabella and {Looser}, Tobias J. and {Pino}, Bruno Rodr{\'\i}guez Del and {Robertson}, Brant and {Parlanti}, Eleonora and {Perna}, Michele and {{\"U}bler}, Hannah and {Venturi}, Giacomo and {Witstok}, Joris},
        title = "{Measurement of the gas consumption history of a massive quiescent galaxy}",
      journal = {Nature Astronomy},
         year = 2026,
        month = mar,
       volume = {10},
        pages = {431-439},
          doi = {10.1038/s41550-025-02751-z},
       adsurl = {https://ui.adsabs.harvard.edu/abs/2026NatAs..10..431S}
}

@ARTICLE{Scholtz:2026_jades,
       author = {{Scholtz}, J. and {Carniani}, S. and {Parlanti}, E. and {D'Eugenio}, F. and {Curtis-Lake}, E. and {Jakobsen}, P. and {Bunker}, A.~J. and {Cameron}, A.~J. and {Arribas}, S. and {Baker}, W.~M. and {Charlot}, S. and {Chevellard}, J. and {Circosta}, C. and {Curti}, M. and {Duan}, Q. and {Eisenstein}, D.~J. and {Hainline}, K. and {Ji}, Z. and {Johnson}, B.~D. and {Jones}, G.~C. and {Kumari}, N. and {Maiolino}, R. and {Maseda}, M.~V. and {Perna}, M. and {P{\'e}rez-Gonz{\'a}lez}, P.~G. and {Rawle}, T. and {Rieke}, M. and {Rinaldi}, P. and {Robertson}, B. and {Saxena}, A. and {Shivaei}, I. and {Silcock}, M.~S. and {Sun}, Y. and {Rodr{\'\i}guez Del Pino}, B. and {Tacchella}, S. and {{\"U}bler}, H. and {Venturi}, G. and {Williams}, C.~C. and {Willmer}, C.~N.~A. and {Willott}, C. and {Witstok}, J.},
        title = "{JADES Data Release 4 ─ Paper II. Data reduction, analysis, and emission-line fluxes of the complete spectroscopic sample}",
      journal = {\mnras},
         year = 2026,
        month = jul,
       volume = {549},
       number = {4},
          eid = {stag939},
        pages = {stag939},
          doi = {10.1093/mnras/stag939},
archivePrefix = {arXiv},
       eprint = {2510.01034},
 primaryClass = {astro-ph.GA},
       adsurl = {https://ui.adsabs.harvard.edu/abs/2026MNRAS.549ag939S}
}

@ARTICLE{Taylor:2025,
       author = {{Taylor}, Anthony J. and {Finkelstein}, Steven L. and {Kocevski}, Dale D. and {Jeon}, Junehyoung and {Bromm}, Volker and {Amor{\'\i}n}, Ricardo O. and {Arrabal Haro}, Pablo and {Backhaus}, Bren E. and {Bagley}, Micaela B. and {Banados}, Eduardo and {Bhatawdekar}, Rachana and {Brooks}, Madisyn and {Calabr{\`o}}, Antonello and {Ch{\'a}vez Ortiz}, {\'O}scar A. and {Cheng}, Yingjie and {Cleri}, Nikko J. and {Cole}, Justin W. and {Davis}, Kelcey and {Dickinson}, Mark and {Donnan}, Callum and {Dunlop}, James S. and {Ellis}, Richard S. and {Fern{\'a}ndez}, Vital and {Fontana}, Adriano and {Fujimoto}, Seiji and {Giavalisco}, Mauro and {Grazian}, Andrea and {Guo}, Jingsong and {Hathi}, Nimish P. and {Holwerda}, Benne W. and {Hirschmann}, Michaela and {Inayoshi}, Kohei and {Kartaltepe}, Jeyhan S. and {Khusanova}, Yana and {Koekemoer}, Anton M. and {Kokorev}, Vasily and {Larson}, Rebecca L. and {Leung}, Gene C.~K. and {Lucas}, Ray A. and {McLeod}, Derek J. and {Napolitano}, Lorenzo and {Onoue}, Masafusa and {Pacucci}, Fabio and {Papovich}, Casey and {P{\'e}rez-Gonz{\'a}lez}, Pablo G. and {Pirzkal}, Nor and {Somerville}, Rachel S. and {Trump}, Jonathan R. and {Wilkins}, Stephen M. and {Yung}, L.~Y. Aaron and {Zhang}, Haowen},
        title = "{Broad-line AGNs at 3.5 < z < 6: The Black Hole Mass Function and a Connection with Little Red Dots}",
      journal = {\apj},
         year = 2025,
        month = jun,
       volume = {986},
       number = {2},
          eid = {165},
        pages = {165},
          doi = {10.3847/1538-4357/add15b},
archivePrefix = {arXiv},
       eprint = {2409.06772},
 primaryClass = {astro-ph.GA},
       adsurl = {https://ui.adsabs.harvard.edu/abs/2025ApJ...986..165T}
}

@ARTICLE{Tacconi:2018,
       author = {{Tacconi}, L.~J. and {Genzel}, R. and {Saintonge}, A. and {Combes}, F. and {Garc{\'\i}a-Burillo}, S. and {Neri}, R. and {Bolatto}, A. and {Contini}, T. and {F{\"o}rster Schreiber}, N.~M. and {Lilly}, S. and {Lutz}, D. and {Wuyts}, S. and {Accurso}, G. and {Boissier}, J. and {Boone}, F. and {Bouch{\'e}}, N. and {Bournaud}, F. and {Burkert}, A. and {Carollo}, M. and {Cooper}, M. and {Cox}, P. and {Feruglio}, C. and {Freundlich}, J. and {Herrera-Camus}, R. and {Juneau}, S. and {Lippa}, M. and {Naab}, T. and {Renzini}, A. and {Salome}, P. and {Sternberg}, A. and {Tadaki}, K. and {{\"U}bler}, H. and {Walter}, F. and {Weiner}, B. and {Weiss}, A.},
        title = "{PHIBSS: Unified Scaling Relations of Gas Depletion Time and Molecular Gas Fractions}",
      journal = {\apj},
         year = 2018,
        month = feb,
       volume = {853},
       number = {2},
          eid = {179},
        pages = {179},
          doi = {10.3847/1538-4357/aaa4b4},
archivePrefix = {arXiv},
       eprint = {1702.01140},
 primaryClass = {astro-ph.GA},
       adsurl = {https://ui.adsabs.harvard.edu/abs/2018ApJ...853..179T}
}

@ARTICLE{Tacconi:2020,
       author = {{Tacconi}, Linda J. and {Genzel}, Reinhard and {Sternberg}, Amiel},
        title = "{The Evolution of the Star-Forming Interstellar Medium Across Cosmic Time}",
      journal = {\araa},
         year = 2020,
        month = aug,
       volume = {58},
        pages = {157-203},
          doi = {10.1146/annurev-astro-082812-141034},
archivePrefix = {arXiv},
       eprint = {2003.06245},
 primaryClass = {astro-ph.GA},
       adsurl = {https://ui.adsabs.harvard.edu/abs/2020ARA&A..58..157T}
}

@ARTICLE{Cataldi:2025,
       author = {{Cataldi}, E. and {Belfiore}, F. and {Curti}, M. and {Moreschini}, B. and {Mannucci}, F. and {D'Amato}, Q. and {Cresci}, G. and {Feltre}, A. and {Ginolfi}, M. and {Marconi}, A. and {Amiri}, A. and {Arnaboldi}, M. and {Bertola}, E. and {Bracci}, C. and {Carniani}, S. and {Ceci}, M. and {Chakraborty}, A. and {Cirasuolo}, M. and {Cullen}, F. and {Kobayashi}, C. and {Kumari}, N. and {Maiolino}, R. and {Marconcini}, C. and {Scialpi}, M. and {Ulivi}, L.},
        title = "{MARTA: Temperature-temperature relationships and strong-line metallicity calibrations from multiple auroral-line detections at cosmic noon}",
      journal = {\aap},
         year = 2025,
        month = nov,
       volume = {703},
          eid = {A208},
        pages = {A208},
          doi = {10.1051/0004-6361/202554843},
archivePrefix = {arXiv},
       eprint = {2504.03839},
 primaryClass = {astro-ph.GA},
       adsurl = {https://ui.adsabs.harvard.edu/abs/2025A&A...703A.208C}
}

@ARTICLE{Napolitano:2025,
       author = {{Napolitano}, Lorenzo and {Castellano}, Marco and {Pentericci}, Laura and {Vignali}, Cristian and {Gilli}, Roberto and {Fontana}, Adriano and {Santini}, Paola and {Treu}, Tommaso and {Calabr{\`o}}, Antonello and {Llerena}, Mario and {Piconcelli}, Enrico and {Zappacosta}, Luca and {Mascia}, Sara and {Tripodi}, Roberta and {Arrabal Haro}, Pablo and {Bergamini}, Pietro and {Bakx}, Tom J.~L.~C. and {Dickinson}, Mark and {Glazebrook}, Karl and {Henry}, Alaina and {Leethochawalit}, Nicha and {Mazzolari}, Giovanni and {Merlin}, Emiliano and {Morishita}, Takahiro and {Nanayakkara}, Themiya and {Paris}, Diego and {Puccetti}, Simonetta and {Roberts-Borsani}, Guido and {Rojas Ruiz}, Sofia and {Rosati}, Piero and {Vanzella}, Eros and {Vito}, Fabio and {Vulcani}, Benedetta and {Wang}, Xin and {Yoon}, Ilsang and {Zavala}, Jorge A.},
        title = "{The Dual Nature of GHZ9: Coexisting Active Galactic Nuclei and Star Formation Activity in a Remote X-Ray Source at z = 10.145}",
      journal = {\apj},
         year = 2025,
        month = aug,
       volume = {989},
       number = {1},
          eid = {75},
        pages = {75},
          doi = {10.3847/1538-4357/ade706},
archivePrefix = {arXiv},
       eprint = {2410.18763},
 primaryClass = {astro-ph.GA},
       adsurl = {https://ui.adsabs.harvard.edu/abs/2025ApJ...989...75N}
}

@ARTICLE{Cenci:2025,
       author = {{Cenci}, Elia and {Habouzit}, Melanie},
        title = "{Little Red Dots as direct-collapse black hole nurseries}",
      journal = {\mnras},
         year = 2025,
        month = sep,
       volume = {542},
       number = {3},
        pages = {2597-2609},
          doi = {10.1093/mnras/staf1362},
archivePrefix = {arXiv},
       eprint = {2508.14897},
 primaryClass = {astro-ph.GA},
       adsurl = {https://ui.adsabs.harvard.edu/abs/2025MNRAS.542.2597C}
}

@ARTICLE{Ma:2026_ndensity,
       author = {{Ma}, Yilun and {Greene}, Jenny E. and {Setton}, David J. and {Goulding}, Andy D. and {Annunziatella}, Marianna and {Fan}, Xiaohui and {Kokorev}, Vasily and {Labbe}, Ivo and {Li}, Jiaxuan and {Lin}, Xiaojing and {Marchesini}, Danilo and {Matthee}, Jorryt and {Robbins}, Luke and {Sajina}, Anna and {Sawicki}, Marcin and {Telford}, O. Grace},
        title = "{Counting Little Red Dots at z < 4 with Ground-based Surveys and Spectroscopic Follow-up}",
      journal = {\apj},
         year = 2026,
        month = mar,
       volume = {1000},
       number = {1},
          eid = {59},
        pages = {59},
          doi = {10.3847/1538-4357/ae4596},
archivePrefix = {arXiv},
       eprint = {2504.08032},
 primaryClass = {astro-ph.GA},
       adsurl = {https://ui.adsabs.harvard.edu/abs/2026ApJ..1000...59M}
}

@ARTICLE{Cappellari:2006,
       author = {{Cappellari}, Michele and {Bacon}, R. and {Bureau}, M. and {Damen}, M.~C. and {Davies}, Roger L. and {de Zeeuw}, P.~T. and {Emsellem}, Eric and {Falc{\'o}n-Barroso}, Jes{\'u}s and {Krajnovi{\'c}}, Davor and {Kuntschner}, Harald and {McDermid}, Richard M. and {Peletier}, Reynier F. and {Sarzi}, Marc and {van den Bosch}, Remco C.~E. and {van de Ven}, Glenn},
        title = "{The SAURON project - IV. The mass-to-light ratio, the virial mass estimator and the Fundamental Plane of elliptical and lenticular galaxies}",
      journal = {\mnras},
         year = 2006,
        month = mar,
       volume = {366},
       number = {4},
        pages = {1126-1150},
          doi = {10.1111/j.1365-2966.2005.09981.x},
archivePrefix = {arXiv},
       eprint = {astro-ph/0505042},
 primaryClass = {astro-ph},
       adsurl = {https://ui.adsabs.harvard.edu/abs/2006MNRAS.366.1126C}
}

@ARTICLE{Cappellari:2023,
    author = {{Cappellari}, M.},
    title = "{Full spectrum fitting with photometry in PPXF: stellar population
        versus dynamical masses, non-parametric star formation history and
        metallicity for 3200 LEGA-C galaxies at redshift $z\approx0.8$}",
    journal = {MNRAS},
    eprint = {2208.14974},
    year = 2023,
    volume = 526,
    pages = {3273-3300},
    doi = {10.1093/mnras/stad2597}
}

@ARTICLE{Wuyts:2016,
       author = {{Wuyts}, Stijn and {F{\"o}rster Schreiber}, Natascha M. and {Wisnioski}, Emily and {Genzel}, Reinhard and {Burkert}, Andreas and {Bandara}, Kaushala and {Beifiori}, Alessandra and {Belli}, Sirio and {Bender}, Ralf and {Brammer}, Gabriel B. and {Chan}, Jeffrey and {Davies}, Ric and {Fossati}, Matteo and {Galametz}, Audrey and {Kulkarni}, Sandesh K. and {Lang}, Philipp and {Lutz}, Dieter and {Mendel}, J. Trevor and {Momcheva}, Ivelina G. and {Naab}, Thorsten and {Nelson}, Erica J. and {Saglia}, Roberto P. and {Seitz}, Stella and {Tacconi}, Linda J. and {Tadaki}, Ken-ichi and {{\"U}bler}, Hannah and {van Dokkum}, Pieter G. and {Wilman}, David J. and {Wuyts}, Eva},
        title = "{KMOS3D: Dynamical Constraints on the Mass Budget in Early Star-forming Disks}",
      journal = {\apj},
         year = 2016,
        month = nov,
       volume = {831},
       number = {2},
          eid = {149},
        pages = {149},
          doi = {10.3847/0004-637X/831/2/149},
archivePrefix = {arXiv},
       eprint = {1603.03432},
 primaryClass = {astro-ph.GA},
       adsurl = {https://ui.adsabs.harvard.edu/abs/2016ApJ...831..149W}
}

@ARTICLE{Stern:2012,
       author = {{Stern}, Jonathan and {Laor}, Ari},
        title = "{Type 1 AGN at low z- I. Emission properties}",
      journal = {\mnras},
         year = 2012,
        month = jun,
       volume = {423},
       number = {1},
        pages = {600-631},
          doi = {10.1111/j.1365-2966.2012.20901.x},
archivePrefix = {arXiv},
       eprint = {1203.3158},
 primaryClass = {astro-ph.CO},
       adsurl = {https://ui.adsabs.harvard.edu/abs/2012MNRAS.423..600S}
}

@ARTICLE{Volonteri:2005,
       author = {{Volonteri}, Marta and {Rees}, Martin J.},
        title = "{Rapid Growth of High-Redshift Black Holes}",
      journal = {\apj},
         year = 2005,
        month = nov,
       volume = {633},
       number = {2},
        pages = {624-629},
          doi = {10.1086/466521},
archivePrefix = {arXiv},
       eprint = {astro-ph/0506040},
 primaryClass = {astro-ph},
       adsurl = {https://ui.adsabs.harvard.edu/abs/2005ApJ...633..624V}
}

@ARTICLE{Volonteri:2021,
       author = {{Volonteri}, Marta and {Habouzit}, M{\'e}lanie and {Colpi}, Monica},
        title = "{The origins of massive black holes}",
      journal = {Nature Reviews Physics},
         year = 2021,
        month = sep,
       volume = {3},
       number = {11},
        pages = {732-743},
          doi = {10.1038/s42254-021-00364-9},
archivePrefix = {arXiv},
       eprint = {2110.10175},
 primaryClass = {astro-ph.GA},
       adsurl = {https://ui.adsabs.harvard.edu/abs/2021NatRP...3..732V}
}

@ARTICLE{Volonteri:2023,
       author = {{Volonteri}, Marta and {Habouzit}, M{\'e}lanie and {Colpi}, Monica},
        title = "{What if young z > 9 JWST galaxies hosted massive black holes?}",
      journal = {\mnras},
         year = 2023,
        month = may,
       volume = {521},
       number = {1},
        pages = {241-250},
          doi = {10.1093/mnras/stad499},
archivePrefix = {arXiv},
       eprint = {2212.04710},
 primaryClass = {astro-ph.GA},
       adsurl = {https://ui.adsabs.harvard.edu/abs/2023MNRAS.521..241V}
}

@ARTICLE{Topping:2024,
       author = {{Topping}, Michael W. and {Stark}, Daniel P. and {Senchyna}, Peter and {Plat}, Adele and {Zitrin}, Adi and {Endsley}, Ryan and {Charlot}, St{\'e}phane and {Furtak}, Lukas J. and {Maseda}, Michael V. and {Smit}, Renske and {Mainali}, Ramesh and {Chevallard}, Jacopo and {Molyneux}, Stephen and {Rigby}, Jane R.},
        title = "{Metal-poor star formation at z > 6 with JWST: new insight into hard radiation fields and nitrogen enrichment on 20 pc scales}",
      journal = {\mnras},
         year = 2024,
        month = apr,
       volume = {529},
       number = {4},
        pages = {3301-3322},
          doi = {10.1093/mnras/stae682},
archivePrefix = {arXiv},
       eprint = {2401.08764},
 primaryClass = {astro-ph.GA},
       adsurl = {https://ui.adsabs.harvard.edu/abs/2024MNRAS.529.3301T}
}

@ARTICLE{Torralba:2026,
       author = {{Torralba}, Alberto and {Matthee}, Jorryt and {Pezzulli}, Gabriele and {Urrutia}, Tanya and {Gronke}, Max and {Mascia}, Sara and {D'Eugenio}, Francesco and {Di Cesare}, Claudia and {Eilers}, Anna-Christina and {Greene}, Jenny E. and {Iani}, Edoardo and {Ishikawa}, Yuzo and {Mackenzie}, Ruari and {Naidu}, Rohan P. and {Navarrete}, Benjam{\'\i}n and {Kotiwale}, Gauri},
        title = "{A weak Ly{\ensuremath{\alpha}} halo for an extremely bright little red dot: Indications of enshrouded supermassive black hole growth}",
      journal = {\aap},
         year = 2026,
        month = jan,
       volume = {705},
          eid = {A147},
        pages = {A147},
          doi = {10.1051/0004-6361/202555596},
archivePrefix = {arXiv},
       eprint = {2505.09542},
 primaryClass = {astro-ph.GA},
       adsurl = {https://ui.adsabs.harvard.edu/abs/2026A&A...705A.147T}
}

@ARTICLE{Kakkad:2020,
       author = {{Kakkad}, D. and {Mainieri}, V. and {Vietri}, G. and {Carniani}, S. and {Harrison}, C.~M. and {Perna}, M. and {Scholtz}, J. and {Circosta}, C. and {Cresci}, G. and {Husemann}, B. and {Bischetti}, M. and {Feruglio}, C. and {Fiore}, F. and {Marconi}, A. and {Padovani}, P. and {Brusa}, M. and {Cicone}, C. and {Comastri}, A. and {Lanzuisi}, G. and {Mannucci}, F. and {Menci}, N. and {Netzer}, H. and {Piconcelli}, E. and {Puglisi}, A. and {Salvato}, M. and {Schramm}, M. and {Silverman}, J. and {Vignali}, C. and {Zamorani}, G. and {Zappacosta}, L.},
        title = "{SUPER. II. Spatially resolved ionised gas kinematics and scaling relations in z {\ensuremath{\sim}} 2 AGN host galaxies}",
      journal = {\aap},
         year = 2020,
        month = oct,
       volume = {642},
          eid = {A147},
        pages = {A147},
          doi = {10.1051/0004-6361/202038551},
archivePrefix = {arXiv},
       eprint = {2008.01728},
 primaryClass = {astro-ph.GA},
       adsurl = {https://ui.adsabs.harvard.edu/abs/2020A&A...642A.147K}
}

@ARTICLE{Kokorev:2023,
       author = {{Kokorev}, Vasily and {Fujimoto}, Seiji and {Labbe}, Ivo and {Greene}, Jenny E. and {Bezanson}, Rachel and {Dayal}, Pratika and {Nelson}, Erica J. and {Atek}, Hakim and {Brammer}, Gabriel and {Caputi}, Karina I. and {Chemerynska}, Iryna and {Cutler}, Sam E. and {Feldmann}, Robert and {Fudamoto}, Yoshinobu and {Furtak}, Lukas J. and {Goulding}, Andy D. and {de Graaff}, Anna and {Leja}, Joel and {Marchesini}, Danilo and {Miller}, Tim B. and {Nanayakkara}, Themiya and {Oesch}, Pascal A. and {Pan}, Richard and {Price}, Sedona H. and {Setton}, David J. and {Smit}, Renske and {Stefanon}, Mauro and {Wang}, Bingjie and {Weaver}, John R. and {Whitaker}, Katherine E. and {Williams}, Christina C. and {Zitrin}, Adi},
        title = "{UNCOVER: A NIRSpec Identification of a Broad-line AGN at z = 8.50}",
      journal = {\apjl},
         year = 2023,
        month = nov,
       volume = {957},
       number = {1},
          eid = {L7},
        pages = {L7},
          doi = {10.3847/2041-8213/ad037a},
archivePrefix = {arXiv},
       eprint = {2308.11610},
 primaryClass = {astro-ph.GA},
       adsurl = {https://ui.adsabs.harvard.edu/abs/2023ApJ...957L...7K}
}

@ARTICLE{Tacchella:2015sins,
       author = {{Tacchella}, S. and {Lang}, P. and {Carollo}, C.~M. and {F{\"o}rster Schreiber}, N.~M. and {Renzini}, A. and {Shapley}, A.~E. and {Wuyts}, S. and {Cresci}, G. and {Genzel}, R. and {Lilly}, S.~J. and {Mancini}, C. and {Newman}, S.~F. and {Tacconi}, L.~J. and {Zamorani}, G. and {Davies}, R.~I. and {Kurk}, J. and {Pozzetti}, L.},
        title = "{SINS/zC-SINF Survey of z {\ensuremath{\sim}} 2 Galaxy Kinematics: Rest-frame Morphology, Structure, and Colors from Near-infrared Hubble Space Telescope Imaging}",
      journal = {\apj},
         year = 2015,
        month = apr,
       volume = {802},
       number = {2},
          eid = {101},
        pages = {101},
          doi = {10.1088/0004-637X/802/2/101},
archivePrefix = {arXiv},
       eprint = {1411.7034},
 primaryClass = {astro-ph.GA},
       adsurl = {https://ui.adsabs.harvard.edu/abs/2015ApJ...802..101T}
}

@ARTICLE{Danhaive:2025,
       author = {{Danhaive}, A. Lola and {Tacchella}, Sandro and {{\"U}bler}, Hannah and {de Graaff}, Anna and {Egami}, Eiichi and {Johnson}, Benjamin D. and {Sun}, Fengwu and {Arribas}, Santiago and {Bunker}, Andrew J. and {Carniani}, Stefano and {Jones}, Gareth C. and {Maiolino}, Roberto and {McClymont}, William and {Parlanti}, Eleonora and {Simmonds}, Charlotte and {Villanueva}, Natalia C. and {Baker}, William M. and {Jaffe}, Daniel T. and {Eisenstein}, Daniel and {Hainline}, Kevin and {Helton}, Jakob M. and {Ji}, Zhiyuan and {Lin}, Xiaojing and {Liu}, Yichen and {Pusk{\'a}s}, D{\'a}vid and {Rieke}, Marcia and {Rinaldi}, Pierluigi and {Robertson}, Brant and {Scholz}, Jan and {Williams}, Christina C. and {Willmer}, Christopher N.~A.},
        title = "{The dawn of discs: unveiling the turbulent ionized gas kinematics of the galaxy population at z {\ensuremath{\sim}} 4─6 with JWST/NIRCam grism spectroscopy}",
      journal = {\mnras},
         year = 2025,
        month = nov,
       volume = {543},
       number = {4},
        pages = {3249-3302},
          doi = {10.1093/mnras/staf1540},
archivePrefix = {arXiv},
       eprint = {2503.21863},
 primaryClass = {astro-ph.GA},
       adsurl = {https://ui.adsabs.harvard.edu/abs/2025MNRAS.543.3249D}
}

@ARTICLE{Fei:2026,
       author = {{Fei}, Qinyue and {Fujimoto}, Seiji and {Naidu}, Rohan P. and {Chisholm}, John and {Atek}, Hakim and {Brammer}, Gabriel and {Asada}, Yoshihisa and {Berg}, Danielle A. and {Bromm}, Volker and {Furtak}, Lukas J. and {Greene}, Jenny E. and {Hsiao}, Tiger Yu-Yang and {Jeon}, Junehyoung and {Kokorev}, Vasily and {Matthee}, Jorryt and {Natarajan}, Priyamvada and {Pan}, Richard and {Richard}, Johan and {Saldana-Lopez}, Alberto and {Schaerer}, Daniel and {Volonteri}, Marta and {Zitrin}, Adi},
        title = "{A GLIMPSE of Intermediate Mass Black Holes in the Epoch of Reionization: Witnessing the Descendants of Direct Collapse?}",
      journal = {\apj},
         year = 2026,
        month = jun,
       volume = {1003},
       number = {2},
          eid = {244},
        pages = {244},
          doi = {10.3847/1538-4357/ae6248},
archivePrefix = {arXiv},
       eprint = {2509.20452},
 primaryClass = {astro-ph.GA},
       adsurl = {https://ui.adsabs.harvard.edu/abs/2026ApJ..1003..244F}
}

@ARTICLE{Feltre:2016,
       author = {{Feltre}, A. and {Charlot}, S. and {Gutkin}, J.},
        title = "{Nuclear activity versus star formation: emission-line diagnostics at ultraviolet and optical wavelengths}",
      journal = {\mnras},
         year = 2016,
        month = mar,
       volume = {456},
       number = {3},
        pages = {3354-3374},
          doi = {10.1093/mnras/stv2794},
archivePrefix = {arXiv},
       eprint = {1511.08217},
 primaryClass = {astro-ph.GA},
       adsurl = {https://ui.adsabs.harvard.edu/abs/2016MNRAS.456.3354F}
}

@ARTICLE{Ferruit:2022,
       author = {{Ferruit}, P. and {Jakobsen}, P. and {Giardino}, G. and {Rawle}, T. and {Alves de Oliveira}, C. and {Arribas}, S. and {Beck}, T.~L. and {Birkmann}, S. and {B{\"o}ker}, T. and {Bunker}, A.~J. and {Charlot}, S. and {de Marchi}, G. and {Franx}, M. and {Henry}, A. and {Karakla}, D. and {Kassin}, S.~A. and {Kumari}, N. and {L{\'o}pez-Caniego}, M. and {L{\"u}tzgendorf}, N. and {Maiolino}, R. and {Manjavacas}, E. and {Marston}, A. and {Moseley}, S.~H. and {Muzerolle}, J. and {Pirzkal}, N. and {Rauscher}, B. and {Rix}, H. -W. and {Sabbi}, E. and {Sirianni}, M. and {te Plate}, M. and {Valenti}, J. and {Willott}, C.~J. and {Zeidler}, P.},
        title = "{The Near-Infrared Spectrograph (NIRSpec) on the James Webb Space Telescope. II. Multi-object spectroscopy (MOS)}",
      journal = {\aap},
         year = 2022,
        month = may,
       volume = {661},
          eid = {A81},
        pages = {A81},
          doi = {10.1051/0004-6361/202142673},
archivePrefix = {arXiv},
       eprint = {2202.03306},
 primaryClass = {astro-ph.IM},
       adsurl = {https://ui.adsabs.harvard.edu/abs/2022A&A...661A..81F}
}

@ARTICLE{Eisenstein:2023,
       author = {{Eisenstein}, Daniel J. and {Johnson}, Benjamin D. and {Robertson}, Brant and {Tacchella}, Sandro and {Hainline}, Kevin and {Jakobsen}, Peter and {Maiolino}, Roberto and {Bonaventura}, Nina and {Bunker}, Andrew J. and {Cameron}, Alex J. and {Cargile}, Phillip A. and {Curtis-Lake}, Emma and {Hausen}, Ryan and {Pusk{\'a}s}, D{\'a}vid and {Rieke}, Marcia and {Sun}, Fengwu and {Willmer}, Christopher N.~A. and {Willott}, Chris and {Alberts}, Stacey and {Arribas}, Santiago and {Baker}, William M. and {Baum}, Stefi and {Bhatawdekar}, Rachana and {Carniani}, Stefano and {Charlot}, Stephane and {Chen}, Zuyi and {Chevallard}, Jacopo and {Curti}, Mirko and {DeCoursey}, Christa and {D'Eugenio}, Francesco and {de Graaff}, Anna and {Egami}, Eiichi and {Helton}, Jakob M. and {Ji}, Zhiyuan and {Jones}, Gareth C. and {Kumari}, Nimisha and {L{\"u}tzgendorf}, Nora and {Laseter}, Isaac and {Looser}, Tobias J. and {Lyu}, Jianwei and {Maseda}, Michael V. and {Nelson}, Erica and {Parlanti}, Eleonora and {Rauscher}, Bernard J. and {Rawle}, Tim and {Rieke}, George and {Rix}, Hans-Walter and {Rujopakarn}, Wiphu and {Sandles}, Lester and {Saxena}, Aayush and {Scholtz}, Jan and {Sharpe}, Katherine and {Shivaei}, Irene and {Simmonds}, Charlotte and {Smit}, Renske and {Topping}, Michael W. and {{\"U}bler}, Hannah and {Venturi}, Giacomo and {Williams}, Christina C. and {Witstok}, Joris and {Woodrum}, Charity},
        title = "{The JADES Origins Field: A New JWST Deep Field in the JADES Second NIRCam Data Release}",
      journal = {arXiv e-prints},
         year = 2023,
        month = oct,
          eid = {arXiv:2310.12340},
        pages = {arXiv:2310.12340},
          doi = {10.48550/arXiv.2310.12340},
archivePrefix = {arXiv},
       eprint = {2310.12340},
 primaryClass = {astro-ph.GA},
       adsurl = {https://ui.adsabs.harvard.edu/abs/2023arXiv231012340E}
}

@ARTICLE{Eisenstein:2023a,
       author = {{Eisenstein}, Daniel J. and {Willott}, Chris and {Alberts}, Stacey and {Arribas}, Santiago and {Bonaventura}, Nina and {Bunker}, Andrew J. and {Cameron}, Alex J. and {Carniani}, Stefano and {Charlot}, Stephane and {Curtis-Lake}, Emma and {D'Eugenio}, Francesco and {Endsley}, Ryan and {Ferruit}, Pierre and {Giardino}, Giovanna and {Hainline}, Kevin and {Hausen}, Ryan and {Jakobsen}, Peter and {Johnson}, Benjamin D. and {Maiolino}, Roberto and {Rieke}, Marcia and {Rieke}, George and {Rix}, Hans-Walter and {Robertson}, Brant and {Stark}, Daniel P. and {Tacchella}, Sandro and {Williams}, Christina C. and {Willmer}, Christopher N.~A. and {Baker}, William M. and {Baum}, Stefi and {Bhatawdekar}, Rachana and {Boyett}, Kristan and {Chen}, Zuyi and {Chevallard}, Jacopo and {Circosta}, Chiara and {Curti}, Mirko and {Danhaive}, A. Lola and {DeCoursey}, Christa and {de Graaff}, Anna and {Dressler}, Alan and {Egami}, Eiichi and {Helton}, Jakob M. and {Hviding}, Raphael E. and {Ji}, Zhiyuan and {Jones}, Gareth C. and {Kumari}, Nimisha and {L{\"u}tzgendorf}, Nora and {Laseter}, Isaac and {Looser}, Tobias J. and {Lyu}, Jianwei and {Maseda}, Michael V. and {Nelson}, Erica and {Parlanti}, Eleonora and {Perna}, Michele and {Pusk{\'a}s}, D{\'a}vid and {Rawle}, Tim and {Rodr{\'\i}guez Del Pino}, Bruno and {Sandles}, Lester and {Saxena}, Aayush and {Scholtz}, Jan and {Sharpe}, Katherine and {Shivaei}, Irene and {Silcock}, Maddie S. and {Simmonds}, Charlotte and {Skarbinski}, Maya and {Smit}, Renske and {Stone}, Meredith and {Suess}, Katherine A. and {Sun}, Fengwu and {Tang}, Mengtao and {Topping}, Michael W. and {{\"U}bler}, Hannah and {Villanueva}, Natalia C. and {Wallace}, Imaan E.~B. and {Whitler}, Lily and {Witstok}, Joris and {Woodrum}, Charity},
        title = "{Overview of the JWST Advanced Deep Extragalactic Survey (JADES)}",
      journal = {arXiv e-prints},
         year = 2023,
        month = jun,
          eid = {arXiv:2306.02465},
        pages = {arXiv:2306.02465},
          doi = {10.48550/arXiv.2306.02465},
archivePrefix = {arXiv},
       eprint = {2306.02465},
 primaryClass = {astro-ph.GA},
       adsurl = {https://ui.adsabs.harvard.edu/abs/2023arXiv230602465E}
}

@ARTICLE{Bunker:2024,
       author = {{Bunker}, Andrew J. and {Cameron}, Alex J. and {Curtis-Lake}, Emma and {Jakobsen}, Peter and {Carniani}, Stefano and {Curti}, Mirko and {Witstok}, Joris and {Maiolino}, Roberto and {D'Eugenio}, Francesco and {Looser}, Tobias J. and {Willott}, Chris and {Bonaventura}, Nina and {Hainline}, Kevin and {{\"U}bler}, Hannah and {Willmer}, Christopher N.~A. and {Saxena}, Aayush and {Smit}, Renske and {Alberts}, Stacey and {Arribas}, Santiago and {Baker}, William M. and {Baum}, Stefi and {Bhatawdekar}, Rachana and {Bowler}, Rebecca A.~A. and {Boyett}, Kristan and {Charlot}, Stephane and {Chen}, Zuyi and {Chevallard}, Jacopo and {Circosta}, Chiara and {DeCoursey}, Christa and {de Graaff}, Anna and {Egami}, Eiichi and {Eisenstein}, Daniel J. and {Endsley}, Ryan and {Ferruit}, Pierre and {Giardino}, Giovanna and {Hausen}, Ryan and {Helton}, Jakob M. and {Hviding}, Raphael E. and {Ji}, Zhiyuan and {Johnson}, Benjamin D. and {Jones}, Gareth C. and {Kumari}, Nimisha and {Laseter}, Isaac and {L{\"u}tzgendorf}, Nora and {Maseda}, Michael V. and {Nelson}, Erica and {Parlanti}, Eleonora and {Perna}, Michele and {Rauscher}, Bernard J. and {Rawle}, Tim and {Rix}, Hans-Walter and {Rieke}, Marcia and {Robertson}, Brant and {Rodr{\'\i}guez Del Pino}, Bruno and {Sandles}, Lester and {Scholtz}, Jan and {Sharpe}, Katherine and {Skarbinski}, Maya and {Stark}, Daniel P. and {Sun}, Fengwu and {Tacchella}, Sandro and {Topping}, Michael W. and {Villanueva}, Natalia C. and {Wallace}, Imaan E.~B. and {Williams}, Christina C. and {Woodrum}, Charity},
        title = "{JADES NIRSpec initial data release for the Hubble Ultra Deep Field: Redshifts and line fluxes of distant galaxies from the deepest JWST Cycle 1 NIRSpec multi-object spectroscopy}",
      journal = {\aap},
         year = 2024,
        month = oct,
       volume = {690},
          eid = {A288},
        pages = {A288},
          doi = {10.1051/0004-6361/202347094},
archivePrefix = {arXiv},
       eprint = {2306.02467},
 primaryClass = {astro-ph.GA},
       adsurl = {https://ui.adsabs.harvard.edu/abs/2024A&A...690A.288B}
}

@ARTICLE{Genzel:2023,
       author = {{Genzel}, R. and {Jolly}, J. -B. and {Liu}, D. and {Price}, S.~H. and {Lee}, L.~L. and {F{\"o}rster Schreiber}, N.~M. and {Tacconi}, L.~J. and {Herrera-Camus}, R. and {Barfety}, C. and {Burkert}, A. and {Cao}, Y. and {Davies}, R.~I. and {Dekel}, A. and {Lee}, M.~M. and {Lutz}, D. and {Naab}, T. and {Neri}, R. and {Nestor Shachar}, A. and {Pastras}, S. and {Pulsoni}, C. and {Renzini}, A. and {Schuster}, K. and {Shimizu}, T.~T. and {Stanley}, F. and {Sternberg}, A. and {{\"U}bler}, H.},
        title = "{Evidence for Large-scale, Rapid Gas Inflows in z   2 Star-forming Disks}",
      journal = {\apj},
         year = 2023,
        month = nov,
       volume = {957},
       number = {1},
          eid = {48},
        pages = {48},
          doi = {10.3847/1538-4357/acef1a},
archivePrefix = {arXiv},
       eprint = {2305.02959},
 primaryClass = {astro-ph.GA},
       adsurl = {https://ui.adsabs.harvard.edu/abs/2023ApJ...957...48G}
}

@ARTICLE{Bertola:2024,
       author = {{Bertola}, E. and {Circosta}, C. and {Ginolfi}, M. and {Mainieri}, V. and {Vignali}, C. and {Calistro Rivera}, G. and {Ward}, S.~R. and {Lopez}, I.~E. and {Pensabene}, A. and {Alexander}, D.~M. and {Bischetti}, M. and {Brusa}, M. and {Cappi}, M. and {Comastri}, A. and {Contursi}, A. and {Cicone}, C. and {Cresci}, G. and {Dadina}, M. and {D'Amato}, Q. and {Feltre}, A. and {Harrison}, C.~M. and {Kakkad}, D. and {Lamperti}, I. and {Lanzuisi}, G. and {Mannucci}, F. and {Marconi}, A. and {Perna}, M. and {Piconcelli}, E. and {Puglisi}, A. and {Ricci}, F. and {Scholtz}, J. and {Tozzi}, G. and {Vietri}, G. and {Zamorani}, G. and {Zappacosta}, L.},
        title = "{KASHz+SUPER: Evidence of cold molecular gas depletion in AGN hosts at cosmic noon}",
      journal = {\aap},
         year = 2024,
        month = nov,
       volume = {691},
          eid = {A178},
        pages = {A178},
          doi = {10.1051/0004-6361/202450420},
archivePrefix = {arXiv},
       eprint = {2408.16821},
 primaryClass = {astro-ph.GA},
       adsurl = {https://ui.adsabs.harvard.edu/abs/2024A&A...691A.178B}
}

@ARTICLE{Somerville:1999,
       author = {{Somerville}, Rachel S. and {Primack}, Joel R.},
        title = "{Semi-analytic modelling of galaxy formation: the local Universe}",
      journal = {\mnras},
         year = 1999,
        month = dec,
       volume = {310},
       number = {4},
        pages = {1087-1110},
          doi = {10.1046/j.1365-8711.1999.03032.x},
archivePrefix = {arXiv},
       eprint = {astro-ph/9802268},
 primaryClass = {astro-ph},
       adsurl = {https://ui.adsabs.harvard.edu/abs/1999MNRAS.310.1087S}
}

@ARTICLE{Somerville:2015_sam,
       author = {{Somerville}, Rachel S. and {Popping}, Gerg{\"o} and {Trager}, Scott C.},
        title = "{Star formation in semi-analytic galaxy formation models with multiphase gas}",
      journal = {\mnras},
         year = 2015,
        month = nov,
       volume = {453},
       number = {4},
        pages = {4337-4367},
          doi = {10.1093/mnras/stv1877},
archivePrefix = {arXiv},
       eprint = {1503.00755},
 primaryClass = {astro-ph.GA},
       adsurl = {https://ui.adsabs.harvard.edu/abs/2015MNRAS.453.4337S}
}

@ARTICLE{Somerville:2021,
       author = {{Somerville}, Rachel S. and {Olsen}, Charlotte and {Yung}, L.~Y. Aaron and {Pacifici}, Camilla and {Ferguson}, Henry C. and {Behroozi}, Peter and {Osborne}, Shannon and {Wechsler}, Risa H. and {Pandya}, Viraj and {Faber}, Sandra M. and {Primack}, Joel R. and {Dekel}, Avishai},
        title = "{Mock light-cones and theory friendly catalogues for the CANDELS survey}",
      journal = {\mnras},
         year = 2021,
        month = apr,
       volume = {502},
       number = {4},
        pages = {4858-4876},
          doi = {10.1093/mnras/stab231},
archivePrefix = {arXiv},
       eprint = {2102.00108},
 primaryClass = {astro-ph.GA},
       adsurl = {https://ui.adsabs.harvard.edu/abs/2021MNRAS.502.4858S}
}

@ARTICLE{Li:2025,
       author = {{Li}, Junyao and {Silverman}, John D. and {Shen}, Yue and {Volonteri}, Marta and {Jahnke}, Knud and {Zhuang}, Ming-Yang and {Scoggins}, Matthew T. and {Ding}, Xuheng and {Harikane}, Yuichi and {Onoue}, Masafusa and {Tanaka}, Takumi S.},
        title = "{Tip of the Iceberg: Overmassive Black Holes at 4 < z < 7 Found by JWST Are Not Inconsistent with the Local  Relation}",
      journal = {\apj},
         year = 2025,
        month = mar,
       volume = {981},
       number = {1},
          eid = {19},
        pages = {19},
          doi = {10.3847/1538-4357/ada603},
archivePrefix = {arXiv},
       eprint = {2403.00074},
 primaryClass = {astro-ph.GA},
       adsurl = {https://ui.adsabs.harvard.edu/abs/2025ApJ...981...19L}
}

@ARTICLE{Pacucci:2023,
       author = {{Pacucci}, Fabio and {Nguyen}, Bao and {Carniani}, Stefano and {Maiolino}, Roberto and {Fan}, Xiaohui},
        title = "{JWST CEERS and JADES Active Galaxies at z = 4-7 Violate the Local M $_{{\textbullet}}$-M $_{{\ensuremath{\star}}}$ Relation at >3{\ensuremath{\sigma}}: Implications for Low-mass Black Holes and Seeding Models}",
      journal = {\apjl},
         year = 2023,
        month = nov,
       volume = {957},
       number = {1},
          eid = {L3},
        pages = {L3},
          doi = {10.3847/2041-8213/ad0158},
archivePrefix = {arXiv},
       eprint = {2308.12331},
 primaryClass = {astro-ph.GA},
       adsurl = {https://ui.adsabs.harvard.edu/abs/2023ApJ...957L...3P}
}

@ARTICLE{Pacucci:2026,
       author = {{Pacucci}, Fabio and {Ferrara}, Andrea and {Kocevski}, Dale D.},
        title = "{The Little Red Dots Are Direct Collapse Black Holes}",
      journal = {arXiv e-prints},
         year = 2026,
        month = jan,
          eid = {arXiv:2601.14368},
        pages = {arXiv:2601.14368},
          doi = {10.48550/arXiv.2601.14368},
archivePrefix = {arXiv},
       eprint = {2601.14368},
 primaryClass = {astro-ph.GA},
       adsurl = {https://ui.adsabs.harvard.edu/abs/2026arXiv260114368P}
}

@ARTICLE{Ziparo:2026,
       author = {{Ziparo}, Francesco and {Carniani}, Stefano and {Gallerani}, Simona and {Trefoloni}, Bartolomeo},
        title = "{A Selection Aware View of Black Hole-Galaxy Coevolution at High Redshift}",
      journal = {arXiv e-prints},
         year = 2026,
        month = mar,
          eid = {arXiv:2603.04358},
        pages = {arXiv:2603.04358},
          doi = {10.48550/arXiv.2603.04358},
archivePrefix = {arXiv},
       eprint = {2603.04358},
 primaryClass = {astro-ph.GA},
       adsurl = {https://ui.adsabs.harvard.edu/abs/2026arXiv260304358Z}
}

@ARTICLE{Curti:2024,
       author = {{Curti}, Mirko and {Maiolino}, Roberto and {Curtis-Lake}, Emma and {Chevallard}, Jacopo and {Carniani}, Stefano and {D'Eugenio}, Francesco and {Looser}, Tobias J. and {Scholtz}, Jan and {Charlot}, Stephane and {Cameron}, Alex and {{\"U}bler}, Hannah and {Witstok}, Joris and {Boyett}, Kristian and {Laseter}, Isaac and {Sandles}, Lester and {Arribas}, Santiago and {Bunker}, Andrew and {Giardino}, Giovanna and {Maseda}, Michael V. and {Rawle}, Tim and {Rodr{\'\i}guez Del Pino}, Bruno and {Smit}, Renske and {Willott}, Chris J. and {Eisenstein}, Daniel J. and {Hausen}, Ryan and {Johnson}, Benjamin and {Rieke}, Marcia and {Robertson}, Brant and {Tacchella}, Sandro and {Williams}, Christina C. and {Willmer}, Christopher and {Baker}, William M. and {Bhatawdekar}, Rachana and {Egami}, Eiichi and {Helton}, Jakob M. and {Ji}, Zhiyuan and {Kumari}, Nimisha and {Perna}, Michele and {Shivaei}, Irene and {Sun}, Fengwu},
        title = "{JADES: Insights into the low-mass end of the mass-metallicity-SFR relation at 3 < z < 10 from deep JWST/NIRSpec spectroscopy}",
      journal = {\aap},
         year = 2024,
        month = apr,
       volume = {684},
          eid = {A75},
        pages = {A75},
          doi = {10.1051/0004-6361/202346698},
archivePrefix = {arXiv},
       eprint = {2304.08516},
 primaryClass = {astro-ph.GA},
       adsurl = {https://ui.adsabs.harvard.edu/abs/2024A&A...684A..75C}
}

@ARTICLE{Curti:2023,
       author = {{Curti}, Mirko and {D'Eugenio}, Francesco and {Carniani}, Stefano and {Maiolino}, Roberto and {Sandles}, Lester and {Witstok}, Joris and {Baker}, William M. and {Bennett}, Jake S. and {Piotrowska}, Joanna M. and {Tacchella}, Sandro and {Charlot}, Stephane and {Nakajima}, Kimihiko and {Maheson}, Gabriel and {Mannucci}, Filippo and {Amiri}, Amirnezam and {Arribas}, Santiago and {Belfiore}, Francesco and {Bonaventura}, Nina R. and {Bunker}, Andrew J. and {Chevallard}, Jacopo and {Cresci}, Giovanni and {Curtis-Lake}, Emma and {Hayden-Pawson}, Connor and {Jones}, Gareth C. and {Kumari}, Nimisha and {Laseter}, Isaac and {Looser}, Tobias J. and {Marconi}, Alessandro and {Maseda}, Michael V. and {Scholtz}, Jan and {Smit}, Renske and {{\"U}bler}, Hannah and {Wallace}, Imaan E.~B.},
        title = "{The chemical enrichment in the early Universe as probed by JWST via direct metallicity measurements at z {\ensuremath{\sim}} 8}",
      journal = {\mnras},
         year = 2023,
        month = jan,
       volume = {518},
       number = {1},
        pages = {425-438},
          doi = {10.1093/mnras/stac2737},
archivePrefix = {arXiv},
       eprint = {2207.12375},
 primaryClass = {astro-ph.GA},
       adsurl = {https://ui.adsabs.harvard.edu/abs/2023MNRAS.518..425C}
}

@ARTICLE{Birkin:2024,
       author = {{Birkin}, Jack E. and {Puglisi}, A. and {Swinbank}, A.~M. and {Smail}, Ian and {An}, Fang Xia and {Chapman}, S.~C. and {Chen}, Chian-Chou and {Conselice}, C.~J. and {Dudzevi{\v{c}}i{\={u}}t{\.{e}}}, U. and {Farrah}, D. and {Gullberg}, B. and {Matsuda}, Y. and {Schinnerer}, E. and {Scott}, D. and {Wardlow}, J.~L. and {van der Werf}, P.},
        title = "{KAOSS: turbulent, but disc-like kinematics in dust-obscured star-forming galaxies at z   1.3-2.6}",
      journal = {\mnras},
         year = 2024,
        month = jun,
       volume = {531},
       number = {1},
        pages = {61-83},
          doi = {10.1093/mnras/stae1089},
archivePrefix = {arXiv},
       eprint = {2301.05720},
 primaryClass = {astro-ph.GA},
       adsurl = {https://ui.adsabs.harvard.edu/abs/2024MNRAS.531...61B}
}

@ARTICLE{Chen:2025,
       author = {{Chen}, Chang-Hao and {Ho}, Luis C. and {Li}, Ruancun and {Inayoshi}, Kohei},
        title = "{The Physical Nature of the Off-center Extended Emission Associated with the Little Red Dots}",
      journal = {\apjl},
         year = 2025,
        month = aug,
       volume = {989},
       number = {1},
          eid = {L12},
        pages = {L12},
          doi = {10.3847/2041-8213/adee0a},
archivePrefix = {arXiv},
       eprint = {2505.03183},
 primaryClass = {astro-ph.GA},
       adsurl = {https://ui.adsabs.harvard.edu/abs/2025ApJ...989L..12C}
}

@ARTICLE{Dors:2021,
       author = {{Dors}, Oli L.},
        title = "{Chemical abundances in Seyfert galaxies - VI. Empirical abundance calibration}",
      journal = {\mnras},
         year = 2021,
        month = oct,
       volume = {507},
       number = {1},
        pages = {466-474},
          doi = {10.1093/mnras/stab2166},
       adsurl = {https://ui.adsabs.harvard.edu/abs/2021MNRAS.507..466D}
}

@ARTICLE{Boquien:2019,
       author = {{Boquien}, M. and {Burgarella}, D. and {Roehlly}, Y. and {Buat}, V. and {Ciesla}, L. and {Corre}, D. and {Inoue}, A.~K. and {Salas}, H.},
        title = "{CIGALE: a python Code Investigating GALaxy Emission}",
      journal = {\aap},
         year = 2019,
        month = feb,
       volume = {622},
          eid = {A103},
        pages = {A103},
          doi = {10.1051/0004-6361/201834156},
archivePrefix = {arXiv},
       eprint = {1811.03094},
 primaryClass = {astro-ph.GA},
       adsurl = {https://ui.adsabs.harvard.edu/abs/2019A&A...622A.103B}
}

@ARTICLE{Chabrier:2003,
       author = {{Chabrier}, Gilles},
        title = "{Galactic Stellar and Substellar Initial Mass Function}",
      journal = {\pasp},
         year = 2003,
        month = jul,
       volume = {115},
       number = {809},
        pages = {763-795},
          doi = {10.1086/376392},
archivePrefix = {arXiv},
       eprint = {astro-ph/0304382},
 primaryClass = {astro-ph},
       adsurl = {https://ui.adsabs.harvard.edu/abs/2003PASP..115..763C}
}

@ARTICLE{Chandrasekhar:1943,
       author = {{Chandrasekhar}, S.},
        title = "{Dynamical Friction. I. General Considerations: the Coefficient of Dynamical Friction.}",
      journal = {\apj},
         year = 1943,
        month = mar,
       volume = {97},
        pages = {255},
          doi = {10.1086/144517},
       adsurl = {https://ui.adsabs.harvard.edu/abs/1943ApJ....97..255C}
}

@ARTICLE{Bruzual:2003,
       author = {{Bruzual}, G. and {Charlot}, S.},
        title = "{Stellar population synthesis at the resolution of 2003}",
      journal = {\mnras},
         year = 2003,
        month = oct,
       volume = {344},
       number = {4},
        pages = {1000-1028},
          doi = {10.1046/j.1365-8711.2003.06897.x},
archivePrefix = {arXiv},
       eprint = {astro-ph/0309134},
 primaryClass = {astro-ph},
       adsurl = {https://ui.adsabs.harvard.edu/abs/2003MNRAS.344.1000B}
}

@ARTICLE{Yang:2020,
       author = {{Yang}, G. and {Boquien}, M. and {Buat}, V. and {Burgarella}, D. and {Ciesla}, L. and {Duras}, F. and {Stalevski}, M. and {Brandt}, W.~N. and {Papovich}, C.},
        title = "{X-CIGALE: Fitting AGN/galaxy SEDs from X-ray to infrared}",
      journal = {\mnras},
         year = 2020,
        month = jan,
       volume = {491},
       number = {1},
        pages = {740-757},
          doi = {10.1093/mnras/stz3001},
archivePrefix = {arXiv},
       eprint = {2001.08263},
 primaryClass = {astro-ph.GA},
       adsurl = {https://ui.adsabs.harvard.edu/abs/2020MNRAS.491..740Y}
}

@ARTICLE{Vazdekis:2010,
       author = {{Vazdekis}, A. and {S{\'a}nchez-Bl{\'a}zquez}, P. and {Falc{\'o}n-Barroso}, J. and {Cenarro}, A.~J. and {Beasley}, M.~A. and {Cardiel}, N. and {Gorgas}, J. and {Peletier}, R.~F.},
        title = "{Evolutionary stellar population synthesis with MILES - I. The base models and a new line index system}",
      journal = {\mnras},
         year = 2010,
        month = jun,
       volume = {404},
       number = {4},
        pages = {1639-1671},
          doi = {10.1111/j.1365-2966.2010.16407.x},
archivePrefix = {arXiv},
       eprint = {1004.4439},
 primaryClass = {astro-ph.CO},
       adsurl = {https://ui.adsabs.harvard.edu/abs/2010MNRAS.404.1639V}
}

@ARTICLE{Zhuang:2024,
       author = {{Zhuang}, Ming-Yang and {Shen}, Yue},
        title = "{Characterization of JWST NIRCam PSFs and Implications for AGN+host Image Decomposition}",
      journal = {\apj},
         year = 2024,
        month = feb,
       volume = {962},
       number = {2},
          eid = {139},
        pages = {139},
          doi = {10.3847/1538-4357/ad1183},
archivePrefix = {arXiv},
       eprint = {2304.13776},
 primaryClass = {astro-ph.GA},
       adsurl = {https://ui.adsabs.harvard.edu/abs/2024ApJ...962..139Z}
}

@ARTICLE{Rantala:2025,
       author = {{Rantala}, Antti and {Lah{\'e}n}, Natalia and {Naab}, Thorsten and {Escobar}, Gast{\'o}n J. and {Iorio}, Giuliano},
        title = "{FROST-CLUSTERS ─ II. Massive stars, binaries, and triples boost supermassive black hole seed formation in assembling star clusters}",
      journal = {\mnras},
         year = 2025,
        month = nov,
       volume = {543},
       number = {3},
        pages = {2130-2158},
          doi = {10.1093/mnras/staf1519},
archivePrefix = {arXiv},
       eprint = {2506.04330},
 primaryClass = {astro-ph.GA},
       adsurl = {https://ui.adsabs.harvard.edu/abs/2025MNRAS.543.2130R}
}

@ARTICLE{Devecchi:2009,
       author = {{Devecchi}, B. and {Volonteri}, M.},
        title = "{Formation of the First Nuclear Clusters and Massive Black Holes at High Redshift}",
      journal = {\apj},
         year = 2009,
        month = mar,
       volume = {694},
       number = {1},
        pages = {302-313},
          doi = {10.1088/0004-637X/694/1/302},
archivePrefix = {arXiv},
       eprint = {0810.1057},
 primaryClass = {astro-ph},
       adsurl = {https://ui.adsabs.harvard.edu/abs/2009ApJ...694..302D}
}

@article{Ubler:2019,
	adsurl = {https://ui.adsabs.harvard.edu/abs/2019ApJ...880...48U},
	archiveprefix = {arXiv},
	author = {{{\"U}bler}, H. and {Genzel}, R. and {Wisnioski}, E. and {F{\"o}rster Schreiber}, N.~M. and {Shimizu}, T.~T. and {Price}, S.~H. and {Tacconi}, L.~J. and {Belli}, S. and {Wilman}, D.~J. and {Fossati}, M. and {Mendel}, J.~T. and {Davies}, R.~L. and {Beifiori}, A. and {Bender}, R. and {Brammer}, G.~B. and {Burkert}, A. and {Chan}, J. and {Davies}, R.~I. and {Fabricius}, M. and {Galametz}, A. and {Herrera-Camus}, R. and {Lang}, P. and {Lutz}, D. and {Momcheva}, I.~G. and {Naab}, T. and {Nelson}, E.~J. and {Saglia}, R.~P. and {Tadaki}, K. and {van Dokkum}, P.~G. and {Wuyts}, S.},
	doi = {10.3847/1538-4357/ab27cc},
	eid = {48},
	eprint = {1906.02737},
	journal = {\apj},
	month = jul,
	number = {1},
	pages = {48},
	primaryclass = {astro-ph.GA},
	title = {{The Evolution and Origin of Ionized Gas Velocity Dispersion from z {\ensuremath{\sim}} 2.6 to z {\ensuremath{\sim}} 0.6 with KMOS$^{3D}$}},
	volume = {880},
	year = 2019}

@ARTICLE{Ubler:2025,
       author = {{{\"U}bler}, Hannah and {Mazzolari}, Giovanni and {Maiolino}, Roberto and {D'Eugenio}, Francesco and {Davari}, Nazanin and {Juod{\v{z}}balis}, Ignas and {Schneider}, Raffaella and {Valiante}, Rosa and {Arribas}, Santiago and {Bertola}, Elena and {Bunker}, Andrew J. and {Bromm}, Volker and {Carniani}, Stefano and {Charlot}, St{\'e}phane and {Cresci}, Giovanni and {Curti}, Mirko and {Davies}, Richard and {Eisenhauer}, Frank and {Fabian}, Andrew and {F{\"o}rster Schreiber}, Natascha M. and {Genzel}, Reinhard and {Inayoshi}, Kohei and {Ivey}, Lucy R. and {Jones}, Gareth C. and {Liu}, Boyuan and {Lutz}, Dieter and {Mackenzie}, Ruari and {Matthee}, Jorryt and {Parlanti}, Eleonora and {Perna}, Michele and {Robertson}, Brant and {Rodr{\'\i}guez del Pino}, Bruno and {Shimizu}, T. Taro and {Sijacki}, Debora and {Sturm}, Eckhard and {Tacchella}, Sandro and {Tacconi}, Linda and {Tozzi}, Giulia and {Trinca}, Alessandro and {Venturi}, Giacomo and {Volonteri}, Marta and {Willot}, Chris and {Zhang}, Saiyang},
        title = "{BlackTHUNDER: evidence for three massive black holes in a z\raisebox{-0.5ex}\textasciitilde5 galaxy}",
      journal = {arXiv e-prints},
         year = 2025,
        month = sep,
          eid = {arXiv:2509.21575},
        pages = {arXiv:2509.21575},
          doi = {10.48550/arXiv.2509.21575},
archivePrefix = {arXiv},
       eprint = {2509.21575},
 primaryClass = {astro-ph.GA},
       adsurl = {https://ui.adsabs.harvard.edu/abs/2025arXiv250921575U}
}

@ARTICLE{Marshall:2025,
       author = {{Marshall}, Madeline A. and {Yue}, Minghao and {Eilers}, Anna-Christina and {Scholtz}, Jan and {Perna}, Michele and {Willott}, Chris J. and {Maiolino}, Roberto and {{\"U}bler}, Hannah and {Arribas}, Santiago and {Bunker}, Andrew J. and {Charlot}, Stephane and {Rodr{\'\i}guez Del Pino}, Bruno and {B{\"o}ker}, Torsten and {Carniani}, Stefano and {Circosta}, Chiara and {Cresci}, Giovanni and {D'Eugenio}, Francesco and {Jones}, Gareth C. and {Venturi}, Giacomo and {Bordoloi}, Rongmon and {Kashino}, Daichi and {Mackenzie}, Ruari and {Matthee}, Jorryt and {Naidu}, Rohan and {Simcoe}, Robert A.},
        title = "{GA-NIFS and EIGER: A merging quasar host at z = 7 with an overmassive black hole}",
      journal = {\aap},
         year = 2025,
        month = oct,
       volume = {702},
          eid = {A50},
        pages = {A50},
          doi = {10.1051/0004-6361/202452650},
archivePrefix = {arXiv},
       eprint = {2410.11035},
 primaryClass = {astro-ph.GA},
       adsurl = {https://ui.adsabs.harvard.edu/abs/2025A&A...702A..50M}
}

@ARTICLE{Davies:2011,
       author = {{Davies}, Melvyn B. and {Miller}, M. Coleman and {Bellovary}, Jillian M.},
        title = "{Supermassive Black Hole Formation Via Gas Accretion in Nuclear Stellar Clusters}",
      journal = {\apjl},
         year = 2011,
        month = oct,
       volume = {740},
       number = {2},
          eid = {L42},
        pages = {L42},
          doi = {10.1088/2041-8205/740/2/L42},
archivePrefix = {arXiv},
       eprint = {1106.5943},
 primaryClass = {astro-ph.CO},
       adsurl = {https://ui.adsabs.harvard.edu/abs/2011ApJ...740L..42D}
}

@ARTICLE{Cresci:2023,
       author = {{Cresci}, G. and {Tozzi}, G. and {Perna}, M. and {Brusa}, M. and {Marconcini}, C. and {Marconi}, A. and {Carniani}, S. and {Brienza}, M. and {Giroletti}, M. and {Belfiore}, F. and {Ginolfi}, M. and {Mannucci}, F. and {Ulivi}, L. and {Scholtz}, J. and {Venturi}, G. and {Arribas}, S. and {{\"U}bler}, H. and {D'Eugenio}, F. and {Mingozzi}, M. and {Balmaverde}, B. and {Capetti}, A. and {Parlanti}, E. and {Zana}, T.},
        title = "{Bubbles and outflows: The novel JWST/NIRSpec view of the z = 1.59 obscured quasar XID2028}",
      journal = {\aap},
         year = 2023,
        month = apr,
       volume = {672},
          eid = {A128},
        pages = {A128},
          doi = {10.1051/0004-6361/202346001},
archivePrefix = {arXiv},
       eprint = {2301.11060},
 primaryClass = {astro-ph.GA},
       adsurl = {https://ui.adsabs.harvard.edu/abs/2023A&A...672A.128C}
}

@ARTICLE{Baldwin:1981,
       author = {{Baldwin}, J.~A. and {Phillips}, M.~M. and {Terlevich}, R.},
        title = "{Classification parameters for the emission-line spectra of extragalactic objects.}",
      journal = {\pasp},
         year = 1981,
        month = feb,
       volume = {93},
        pages = {5-19},
          doi = {10.1086/130766},
       adsurl = {https://ui.adsabs.harvard.edu/abs/1981PASP...93....5B}
}

@ARTICLE{Peres:1962,
       author = {{Peres}, Asher},
        title = "{Classical Radiation Recoil}",
      journal = {Physical Review},
         year = 1962,
        month = dec,
       volume = {128},
       number = {5},
        pages = {2471-2475},
          doi = {10.1103/PhysRev.128.2471},
       adsurl = {https://ui.adsabs.harvard.edu/abs/1962PhRv..128.2471P}
}

@ARTICLE{Perez-Gonzalez:2026,
       author = {{P{\'e}rez-Gonz{\'a}lez}, Pablo G. and {Barro}, Guillermo and {Carniani}, Stefano and {D'Eugenio}, Francesco and {Rieke}, George H. and {Tripodi}, Roberta and {Bunker}, Andrew J. and {Ji}, Xihan and {Marques-Chaves}, Rui and {Schaerer}, Daniel and {Venturi}, Giacomo and {Ar{\'e}valo-Gonz{\'a}lez}, Flor and {Arribas}, Santiago and {Rinaldi}, Pierluigi and {Rodr{\'\i}guez Del Pino}, Bruno and {Witstok}, Joris and {Bhatawdekar}, Rachana and {Boogaard}, Leindert A. and {Charlot}, Stephane and {Chevallard}, Jacopo and {Costantin}, Luca and {Curti}, Mirko and {Curtis-Lake}, Emma and {Daddi}, Emanuele and {Davis}, Kelcey and {Dickinson}, Mark and {Donnan}, Callum T. and {Donnan}, Fergus R. and {Dunlop}, James S. and {Eisenstein}, Daniel J. and {Ferguson}, Henry C. and {Fern{\'a}ndez Aranda}, Rom{\'a}n and {Finkelstein}, Steven L. and {Fujimoto}, Seiji and {Gandolfi}, Giovanni and {Giavalisco}, Mauro and {Grogin}, Norman A. and {Hamed}, Mahmoud and {Hirschmann}, Michaela and {Kartaltepe}, Jeyhan S. and {Kocevski}, Dale D. and {Koekemoer}, Anton M. and {Leung}, Gene C.~K. and {Lofaro}, Cristina M. and {Lucas}, Ray A. and {McLeod}, Derek J. and {Melinder}, Jens and {{\"O}stlin}, Goran and {Papovich}, Casey and {Pentericci}, Laura and {P{\'e}rez-D{\'\i}az}, Borja and {Rieke}, Marcia and {Scholtz}, Jan and {Somerville}, Rachel S. and {Stanton}, Thomas M. and {Stevenson}, Struan D. and {Shivaei}, Irene and {Tacchella}, Sandro and {Trump}, Jonathan R. and {{\"U}bler}, Hannah and {Wang}, Xin and {Williams}, Christina C. and {Willmer}, Christopher N.~A. and {Yung}, L.~Y. Aaron and {Zhu}, Yongda},
        title = "{Little Red Dots: One Photometric Tag Concealing Diverse Spectroscopic Flavors of Massive Star Formation and Black Hole Activity}",
      journal = {arXiv e-prints},
         year = 2026,
        month = feb,
          eid = {arXiv:2602.20247},
        pages = {arXiv:2602.20247},
          doi = {10.48550/arXiv.2602.20247},
archivePrefix = {arXiv},
       eprint = {2602.20247},
 primaryClass = {astro-ph.GA},
       adsurl = {https://ui.adsabs.harvard.edu/abs/2026arXiv260220247P}
}

@ARTICLE{Perna:2023,
       author = {{Perna}, M. and {Arribas}, S. and {Marshall}, M. and {D'Eugenio}, F. and {{\"U}bler}, H. and {Bunker}, A. and {Charlot}, S. and {Carniani}, S. and {Jakobsen}, P. and {Maiolino}, R. and {Rodr{\'\i}guez Del Pino}, B. and {Willott}, C.~J. and {B{\"o}ker}, T. and {Circosta}, C. and {Cresci}, G. and {Curti}, M. and {Husemann}, B. and {Kumari}, N. and {Lamperti}, I. and {P{\'e}rez-Gonz{\'a}lez}, P.~G. and {Scholtz}, J.},
        title = "{GA-NIFS: The ultra-dense, interacting environment of a dual AGN at z {\ensuremath{\sim}} 3.3 revealed by JWST/NIRSpec IFS}",
      journal = {\aap},
         year = 2023,
        month = nov,
       volume = {679},
          eid = {A89},
        pages = {A89},
          doi = {10.1051/0004-6361/202346649},
archivePrefix = {arXiv},
       eprint = {2304.06756},
 primaryClass = {astro-ph.GA},
       adsurl = {https://ui.adsabs.harvard.edu/abs/2023A&A...679A..89P}
}

@ARTICLE{Perna:2025,
       author = {{Perna}, Michele and {Arribas}, Santiago and {Lamperti}, Isabella and {Circosta}, Chiara and {Bertola}, Elena and {P{\'e}rez-Gonz{\'a}lez}, Pablo G. and {D'Eugenio}, Francesco and {{\"U}bler}, Hannah and {Cresci}, Giovanni and {Volonteri}, Marta and {Mannucci}, Filippo and {Maiolino}, Roberto and {Rodr{\'\i}guez Del Pino}, Bruno and {B{\"o}ker}, Torsten and {Bunker}, Andrew J. and {Charlot}, St{\'e}phane and {Willott}, Chris J. and {Carniani}, Stefano and {Curti}, Mirko and {Jones}, Gareth C. and {Kumari}, Nimisha and {Marshall}, Madeline A. and {Venturi}, Giacomo and {Saxena}, Aayush and {Scholtz}, Jan and {Witstok}, Joris},
        title = "{GA-NIFS: High number of dual active galactic nuclei at z {\ensuremath{\sim}} 3}",
      journal = {\aap},
         year = 2025,
        month = apr,
       volume = {696},
          eid = {A59},
        pages = {A59},
          doi = {10.1051/0004-6361/202453430},
archivePrefix = {arXiv},
       eprint = {2310.03067},
 primaryClass = {astro-ph.GA},
       adsurl = {https://ui.adsabs.harvard.edu/abs/2025A&A...696A..59P}
}

@article{Wisnioski:2015,
	adsurl = {https://ui.adsabs.harvard.edu/abs/2015ApJ...799..209W},
	archiveprefix = {arXiv},
	author = {{Wisnioski}, E. and {F{\"o}rster Schreiber}, N.~M. and {Wuyts}, S. and {Wuyts}, E. and {Bandara}, K. and {Wilman}, D. and {Genzel}, R. and {Bender}, R. and {Davies}, R. and {Fossati}, M. and {Lang}, P. and {Mendel}, J.~T. and {Beifiori}, A. and {Brammer}, G. and {Chan}, J. and {Fabricius}, M. and {Fudamoto}, Y. and {Kulkarni}, S. and {Kurk}, J. and {Lutz}, D. and {Nelson}, E.~J. and {Momcheva}, I. and {Rosario}, D. and {Saglia}, R. and {Seitz}, S. and {Tacconi}, L.~J. and {van Dokkum}, P.~G.},
	doi = {10.1088/0004-637X/799/2/209},
	eid = {209},
	eprint = {1409.6791},
	journal = {\apj},
	month = feb,
	number = {2},
	pages = {209},
	primaryclass = {astro-ph.GA},
	title = {{The KMOS$^{3D}$ Survey: Design, First Results, and the Evolution of Galaxy Kinematics from 0.7 <= z <= 2.7}},
	volume = {799},
	year = 2015}

@ARTICLE{Pensabene:2020,
       author = {{Pensabene}, A. and {Carniani}, S. and {Perna}, M. and {Cresci}, G. and {Decarli}, R. and {Maiolino}, R. and {Marconi}, A.},
        title = "{The ALMA view of the high-redshift relation between supermassive black holes and their host galaxies}",
      journal = {\aap},
         year = 2020,
        month = may,
       volume = {637},
          eid = {A84},
        pages = {A84},
          doi = {10.1051/0004-6361/201936634},
archivePrefix = {arXiv},
       eprint = {2002.00958},
 primaryClass = {astro-ph.GA},
       adsurl = {https://ui.adsabs.harvard.edu/abs/2020A&A...637A..84P}
}

@ARTICLE{Hviding:2025,
       author = {{Hviding}, Raphael E. and {de Graaff}, Anna and {Miller}, Tim B. and {Setton}, David J. and {Greene}, Jenny E. and {Labb{\'e}}, Ivo and {Brammer}, Gabriel and {Bezanson}, Rachel and {Boogaard}, Leindert A. and {Cleri}, Nikko J. and {Leja}, Joel and {Maseda}, Michael V. and {McConachie}, Ian and {Matthee}, Jorryt and {Naidu}, Rohan P. and {Oesch}, Pascal A. and {Wang}, Bingjie and {Whitaker}, Katherine E. and {Williams}, Christina C.},
        title = "{RUBIES: A spectroscopic census of little red dots: All point sources with v-shaped continua have broad lines}",
      journal = {\aap},
         year = 2025,
        month = oct,
       volume = {702},
          eid = {A57},
        pages = {A57},
          doi = {10.1051/0004-6361/202555816},
archivePrefix = {arXiv},
       eprint = {2506.05459},
 primaryClass = {astro-ph.GA},
       adsurl = {https://ui.adsabs.harvard.edu/abs/2025A&A...702A..57H}
}

@BOOK{Osterbrock:2006,
       author = {{Osterbrock}, Donald E. and {Ferland}, Gary J.},
        title = "{Astrophysics of gaseous nebulae and active galactic nuclei}",
         year = 2006,
       adsurl = {https://ui.adsabs.harvard.edu/abs/2006agna.book.....O}
}

@ARTICLE{Solanes:2018,
       author = {{Solanes}, J.~M. and {Perea}, J.~D. and {Valent{\'\i}-Rojas}, G.},
        title = "{Timescales of major mergers from simulations of isolated binary galaxy collisions}",
      journal = {\aap},
         year = 2018,
        month = jun,
       volume = {614},
          eid = {A66},
        pages = {A66},
          doi = {10.1051/0004-6361/201832855},
archivePrefix = {arXiv},
       eprint = {1804.10559},
 primaryClass = {astro-ph.GA},
       adsurl = {https://ui.adsabs.harvard.edu/abs/2018A&A...614A..66S}
}

@ARTICLE{Accard:2025,
       author = {{Accard}, C. and {B{\'e}thermin}, M. and {Boquien}, M. and {Buat}, V. and {Vallini}, L. and {Renaud}, F. and {Kraljic}, K. and {Aravena}, M. and {Cassata}, P. and {da Cunha}, E. and {Dam}, P. and {de Looze}, I. and {Dessauges-Zavadsky}, M. and {Dubois}, Y. and {Faisst}, A. and {Fudamoto}, Y. and {Ginolfi}, M. and {Gruppioni}, C. and {Han}, S. and {Herrera-Camus}, R. and {Inami}, H. and {Koekemoer}, A.~M. and {Lemaux}, B.~C. and {Li}, J. and {Li}, Y. and {Mobasher}, B. and {Molina}, J. and {Nanni}, A. and {Palla}, M. and {Pozzi}, F. and {Rela{\~n}o}, M. and {Romano}, M. and {Sawant}, P. and {Spilker}, J. and {Tsujita}, A. and {Veraldi}, E. and {Villanueva}, V. and {Wang}, W. and {Yi}, S.~K. and {Zamorani}, G.},
        title = "{The ALPINE-CRISTAL-JWST survey: Spatially resolved star formation relations at z {\ensuremath{\sim}} 5}",
      journal = {\aap},
         year = 2025,
        month = oct,
       volume = {702},
          eid = {A206},
        pages = {A206},
          doi = {10.1051/0004-6361/202556140},
archivePrefix = {arXiv},
       eprint = {2508.13136},
 primaryClass = {astro-ph.GA},
       adsurl = {https://ui.adsabs.harvard.edu/abs/2025A&A...702A.206A}
}

@ARTICLE{Allen:2008,
       author = {{Allen}, Mark G. and {Groves}, Brent A. and {Dopita}, Michael A. and {Sutherland}, Ralph S. and {Kewley}, Lisa J.},
        title = "{The MAPPINGS III Library of Fast Radiative Shock Models}",
      journal = {\apjs},
         year = 2008,
        month = sep,
       volume = {178},
       number = {1},
        pages = {20-55},
          doi = {10.1086/589652},
archivePrefix = {arXiv},
       eprint = {0805.0204},
 primaryClass = {astro-ph},
       adsurl = {https://ui.adsabs.harvard.edu/abs/2008ApJS..178...20A}
}

@ARTICLE{Allevato:2012,
       author = {{Allevato}, V. and {Finoguenov}, A. and {Hasinger}, G. and {Miyaji}, T. and {Cappelluti}, N. and {Salvato}, M. and {Zamorani}, G. and {Gilli}, R. and {George}, M.~R. and {Tanaka}, M. and {Brusa}, M. and {Silverman}, J. and {Civano}, F. and {Elvis}, M. and {Shankar}, F.},
        title = "{Occupation of X-Ray-selected Galaxy Groups by X-Ray Active Galactic Nuclei}",
      journal = {\apj},
         year = 2012,
        month = oct,
       volume = {758},
       number = {1},
          eid = {47},
        pages = {47},
          doi = {10.1088/0004-637X/758/1/47},
archivePrefix = {arXiv},
       eprint = {1209.2420},
 primaryClass = {astro-ph.CO},
       adsurl = {https://ui.adsabs.harvard.edu/abs/2012ApJ...758...47A}
}

@ARTICLE{Veilleux:1987,
       author = {{Veilleux}, Sylvain and {Osterbrock}, Donald E.},
        title = "{Spectral Classification of Emission-Line Galaxies}",
      journal = {\apjs},
         year = 1987,
        month = feb,
       volume = {63},
        pages = {295},
          doi = {10.1086/191166},
       adsurl = {https://ui.adsabs.harvard.edu/abs/1987ApJS...63..295V}
}

@ARTICLE{Vergara:2025,
       author = {{Vergara}, Marcelo C. and {Askar}, Abbas and {Kamlah}, Albrecht W.~H. and {Spurzem}, Rainer and {Flammini Dotti}, Francesco and {Schleicher}, Dominik R.~G. and {Arca Sedda}, Manuel and {Hypki}, Arkadiusz and {Giersz}, Mirek and {Hurley}, Jarrod and {Berczik}, Peter and {Escala}, Andres and {Hoyer}, Nils and {Neumayer}, Nadine and {Pang}, Xiaoying and {Tanikawa}, Ataru and {Cen}, Renyue and {Naab}, Thorsten},
        title = "{Rapid formation of a very massive star (>50000 M$_{{\ensuremath{\odot}}}$), and subsequently, of an IMBH, from runaway collisions: Direct N-body and Monte Carlo simulations of dense star clusters}",
      journal = {\aap},
         year = 2025,
        month = dec,
       volume = {704},
          eid = {A321},
        pages = {A321},
          doi = {10.1051/0004-6361/202555307},
archivePrefix = {arXiv},
       eprint = {2505.07491},
 primaryClass = {astro-ph.GA},
       adsurl = {https://ui.adsabs.harvard.edu/abs/2025A&A...704A.321V}
}

@ARTICLE{Reines:2013,
       author = {{Reines}, Amy E. and {Greene}, Jenny E. and {Geha}, Marla},
        title = "{Dwarf Galaxies with Optical Signatures of Active Massive Black Holes}",
      journal = {\apj},
         year = 2013,
        month = oct,
       volume = {775},
       number = {2},
          eid = {116},
        pages = {116},
          doi = {10.1088/0004-637X/775/2/116},
archivePrefix = {arXiv},
       eprint = {1308.0328},
 primaryClass = {astro-ph.CO},
       adsurl = {https://ui.adsabs.harvard.edu/abs/2013ApJ...775..116R}
}

@ARTICLE{Simmonds:2025,
       author = {{Simmonds}, C. and {Tacchella}, S. and {McClymont}, W. and {Curtis-Lake}, E. and {D'Eugenio}, F. and {Hainline}, K. and {Johnson}, B.~D. and {Kravtsov}, A. and {Pusk{\'a}s}, D. and {Robertson}, B. and {Stoffers}, A. and {Willott}, C. and {Baker}, W.~M. and {Belokurov}, V.~A. and {Bhatawdekar}, R. and {Bunker}, A.~J. and {Carniani}, S. and {Chevallard}, J. and {Curti}, M. and {Duan}, Q. and {Helton}, J.~M. and {Ji}, Z. and {Looser}, T.~J. and {Maiolino}, R. and {Maseda}, M.~V. and {Shivaei}, I. and {Williams}, C.~C.},
        title = "{Bursting at the seams: the star-forming main sequence and its scatter at z = 3─9 using NIRCam photometry from JADES}",
      journal = {\mnras},
         year = 2025,
        month = dec,
       volume = {544},
       number = {4},
        pages = {4551-4575},
          doi = {10.1093/mnras/staf1950},
archivePrefix = {arXiv},
       eprint = {2508.04410},
 primaryClass = {astro-ph.GA},
       adsurl = {https://ui.adsabs.harvard.edu/abs/2025MNRAS.544.4551S}
}

@ARTICLE{Pollock:2026,
       author = {{Pollock}, Clara L. and {Gottumukkala}, Rashmi and {Heintz}, Kasper E. and {Brammer}, Gabriel B. and {Roberts-Borsani}, Guido and {Oesch}, Pascal A. and {Witstok}, Joris and {Arellano-C{\'o}rdova}, Karla Z. and {Cullen}, Fergus and {Scholte}, Dirk and {Terp}, Chamilla and {Rowland}, Lucie and {Sneppen}, Albert and {Ito}, Kei and {Valentino}, Francesco and {Matthee}, Jorryt and {Watson}, Darach and {Toft}, Sune},
        title = "{Novel z {\ensuremath{\sim}} 10 auroral line measurements extend the gradual offset of the fundamental metallicity relation deep into the first gigayear of cosmic time}",
      journal = {\aap},
         year = 2026,
        month = apr,
       volume = {708},
          eid = {A203},
        pages = {A203},
          doi = {10.1051/0004-6361/202556032},
archivePrefix = {arXiv},
       eprint = {2506.15779},
 primaryClass = {astro-ph.GA},
       adsurl = {https://ui.adsabs.harvard.edu/abs/2026A&A...708A.203P}
}

@ARTICLE{Powell:2018,
       author = {{Powell}, M.~C. and {Cappelluti}, N. and {Urry}, C.~M. and {Koss}, M. and {Finoguenov}, A. and {Ricci}, C. and {Trakhtenbrot}, B. and {Allevato}, V. and {Ajello}, M. and {Oh}, K. and {Schawinski}, K. and {Secrest}, N.},
        title = "{The Swift/BAT AGN Spectroscopic Survey. IX. The Clustering Environments of an Unbiased Sample of Local AGNs}",
      journal = {\apj},
         year = 2018,
        month = may,
       volume = {858},
       number = {2},
          eid = {110},
        pages = {110},
          doi = {10.3847/1538-4357/aabd7f},
archivePrefix = {arXiv},
       eprint = {1803.07589},
 primaryClass = {astro-ph.GA},
       adsurl = {https://ui.adsabs.harvard.edu/abs/2018ApJ...858..110P}
}

@ARTICLE{Bekenstein:1973,
       author = {{Bekenstein}, Jacob D.},
        title = "{Gravitational-Radiation Recoil and Runaway Black Holes}",
      journal = {\apj},
         year = 1973,
        month = jul,
       volume = {183},
        pages = {657-664},
          doi = {10.1086/152255},
       adsurl = {https://ui.adsabs.harvard.edu/abs/1973ApJ...183..657B}
}

@ARTICLE{Bromm:2002,
       author = {{Bromm}, Volker and {Coppi}, Paolo S. and {Larson}, Richard B.},
        title = "{The Formation of the First Stars. I. The Primordial Star-forming Cloud}",
      journal = {\apj},
         year = 2002,
        month = jan,
       volume = {564},
       number = {1},
        pages = {23-51},
          doi = {10.1086/323947},
archivePrefix = {arXiv},
       eprint = {astro-ph/0102503},
 primaryClass = {astro-ph},
       adsurl = {https://ui.adsabs.harvard.edu/abs/2002ApJ...564...23B}
}

@ARTICLE{Bromm:2003,
       author = {{Bromm}, Volker and {Loeb}, Abraham},
        title = "{Formation of the First Supermassive Black Holes}",
      journal = {\apj},
         year = 2003,
        month = oct,
       volume = {596},
       number = {1},
        pages = {34-46},
          doi = {10.1086/377529},
archivePrefix = {arXiv},
       eprint = {astro-ph/0212400},
 primaryClass = {astro-ph},
       adsurl = {https://ui.adsabs.harvard.edu/abs/2003ApJ...596...34B}
}

@ARTICLE{Sanders:2026,
       author = {{Sanders}, Ryan L. and {Shapley}, Alice E. and {Topping}, Michael W. and {Reddy}, Naveen A. and {Berg}, Danielle A. and {Khostovan}, Ali Ahmad and {Bouwens}, Rychard J. and {Brammer}, Gabriel and {Carnall}, Adam C. and {Cullen}, Fergus and {Dav{\'e}}, Romeel and {Dunlop}, James S. and {Ellis}, Richard S. and {F{\"o}rster Schreiber}, N.~M. and {Furlanetto}, Steven R. and {Glazebrook}, Karl and {Illingworth}, Garth D. and {Jones}, Tucker and {Kriek}, Mariska and {McLeod}, Derek J. and {McLure}, Ross J. and {Narayanan}, Desika and {Oesch}, Pascal A. and {Pahl}, Anthony J. and {Pettini}, Max and {Schaerer}, Daniel and {Stark}, Daniel P. and {Steidel}, Charles C. and {Tang}, Mengtao and {Clarke}, Leonardo and {Donnan}, Callum T. and {Kehoe}, Emily},
        title = "{The AURORA Survey: High-redshift Empirical Metallicity Calibrations from Electron Temperature Measurements at z = 2─10}",
      journal = {\apj},
         year = 2026,
        month = jun,
       volume = {1003},
       number = {2},
          eid = {228},
        pages = {228},
          doi = {10.3847/1538-4357/ae66e2},
archivePrefix = {arXiv},
       eprint = {2508.10099},
 primaryClass = {astro-ph.GA},
       adsurl = {https://ui.adsabs.harvard.edu/abs/2026ApJ..1003..228S}
}

@ARTICLE{Nakajima:2022,
       author = {{Nakajima}, K. and {Maiolino}, R.},
        title = "{Diagnostics for PopIII galaxies and direct collapse black holes in the early universe}",
      journal = {\mnras},
         year = 2022,
        month = jul,
       volume = {513},
       number = {4},
        pages = {5134-5147},
          doi = {10.1093/mnras/stac1242},
archivePrefix = {arXiv},
       eprint = {2204.11870},
 primaryClass = {astro-ph.GA},
       adsurl = {https://ui.adsabs.harvard.edu/abs/2022MNRAS.513.5134N}
}

@ARTICLE{Labbe:2024,
       author = {{Labbe}, Ivo and {Greene}, Jenny E. and {Matthee}, Jorryt and {Treiber}, Helena and {Kokorev}, Vasily and {Miller}, Tim B. and {Kramarenko}, Ivan and {Setton}, David J. and {Ma}, Yilun and {Goulding}, Andy D. and {Bezanson}, Rachel and {Naidu}, Rohan P. and {Williams}, Christina C. and {Atek}, Hakim and {Brammer}, Gabriel and {Cutler}, Sam E. and {Chemerynska}, Iryna and {Cloonan}, Aidan P. and {Dayal}, Pratika and {de Graaff}, Anna and {Fudamoto}, Yoshinobu and {Fujimoto}, Seiji and {Furtak}, Lukas J. and {Glazebrook}, Karl and {Heintz}, Kasper E. and {Leja}, Joel and {Marchesini}, Danilo and {Nanayakkara}, Themiya and {Nelson}, Erica J. and {Oesch}, Pascal A. and {Pan}, Richard and {Price}, Sedona H. and {Shivaei}, Irene and {Sobral}, David and {Suess}, Katherine A. and {van Dokkum}, Pieter and {Wang}, Bingjie and {Weaver}, John R. and {Whitaker}, Katherine E. and {Zitrin}, Adi},
        title = "{An unambiguous AGN and a Balmer break in an Ultraluminous Little Red Dot at z=4.47 from Ultradeep UNCOVER and All the Little Things Spectroscopy}",
      journal = {arXiv e-prints},
         year = 2024,
        month = dec,
          eid = {arXiv:2412.04557},
        pages = {arXiv:2412.04557},
          doi = {10.48550/arXiv.2412.04557},
archivePrefix = {arXiv},
       eprint = {2412.04557},
 primaryClass = {astro-ph.GA},
       adsurl = {https://ui.adsabs.harvard.edu/abs/2024arXiv241204557L}
}

@ARTICLE{Lacey:1993,
       author = {{Lacey}, Cedric and {Cole}, Shaun},
        title = "{Merger rates in hierarchical models of galaxy formation}",
      journal = {\mnras},
         year = 1993,
        month = jun,
       volume = {262},
       number = {3},
        pages = {627-649},
          doi = {10.1093/mnras/262.3.627},
       adsurl = {https://ui.adsabs.harvard.edu/abs/1993MNRAS.262..627L}
}

@ARTICLE{Lambrides:2026,
       author = {{Lambrides}, Erini and {Larson}, Rebecca L. and {Garofali}, Kristen and {Ptak}, Andrew and {Chiaberge}, Marco and {Long}, Arianna S. and {Hutchison}, Taylor A. and {Norman}, Colin and {McKinney}, Jed and {Akins}, Hollis B. and {Berg}, Danielle A. and {Chisholm}, John and {Civano}, Francesca and {Cloonan}, Aidan P. and {Endsley}, Ryan and {Faisst}, Andreas L. and {Gilli}, Roberto and {Gillman}, Steven and {Hirschmann}, Michaela and {Kartaltepe}, Jeyhan S. and {Kocevski}, Dale D. and {Kokorev}, Vasily and {Pacucci}, Fabio and {Richardson}, Chris T. and {Stiavelli}, Massimo and {Whalen}, Kelly E.},
        title = "{The case for super-Eddington accretion in JWST broad-line active galactic nuclei during the first billion years}",
      journal = {Nature Astronomy},
         year = 2026,
        month = jun,
       volume = {10},
        pages = {868-879},
          doi = {10.1038/s41550-026-02813-w},
archivePrefix = {arXiv},
       eprint = {2409.13047},
 primaryClass = {astro-ph.HE},
       adsurl = {https://ui.adsabs.harvard.edu/abs/2026NatAs..10..868L}
}

@INPROCEEDINGS{Davies:2004,
       author = {{Davies}, R.~I. and {Tacconi}, L.~J. and {Genzel}, R.},
        title = "{Star formation and dynamics in the nuclei of AGN}",
    booktitle = {The Interplay Among Black Holes, Stars and ISM in Galactic Nuclei},
         year = 2004,
       editor = {{Storchi-Bergmann}, Thaisa and {Ho}, Luis C. and {Schmitt}, Henrique R.},
       series = {IAU Symposium},
       volume = {222},
        month = nov,
        pages = {133-136},
          doi = {10.1017/S1743921304001759},
archivePrefix = {arXiv},
       eprint = {astro-ph/0404297},
 primaryClass = {astro-ph},
       adsurl = {https://ui.adsabs.harvard.edu/abs/2004IAUS..222..133D}
}

@ARTICLE{Yue:2024,
         author = {{Yue}, Minghao and {Eilers}, Anna-Christina and {Ananna}, Tonima Tasnim and {Panagiotou}, Christos and {Kara}, Erin and {Miyaji}, Takamitsu},
        title = "{Stacking X-Ray Observations of ``Little Red Dots'': Implications for Their Active Galactic Nucleus Properties}",
      journal = {\apjl},
         year = 2024,
        month = oct,
       volume = {974},
       number = {2},
          eid = {L26},
        pages = {L26},
          doi = {10.3847/2041-8213/ad7eba},
archivePrefix = {arXiv},
       eprint = {2404.13290},
 primaryClass = {astro-ph.GA},
       adsurl = {https://ui.adsabs.harvard.edu/abs/2024ApJ...974L..26Y}
}

@ARTICLE{Yung:2023,
       author = {{Yung}, L.~Y. Aaron and {Somerville}, Rachel S. and {Finkelstein}, Steven L. and {Behroozi}, Peter and {Dav{\'e}}, Romeel and {Ferguson}, Henry C. and {Gardner}, Jonathan P. and {Popping}, Gerg{\"o} and {Malhotra}, Sangeeta and {Papovich}, Casey and {Rhoads}, James E. and {Bagley}, Micaela B. and {Hirschmann}, Michaela and {Koekemoer}, Anton M.},
        title = "{Semi-analytic forecasts for Roman - the beginning of a new era of deep-wide galaxy surveys}",
      journal = {\mnras},
         year = 2023,
        month = feb,
       volume = {519},
       number = {1},
        pages = {1578-1600},
          doi = {10.1093/mnras/stac3595},
archivePrefix = {arXiv},
       eprint = {2210.04902},
 primaryClass = {astro-ph.GA},
       adsurl = {https://ui.adsabs.harvard.edu/abs/2023MNRAS.519.1578Y}
}

@ARTICLE{Golubchik:2026,
       author = {{Golubchik}, Miriam and {Furtak}, Lukas J. and {Allingham}, Joseph F.~V. and {Zitrin}, Adi and {Akins}, Hollis B. and {Kokorev}, Vasily and {Fujimoto}, Seiji and {Abdurro'uf} and {Amor{\'\i}n}, Ricardo O. and {Bauer}, Franz E. and {Bezanson}, Rachel and {Brada{\v{c}}}, Marusa and {Bradley}, Larry D. and {Brammer}, Gabriel B. and {Chisholm}, John and {Coe}, Dan and {Conselice}, Christopher J. and {Dayal}, Pratika and {Dessauges-Zavadsky}, Miroslava and {Diego}, Jose M. and {Faisst}, Andreas L. and {Fei}, Qinyue and {Ferguson}, Henry C. and {Finkelstein}, Steven L. and {Frye}, Brenda L. and {Gonz{\'a}lez-Otero}, Mauro and {Greene}, Jenny E. and {Harikane}, Yuichi and {Hsiao}, Tiger Yu-Yang and {Inayoshi}, Kohei and {Jim{\'e}nez-Teja}, Yolanda and {Knudsen}, Kirsten and {Koekemoer}, Anton M. and {Labb{\'e}}, Ivo and {Lucas}, Ray A. and {Magdis}, Georgios E. and {Matthee}, Jorryt and {Messa}, Matteo and {Naidu}, Rohan P. and {Nakane}, Minami and {Noirot}, Ga{\"e}l and {Pan}, Richard and {Papovich}, Casey and {Richard}, Johan and {Ricotti}, Massimo and {Robbins}, Luke and {Stark}, Daniel P. and {Sun}, Fengwu and {Treu}, Tommaso and {Tripodi}, Roberta and {Vanzella}, Eros and {Willott}, Chris and {Windhorst}, Rogier A.},
        title = "{VENUS: When red meets blue: A multiply imaged little red dot with an apparent blue companion behind the galaxy cluster Abell 383}",
      journal = {\aap},
         year = 2026,
        month = jun,
       volume = {710},
          eid = {A226},
        pages = {A226},
          doi = {10.1051/0004-6361/202558407},
archivePrefix = {arXiv},
       eprint = {2512.02117},
 primaryClass = {astro-ph.GA},
       adsurl = {https://ui.adsabs.harvard.edu/abs/2026A&A...710A.226G}
}

@ARTICLE{Nakajima:2023,
       author = {{Nakajima}, Kimihiko and {Ouchi}, Masami and {Isobe}, Yuki and {Harikane}, Yuichi and {Zhang}, Yechi and {Ono}, Yoshiaki and {Umeda}, Hiroya and {Oguri}, Masamune},
        title = "{JWST Census for the Mass-Metallicity Star Formation Relations at z = 4-10 with Self-consistent Flux Calibration and Proper Metallicity Calibrators}",
      journal = {\apjs},
         year = 2023,
        month = dec,
       volume = {269},
       number = {2},
          eid = {33},
        pages = {33},
          doi = {10.3847/1538-4365/acd556},
archivePrefix = {arXiv},
       eprint = {2301.12825},
 primaryClass = {astro-ph.GA},
       adsurl = {https://ui.adsabs.harvard.edu/abs/2023ApJS..269...33N}
}

@ARTICLE{Abazajian:2009,
       author = {{Abazajian}, Kevork N. and {Adelman-McCarthy}, Jennifer K. and {Ag{\"u}eros}, Marcel A. and {Allam}, Sahar S. and {Allende Prieto}, Carlos and {An}, Deokkeun and {Anderson}, Kurt S.~J. and {Anderson}, Scott F. and {Annis}, James and {Bahcall}, Neta A. and {Bailer-Jones}, C.~A.~L. and {Barentine}, J.~C. and {Bassett}, Bruce A. and {Becker}, Andrew C. and {Beers}, Timothy C. and {Bell}, Eric F. and {Belokurov}, Vasily and {Berlind}, Andreas A. and {Berman}, Eileen F. and {Bernardi}, Mariangela and {Bickerton}, Steven J. and {Bizyaev}, Dmitry and {Blakeslee}, John P. and {Blanton}, Michael R. and {Bochanski}, John J. and {Boroski}, William N. and {Brewington}, Howard J. and {Brinchmann}, Jarle and {Brinkmann}, J. and {Brunner}, Robert J. and {Budav{\'a}ri}, Tam{\'a}s and {Carey}, Larry N. and {Carliles}, Samuel and {Carr}, Michael A. and {Castander}, Francisco J. and {Cinabro}, David and {Connolly}, A.~J. and {Csabai}, Istv{\'a}n and {Cunha}, Carlos E. and {Czarapata}, Paul C. and {Davenport}, James R.~A. and {de Haas}, Ernst and {Dilday}, Ben and {Doi}, Mamoru and {Eisenstein}, Daniel J. and {Evans}, Michael L. and {Evans}, N.~W. and {Fan}, Xiaohui and {Friedman}, Scott D. and {Frieman}, Joshua A. and {Fukugita}, Masataka and {G{\"a}nsicke}, Boris T. and {Gates}, Evalyn and {Gillespie}, Bruce and {Gilmore}, G. and {Gonzalez}, Belinda and {Gonzalez}, Carlos F. and {Grebel}, Eva K. and {Gunn}, James E. and {Gy{\"o}ry}, Zsuzsanna and {Hall}, Patrick B. and {Harding}, Paul and {Harris}, Frederick H. and {Harvanek}, Michael and {Hawley}, Suzanne L. and {Hayes}, Jeffrey J.~E. and {Heckman}, Timothy M. and {Hendry}, John S. and {Hennessy}, Gregory S. and {Hindsley}, Robert B. and {Hoblitt}, J. and {Hogan}, Craig J. and {Hogg}, David W. and {Holtzman}, Jon A. and {Hyde}, Joseph B. and {Ichikawa}, Shin-ichi and {Ichikawa}, Takashi and {Im}, Myungshin and {Ivezi{\'c}}, {\v{Z}}eljko and {Jester}, Sebastian and {Jiang}, Linhua and {Johnson}, Jennifer A. and {Jorgensen}, Anders M. and {Juri{\'c}}, Mario and {Kent}, Stephen M. and {Kessler}, R. and {Kleinman}, S.~J. and {Knapp}, G.~R. and {Konishi}, Kohki and {Kron}, Richard G. and {Krzesinski}, Jurek and {Kuropatkin}, Nikolay and {Lampeitl}, Hubert and {Lebedeva}, Svetlana and {Lee}, Myung Gyoon and {Lee}, Young Sun and {French Leger}, R. and {L{\'e}pine}, S{\'e}bastien and {Li}, Nolan and {Lima}, Marcos and {Lin}, Huan and {Long}, Daniel C. and {Loomis}, Craig P. and {Loveday}, Jon and {Lupton}, Robert H. and {Magnier}, Eugene and {Malanushenko}, Olena and {Malanushenko}, Viktor and {Mandelbaum}, Rachel and {Margon}, Bruce and {Marriner}, John P. and {Mart{\'\i}nez-Delgado}, David and {Matsubara}, Takahiko and {McGehee}, Peregrine M. and {McKay}, Timothy A. and {Meiksin}, Avery and {Morrison}, Heather L. and {Mullally}, Fergal and {Munn}, Jeffrey A. and {Murphy}, Tara and {Nash}, Thomas and {Nebot}, Ada and {Neilsen}, Jr., Eric H. and {Newberg}, Heidi Jo and {Newman}, Peter R. and {Nichol}, Robert C. and {Nicinski}, Tom and {Nieto-Santisteban}, Maria and {Nitta}, Atsuko and {Okamura}, Sadanori and {Oravetz}, Daniel J. and {Ostriker}, Jeremiah P. and {Owen}, Russell and {Padmanabhan}, Nikhil and {Pan}, Kaike and {Park}, Changbom and {Pauls}, George and {Peoples}, Jr., John and {Percival}, Will J. and {Pier}, Jeffrey R. and {Pope}, Adrian C. and {Pourbaix}, Dimitri and {Price}, Paul A. and {Purger}, Norbert and {Quinn}, Thomas and {Raddick}, M. Jordan and {Re Fiorentin}, Paola and {Richards}, Gordon T. and {Richmond}, Michael W. and {Riess}, Adam G. and {Rix}, Hans-Walter and {Rockosi}, Constance M. and {Sako}, Masao and {Schlegel}, David J. and {Schneider}, Donald P. and {Scholz}, Ralf-Dieter and {Schreiber}, Matthias R. and {Schwope}, Axel D. and {Seljak}, Uro{\v{s}} and {Sesar}, Branimir and {Sheldon}, Erin and {Shimasaku}, Kazu and {Sibley}, Valena C. and {Simmons}, A.~E. and {Sivarani}, Thirupathi and {Allyn Smith}, J. and {Smith}, Martin C. and {Smol{\v{c}}i{\'c}}, Vernesa and {Snedden}, Stephanie A. and {Stebbins}, Albert and {Steinmetz}, Matthias and {Stoughton}, Chris and {Strauss}, Michael A. and {SubbaRao}, Mark and {Suto}, Yasushi and {Szalay}, Alexander S. and {Szapudi}, Istv{\'a}n and {Szkody}, Paula and {Tanaka}, Masayuki and {Tegmark}, Max and {Teodoro}, Luis F.~A. and {Thakar}, Aniruddha R. and {Tremonti}, Christy A. and {Tucker}, Douglas L. and {Uomoto}, Alan and {Vanden Berk}, Daniel E. and {Vandenberg}, Jan and {Vidrih}, S. and {Vogeley}, Michael S. and {Voges}, Wolfgang and {Vogt}, Nicole P. and {Wadadekar}, Yogesh and {Watters}, Shannon and {Weinberg}, David H. and {West}, Andrew A. and {White}, Simon D.~M. and {Wilhite}, Brian C. and {Wonders}, Alainna C. and {Yanny}, Brian and {Yocum}, D.~R.},
        title = "{The Seventh Data Release of the Sloan Digital Sky Survey}",
      journal = {\apjs},
         year = 2009,
        month = jun,
       volume = {182},
       number = {2},
        pages = {543-558},
          doi = {10.1088/0067-0049/182/2/543},
archivePrefix = {arXiv},
       eprint = {0812.0649},
 primaryClass = {astro-ph},
       adsurl = {https://ui.adsabs.harvard.edu/abs/2009ApJS..182..543A}
}

@ARTICLE{Abuter:2024,
       author = {{Abuter}, R. and {Allouche}, F. and {Amorim}, A. and {Bailet}, C. and {Berdeu}, A. and {Berger}, J.-P. and {Berio}, P. and {Bigioli}, A. and {Boebion}, O. and {Bolzer}, M.-L. and {Bonnet}, H. and {Bourdarot}, G. and {Bourget}, P. and {Brandner}, W. and {Cao}, Y. and {Conzelmann}, R. and {Comin}, M. and {Cl{\'e}net}, Y. and {Courtney-Barrer}, B. and {Davies}, R. and {Defr{\`e}re}, D. and {Delboulb{\'e}}, A. and {Delplancke-Str{\"o}bele}, F. and {Dembet}, R. and {Dexter}, J. and {de Zeeuw}, P.~T. and {Drescher}, A. and {Eckart}, A. and {{\'E}douard}, C. and {Eisenhauer}, F. and {Fabricius}, M. and {Feuchtgruber}, H. and {Finger}, G. and {F{\"o}rster Schreiber}, N.~M. and {Garcia}, P. and {Garcia Lopez}, R. and {Gao}, F. and {Gendron}, E. and {Genzel}, R. and {Gil}, J.~P. and {Gillessen}, S. and {Gomes}, T. and {Gont{\'e}}, F. and {Gouvret}, C. and {Guajardo}, P. and {Guieu}, S. and {Hackenberg}, W. and {Haddad}, N. and {Hartl}, M. and {Haubois}, X. and {Hau{\ss}mann}, F. and {Hei{\ss}el}, G. and {Henning}, Th. and {Hippler}, S. and {H{\"o}nig}, S.~F. and {Horrobin}, M. and {Hubin}, N. and {Jacqmart}, E. and {Jocou}, L. and {Kaufer}, A. and {Kervella}, P. and {Kolb}, J. and {Korhonen}, H. and {Lacour}, S. and {Lagarde}, S. and {Lai}, O. and {Lapeyr{\`e}re}, V. and {Laugier}, R. and {Le Bouquin}, J.-B. and {Leftley}, J. and {L{\'e}na}, P. and {Lewis}, S. and {Liu}, D. and {Lopez}, B. and {Lutz}, D. and {Magnard}, Y. and {Mang}, F. and {Marcotto}, A. and {Maurel}, D. and {M{\'e}rand}, A. and {Millour}, F. and {More}, N. and {Netzer}, H. and {Nowacki}, H. and {Nowak}, M. and {Oberti}, S. and {Ott}, T. and {Pallanca}, L. and {Paumard}, T. and {Perraut}, K. and {Perrin}, G. and {Petrov}, R. and {Pfuhl}, O. and {Pourr{\'e}}, N. and {Rabien}, S. and {Rau}, C. and {Riquelme}, M. and {Robbe-Dubois}, S. and {Rochat}, S. and {Salman}, M. and {Sanchez-Bermudez}, J. and {Santos}, D.~J.~D. and {Scheithauer}, S. and {Sch{\"o}ller}, M. and {Schubert}, J. and {Schuhler}, N. and {Shangguan}, J. and {Shchekaturov}, P. and {Shimizu}, T.~T. and {Sevin}, A. and {Soulez}, F. and {Spang}, A. and {Stadler}, E. and {Sternberg}, A. and {Straubmeier}, C. and {Sturm}, E. and {Sykes}, C. and {Tacconi}, L.~J. and {Tristram}, K.~R.~W. and {Vincent}, F. and {von Fellenberg}, S. and {Uysal}, S. and {Widmann}, F. and {Wieprecht}, E. and {Wiezorrek}, E. and {Woillez}, J. and {Zins}, G.},
        title = "{A dynamical measure of the black hole mass in a quasar 11 billion years ago}",
      journal = {\nat},
         year = 2024,
        month = mar,
       volume = {627},
       number = {8003},
        pages = {281-285},
          doi = {10.1038/s41586-024-07053-4},
archivePrefix = {arXiv},
       eprint = {2401.14567},
 primaryClass = {astro-ph.GA},
       adsurl = {https://ui.adsabs.harvard.edu/abs/2024Natur.627..281A}
}

@ARTICLE{Spilker:2016,
       author = {{Spilker}, Justin S. and {Bezanson}, Rachel and {Marrone}, Daniel P. and {Weiner}, Benjamin J. and {Whitaker}, Katherine E. and {Williams}, Christina C.},
        title = "{Low Gas Fractions Connect Compact Star-forming Galaxies to Their z \raisebox{-0.5ex}\textasciitilde 2 Quiescent Descendants}",
      journal = {\apj},
         year = 2016,
        month = nov,
       volume = {832},
       number = {1},
          eid = {19},
        pages = {19},
          doi = {10.3847/0004-637X/832/1/19},
archivePrefix = {arXiv},
       eprint = {1607.01785},
 primaryClass = {astro-ph.GA},
       adsurl = {https://ui.adsabs.harvard.edu/abs/2016ApJ...832...19S}
}

@ARTICLE{Larson:2023,
       author = {{Larson}, Rebecca L. and {Finkelstein}, Steven L. and {Kocevski}, Dale D. and {Hutchison}, Taylor A. and {Trump}, Jonathan R. and {Arrabal Haro}, Pablo and {Bromm}, Volker and {Cleri}, Nikko J. and {Dickinson}, Mark and {Fujimoto}, Seiji and {Kartaltepe}, Jeyhan S. and {Koekemoer}, Anton M. and {Papovich}, Casey and {Pirzkal}, Nor and {Tacchella}, Sandro and {Zavala}, Jorge A. and {Bagley}, Micaela and {Behroozi}, Peter and {Champagne}, Jaclyn B. and {Cole}, Justin W. and {Jung}, Intae and {Morales}, Alexa M. and {Yang}, Guang and {Zhang}, Haowen and {Zitrin}, Adi and {Amor{\'\i}n}, Ricardo O. and {Burgarella}, Denis and {Casey}, Caitlin M. and {Ch{\'a}vez Ortiz}, {\'O}scar A. and {Cox}, Isabella G. and {Chworowsky}, Katherine and {Fontana}, Adriano and {Gawiser}, Eric and {Grazian}, Andrea and {Grogin}, Norman A. and {Harish}, Santosh and {Hathi}, Nimish P. and {Hirschmann}, Michaela and {Holwerda}, Benne W. and {Juneau}, St{\'e}phanie and {Leung}, Gene C.~K. and {Lucas}, Ray A. and {McGrath}, Elizabeth J. and {P{\'e}rez-Gonz{\'a}lez}, Pablo G. and {Rigby}, Jane R. and {Seill{\'e}}, Lise-Marie and {Simons}, Raymond C. and {de La Vega}, Alexander and {Weiner}, Benjamin J. and {Wilkins}, Stephen M. and {Yung}, L.~Y. Aaron and {Ceers Team}},
        title = "{A CEERS Discovery of an Accreting Supermassive Black Hole 570 Myr after the Big Bang: Identifying a Progenitor of Massive z > 6 Quasars}",
      journal = {\apjl},
         year = 2023,
        month = aug,
       volume = {953},
       number = {2},
          eid = {L29},
        pages = {L29},
          doi = {10.3847/2041-8213/ace619},
archivePrefix = {arXiv},
       eprint = {2303.08918},
 primaryClass = {astro-ph.GA},
       adsurl = {https://ui.adsabs.harvard.edu/abs/2023ApJ...953L..29L}
}

@ARTICLE{Ananna:2024,
       author = {{Ananna}, Tonima Tasnim and {Bogd{\'a}n}, {\'A}kos and {Kov{\'a}cs}, Orsolya E. and {Natarajan}, Priyamvada and {Hickox}, Ryan C.},
        title = "{X-Ray View of Little Red Dots: Do They Host Supermassive Black Holes?}",
      journal = {\apjl},
         year = 2024,
        month = jul,
       volume = {969},
       number = {1},
          eid = {L18},
        pages = {L18},
          doi = {10.3847/2041-8213/ad5669},
archivePrefix = {arXiv},
       eprint = {2404.19010},
 primaryClass = {astro-ph.GA},
       adsurl = {https://ui.adsabs.harvard.edu/abs/2024ApJ...969L..18A}
}

@ARTICLE{DEugenio:2024,
       author = {{D'Eugenio}, Francesco and {P{\'e}rez-Gonz{\'a}lez}, Pablo G. and {Maiolino}, Roberto and {Scholtz}, Jan and {Perna}, Michele and {Circosta}, Chiara and {{\"U}bler}, Hannah and {Arribas}, Santiago and {B{\"o}ker}, Torsten and {Bunker}, Andrew J. and {Carniani}, Stefano and {Charlot}, Stephane and {Chevallard}, Jacopo and {Cresci}, Giovanni and {Curtis-Lake}, Emma and {Jones}, Gareth C. and {Kumari}, Nimisha and {Lamperti}, Isabella and {Looser}, Tobias J. and {Parlanti}, Eleonora and {Rix}, Hans-Walter and {Robertson}, Brant and {Rodr{\'\i}guez Del Pino}, Bruno and {Tacchella}, Sandro and {Venturi}, Giacomo and {Willott}, Chris J.},
        title = "{A fast-rotator post-starburst galaxy quenched by supermassive black-hole feedback at z = 3}",
      journal = {Nature Astronomy},
         year = 2024,
        month = nov,
       volume = {8},
        pages = {1443-1456},
          doi = {10.1038/s41550-024-02345-1},
archivePrefix = {arXiv},
       eprint = {2308.06317},
 primaryClass = {astro-ph.GA},
       adsurl = {https://ui.adsabs.harvard.edu/abs/2024NatAs...8.1443D}
}

@ARTICLE{DEugenio:2026_abel,
       author = {{D'Eugenio}, Francesco and {Maiolino}, Roberto and {Perna}, Michele and {{\"U}bler}, Hannah and {Ji}, Xihan and {McClymont}, William and {Koudmani}, Sophie and {Sijacki}, Debora and {Juod{\v{z}}balis}, Ignas and {Scholtz}, Jan and {Bennett}, Jake S. and {Bunker}, Andrew J. and {Carniani}, Stefano and {Charlot}, St{\'e}phane and {Cresci}, Giovanni and {Curtis-Lake}, Emma and {Bont{\`a}}, Elena Dalla and {Inayoshi}, Kohei and {Jones}, Gareth C. and {Lyu}, Jianwei and {Marconi}, Alessandro and {Mazzolari}, Giovanni and {Nelson}, Erica J. and {Parlanti}, Eleonora and {Robertson}, Brant E. and {Schneider}, Raffaella and {Simmonds}, Charlotte and {Tacchella}, Sandro and {Venturi}, Giacomo and {Willott}, Chris and {Witstok}, Joris and {Witten}, Callum},
        title = "{BlackTHUNDER strikes twice: Balmer-line absorption in an overmassive Little Red Dot at z = 7.04}",
      journal = {\mnras},
         year = 2026,
        month = apr,
       volume = {547},
       number = {4},
          eid = {stag401},
        pages = {stag401},
          doi = {10.1093/mnras/stag401},
archivePrefix = {arXiv},
       eprint = {2503.11752},
 primaryClass = {astro-ph.GA},
       adsurl = {https://ui.adsabs.harvard.edu/abs/2026MNRAS.547ag401D}
}

@ARTICLE{DEugenio:2026,
       author = {{D'Eugenio}, Francesco and {Juod{\v{z}}balis}, Ignas and {Ji}, Xihan and {Scholtz}, Jan and {Maiolino}, Roberto and {Carniani}, Stefano and {Perna}, Michele and {Mazzolari}, Giovanni and {{\"U}bler}, Hannah and {Arribas}, Santiago and {Bhatawdekar}, Rachana and {Bunker}, Andrew J. and {Cresci}, Giovanni and {Curtis-Lake}, Emma and {Hainline}, Kevin and {Inayoshi}, Kohei and {Isobe}, Yuki and {Ji}, Zhiyuan and {Johnson}, Benjamin D. and {Jones}, Gareth C. and {Looser}, Tobias J. and {Nelson}, Erica J. and {Parlanti}, Eleonora and {Pusk{\'a}s}, D{\'a}vid and {Rinaldi}, Pierluigi and {Robertson}, Brant and {Rodr{\'\i}guez Del Pino}, Bruno and {Shivaei}, Irene and {Sun}, Fengwu and {Tacchella}, Sandro and {Venturi}, Giacomo and {Volonteri}, Marta and {Williams}, Christina C. and {Willmer}, Christopher N.~A. and {Willott}, Chris and {Witstok}, Joris},
        title = "{JADES and BlackTHUNDER: rest-frame Balmer-line absorption and the local environment in a Little Red Dot at z = 5}",
      journal = {\mnras},
         year = 2026,
        month = jan,
       volume = {545},
       number = {3},
          eid = {staf2117},
        pages = {staf2117},
          doi = {10.1093/mnras/staf2117},
archivePrefix = {arXiv},
       eprint = {2506.14870},
 primaryClass = {astro-ph.GA},
       adsurl = {https://ui.adsabs.harvard.edu/abs/2026MNRAS.545f2117D}
}

@ARTICLE{Marasco:2020,
       author = {{Marasco}, A. and {Cresci}, G. and {Nardini}, E. and {Mannucci}, F. and {Marconi}, A. and {Tozzi}, P. and {Tozzi}, G. and {Amiri}, A. and {Venturi}, G. and {Piconcelli}, E. and {Lanzuisi}, G. and {Tombesi}, F. and {Mingozzi}, M. and {Perna}, M. and {Carniani}, S. and {Brusa}, M. and {di Serego Alighieri}, S.},
        title = "{Galaxy-scale ionised winds driven by ultra-fast outflows in two nearby quasars}",
      journal = {\aap},
         year = 2020,
        month = dec,
       volume = {644},
          eid = {A15},
        pages = {A15},
          doi = {10.1051/0004-6361/202038889},
archivePrefix = {arXiv},
       eprint = {2009.11294},
 primaryClass = {astro-ph.GA},
       adsurl = {https://ui.adsabs.harvard.edu/abs/2020A&A...644A..15M}
}

@ARTICLE{McCavana:2012,
       author = {{McCavana}, Tom and {Micic}, Miroslav and {Lewis}, Geraint F. and {Sinha}, Manodeep and {Sharma}, Sanjib and {Holley-Bockelmann}, Kelly and {Bland-Hawthorn}, Joss},
        title = "{The lives of high-redshift mergers}",
      journal = {\mnras},
         year = 2012,
        month = jul,
       volume = {424},
       number = {1},
        pages = {361-371},
          doi = {10.1111/j.1365-2966.2012.21202.x},
archivePrefix = {arXiv},
       eprint = {1204.6319},
 primaryClass = {astro-ph.CO},
       adsurl = {https://ui.adsabs.harvard.edu/abs/2012MNRAS.424..361M}
}

@ARTICLE{Cole:2000,
       author = {{Cole}, Shaun and {Lacey}, Cedric G. and {Baugh}, Carlton M. and {Frenk}, Carlos S.},
        title = "{Hierarchical galaxy formation}",
      journal = {\mnras},
         year = 2000,
        month = nov,
       volume = {319},
       number = {1},
        pages = {168-204},
          doi = {10.1046/j.1365-8711.2000.03879.x},
archivePrefix = {arXiv},
       eprint = {astro-ph/0007281},
 primaryClass = {astro-ph},
       adsurl = {https://ui.adsabs.harvard.edu/abs/2000MNRAS.319..168C}
}

@ARTICLE{Hawking:1971,
       author = {{Hawking}, Stephen},
        title = "{Gravitationally collapsed objects of very low mass}",
      journal = {\mnras},
         year = 1971,
        month = jan,
       volume = {152},
        pages = {75},
          doi = {10.1093/mnras/152.1.75},
       adsurl = {https://ui.adsabs.harvard.edu/abs/1971MNRAS.152...75H}
}

@ARTICLE{Begelman:2006,
       author = {{Begelman}, Mitchell C. and {Volonteri}, Marta and {Rees}, Martin J.},
        title = "{Formation of supermassive black holes by direct collapse in pre-galactic haloes}",
      journal = {\mnras},
         year = 2006,
        month = jul,
       volume = {370},
       number = {1},
        pages = {289-298},
          doi = {10.1111/j.1365-2966.2006.10467.x},
archivePrefix = {arXiv},
       eprint = {astro-ph/0602363},
 primaryClass = {astro-ph},
       adsurl = {https://ui.adsabs.harvard.edu/abs/2006MNRAS.370..289B}
}

@ARTICLE{Bezanson:2018,
       author = {{Bezanson}, Rachel and {van der Wel}, Arjen and {Straatman}, Caroline and {Pacifici}, Camilla and {Wu}, Po-Feng and {Bari{\v{s}}i{\'c}}, Ivana and {Bell}, Eric F. and {Conroy}, Charlie and {D'Eugenio}, Francesco and {Franx}, Marijn and {Gallazzi}, Anna and {van Houdt}, Josha and {Maseda}, Michael V. and {Muzzin}, Adam and {van de Sande}, Jesse and {Sobral}, David and {Spilker}, Justin},
        title = "{1D Kinematics from Stars and Ionized Gas at z {\ensuremath{\sim}} 0.8 from the LEGA-C Spectroscopic Survey of Massive Galaxies}",
      journal = {\apjl},
         year = 2018,
        month = dec,
       volume = {868},
       number = {2},
          eid = {L36},
        pages = {L36},
          doi = {10.3847/2041-8213/aaf16b},
archivePrefix = {arXiv},
       eprint = {1811.07900},
 primaryClass = {astro-ph.GA},
       adsurl = {https://ui.adsabs.harvard.edu/abs/2018ApJ...868L..36B}
}

@ARTICLE{Akins:2025,
       author = {{Akins}, Hollis B. and {Casey}, Caitlin M. and {Berg}, Danielle A. and {Chisholm}, John and {Cloonan}, Aidan P. and {Franco}, Maximilien and {Finkelstein}, Steven L. and {Fujimoto}, Seiji and {Koekemoer}, Anton M. and {Kokorev}, Vasily and {Lambrides}, Erini and {Robertson}, Brant E. and {Taylor}, Anthony J. and {Coulter}, David A. and {Fox}, Ori and {Karmen}, Mitchell},
        title = "{Strong Rest-UV Emission Lines in a ``Little Red Dot'' Active Galactic Nucleus at z = 7: Early Supermassive Black Hole Growth alongside Compact Massive Star Formation?}",
      journal = {\apjl},
         year = 2025,
        month = feb,
       volume = {980},
       number = {2},
          eid = {L29},
        pages = {L29},
          doi = {10.3847/2041-8213/adab76},
archivePrefix = {arXiv},
       eprint = {2410.00949},
 primaryClass = {astro-ph.GA},
       adsurl = {https://ui.adsabs.harvard.edu/abs/2025ApJ...980L..29A}
}

@ARTICLE{Akins:2026,
       author = {{Akins}, Hollis B. and {Casey}, Caitlin M. and {Chisholm}, John and {Berg}, Danielle A. and {Cooper}, Olivia and {Franco}, Maximilien and {Fujimoto}, Seiji and {Lambrides}, Erini and {Long}, Arianna S. and {McKinney}, Jed},
        title = "{Tentative Detection of Neutral Gas in a Little Red Dot at z = 4.46}",
      journal = {\apj},
         year = 2026,
        month = feb,
       volume = {997},
       number = {2},
          eid = {218},
        pages = {218},
          doi = {10.3847/1538-4357/ae2c77},
archivePrefix = {arXiv},
       eprint = {2503.00998},
 primaryClass = {astro-ph.GA},
       adsurl = {https://ui.adsabs.harvard.edu/abs/2026ApJ...997..218A}
}

@ARTICLE{Hirschmann:2019,
       author = {{Hirschmann}, Michaela and {Charlot}, Stephane and {Feltre}, Anna and {Naab}, Thorsten and {Somerville}, Rachel S. and {Choi}, Ena},
        title = "{Synthetic nebular emission from massive galaxies - II. Ultraviolet-line diagnostics of dominant ionizing sources}",
      journal = {\mnras},
         year = 2019,
        month = jul,
       volume = {487},
       number = {1},
        pages = {333-353},
          doi = {10.1093/mnras/stz1256},
archivePrefix = {arXiv},
       eprint = {1811.07909},
 primaryClass = {astro-ph.GA},
       adsurl = {https://ui.adsabs.harvard.edu/abs/2019MNRAS.487..333H}
}

@ARTICLE{Perez-Gonzalez:2024,
       author = {{P{\'e}rez-Gonz{\'a}lez}, Pablo G. and {Barro}, Guillermo and {Rieke}, George H. and {Lyu}, Jianwei and {Rieke}, Marcia and {Alberts}, Stacey and {Williams}, Christina C. and {Hainline}, Kevin and {Sun}, Fengwu and {Pusk{\'a}s}, D{\'a}vid and {Annunziatella}, Marianna and {Baker}, William M. and {Bunker}, Andrew J. and {Egami}, Eiichi and {Ji}, Zhiyuan and {Johnson}, Benjamin D. and {Robertson}, Brant and {Rodr{\'\i}guez Del Pino}, Bruno and {Rujopakarn}, Wiphu and {Shivaei}, Irene and {Tacchella}, Sandro and {Willmer}, Christopher N.~A. and {Willott}, Chris},
        title = "{What Is the Nature of Little Red Dots and what Is Not, MIRI SMILES Edition}",
      journal = {\apj},
         year = 2024,
        month = jun,
       volume = {968},
       number = {1},
          eid = {4},
        pages = {4},
          doi = {10.3847/1538-4357/ad38bb},
archivePrefix = {arXiv},
       eprint = {2401.08782},
 primaryClass = {astro-ph.GA},
       adsurl = {https://ui.adsabs.harvard.edu/abs/2024ApJ...968....4P}
}

@ARTICLE{Shajib:2025,
       author = {{Shajib}, Anowar J. and {Treu}, Tommaso and {Melo}, Alejandra and {Roberts-Borsani}, Guido and {Knabel}, Shawn and {Cappellari}, Michele and {Frieman}, Joshua A.},
        title = "{An accurate measurement of the spectral resolution of the JWST Near Infrared Spectrograph}",
      journal = {\aap},
         year = 2025,
        month = oct,
       volume = {702},
          eid = {L12},
        pages = {L12},
          doi = {10.1051/0004-6361/202556281},
archivePrefix = {arXiv},
       eprint = {2507.03746},
 primaryClass = {astro-ph.IM},
       adsurl = {https://ui.adsabs.harvard.edu/abs/2025A&A...702L..12S}
}

@ARTICLE{Ma:2026,
       author = {{Ma}, Zheng and {Egami}, Eichi and {Zhu}, Yongda and {Sun}, Fengwu and {Lyu}, Jianwei and {Zhang}, Junyu and {Willmer}, Christopher N.~A. and {Bunker}, Andrew J. and {Carniani}, Stefano and {Curtis-Lake}, Emma and {Hausen}, Ryan and {Ji}, Xihan and {Ji}, Zhiyuan and {Juod{\v{z}}balis}, Ignas and {Maiolino}, Roberto and {Rieke}, George H. and {Rinaldi}, Pierluigi and {Sun}, Yang and {Tacchella}, Sandro and {{\"U}bler}, Hannah and {Williams}, Christina C.},
        title = "{Undermassive Hosts of $z = 4-6 $ AGN from JWST/NIRCam Image Decomposition with CONGRESS, FRESCO, and JADES}",
      journal = {arXiv e-prints},
         year = 2026,
        month = jan,
          eid = {arXiv:2601.15962},
        pages = {arXiv:2601.15962},
          doi = {10.48550/arXiv.2601.15962},
archivePrefix = {arXiv},
       eprint = {2601.15962},
 primaryClass = {astro-ph.GA},
       adsurl = {https://ui.adsabs.harvard.edu/abs/2026arXiv260115962M}
}

@ARTICLE{Rieke:2023,
       author = {{Rieke}, Marcia J. and {Kelly}, Douglas M. and {Misselt}, Karl and {Stansberry}, John and {Boyer}, Martha and {Beatty}, Thomas and {Egami}, Eiichi and {Florian}, Michael and {Greene}, Thomas P. and {Hainline}, Kevin and {Leisenring}, Jarron and {Roellig}, Thomas and {Schlawin}, Everett and {Sun}, Fengwu and {Tinnin}, Lee and {Williams}, Christina C. and {Willmer}, Christopher N.~A. and {Wilson}, Debra and {Clark}, Charles R. and {Rohrbach}, Scott and {Brooks}, Brian and {Canipe}, Alicia and {Correnti}, Matteo and {DiFelice}, Audrey and {Gennaro}, Mario and {Girard}, Julien H. and {Hartig}, George and {Hilbert}, Bryan and {Koekemoer}, Anton M. and {Nikolov}, Nikolay K. and {Pirzkal}, Norbert and {Rest}, Armin and {Robberto}, Massimo and {Sunnquist}, Ben and {Telfer}, Randal and {Wu}, Chi Rai and {Ferry}, Malcolm and {Lewis}, Dan and {Baum}, Stefi and {Beichman}, Charles and {Doyon}, Ren{\'e} and {Dressler}, Alan and {Eisenstein}, Daniel J. and {Ferrarese}, Laura and {Hodapp}, Klaus and {Horner}, Scott and {Jaffe}, Daniel T. and {Johnstone}, Doug and {Krist}, John and {Martin}, Peter and {McCarthy}, Donald W. and {Meyer}, Michael and {Rieke}, George H. and {Trauger}, John and {Young}, Erick T.},
        title = "{Performance of NIRCam on JWST in Flight}",
      journal = {\pasp},
         year = 2023,
        month = feb,
       volume = {135},
       number = {1044},
          eid = {028001},
        pages = {028001},
          doi = {10.1088/1538-3873/acac53},
archivePrefix = {arXiv},
       eprint = {2212.12069},
 primaryClass = {astro-ph.IM},
       adsurl = {https://ui.adsabs.harvard.edu/abs/2023PASP..135b8001R}
}

@ARTICLE{Loeb:1994,
       author = {{Loeb}, Abraham and {Rasio}, Frederic A.},
        title = "{Collapse of Primordial Gas Clouds and the Formation of Quasar Black Holes}",
      journal = {\apj},
         year = 1994,
        month = sep,
       volume = {432},
        pages = {52},
          doi = {10.1086/174548},
archivePrefix = {arXiv},
       eprint = {astro-ph/9401026},
 primaryClass = {astro-ph},
       adsurl = {https://ui.adsabs.harvard.edu/abs/1994ApJ...432...52L}
}

@ARTICLE{Lotz:2008,
       author = {{Lotz}, Jennifer M. and {Jonsson}, Patrik and {Cox}, T.~J. and {Primack}, Joel R.},
        title = "{Galaxy merger morphologies and time-scales from simulations of equal-mass gas-rich disc mergers}",
      journal = {\mnras},
         year = 2008,
        month = dec,
       volume = {391},
       number = {3},
        pages = {1137-1162},
          doi = {10.1111/j.1365-2966.2008.14004.x},
archivePrefix = {arXiv},
       eprint = {0805.1246},
 primaryClass = {astro-ph},
       adsurl = {https://ui.adsabs.harvard.edu/abs/2008MNRAS.391.1137L}
}

@ARTICLE{Juodzbalis:2024,
       author = {{Juod{\v{z}}balis}, Ignas and {Maiolino}, Roberto and {Baker}, William M. and {Tacchella}, Sandro and {Scholtz}, Jan and {D'Eugenio}, Francesco and {Witstok}, Joris and {Schneider}, Raffaella and {Trinca}, Alessandro and {Valiante}, Rosa and {DeCoursey}, Christa and {Curti}, Mirko and {Carniani}, Stefano and {Chevallard}, Jacopo and {de Graaff}, Anna and {Arribas}, Santiago and {Bennett}, Jake S. and {Bourne}, Martin A. and {Bunker}, Andrew J. and {Charlot}, St{\'e}phane and {Jiang}, Brian and {Koudmani}, Sophie and {Perna}, Michele and {Robertson}, Brant and {Sijacki}, Debora and {{\"U}bler}, Hannah and {Williams}, Christina C. and {Willott}, Chris},
        title = "{A dormant overmassive black hole in the early Universe}",
      journal = {\nat},
         year = 2024,
        month = dec,
       volume = {636},
       number = {8043},
        pages = {594-597},
          doi = {10.1038/s41586-024-08210-5},
archivePrefix = {arXiv},
       eprint = {2403.03872},
 primaryClass = {astro-ph.GA},
       adsurl = {https://ui.adsabs.harvard.edu/abs/2024Natur.636..594J}
}

@ARTICLE{Juodzbalis:2026_moka,
       author = {{Juod{\v{z}}balis}, Ignas and {Marconcini}, Cosimo and {D'Eugenio}, Francesco and {Maiolino}, Roberto and {Marconi}, Alessandro and {{\"U}bler}, Hannah and {Scholtz}, Jan and {Ji}, Xihan and {Jones}, Gareth C. and {Perna}, Michele and {Arribas}, Santiago and {Bennett}, Jake S. and {Bromm}, Volker and {Bunker}, Andrew J. and {Carniani}, Stefano and {Charlot}, St{\'e}phane and {Cresci}, Giovanni and {Dayal}, Pratika and {Egami}, Eiichi and {Fabian}, Andrew and {Inayoshi}, Kohei and {Isobe}, Yuki and {Ivey}, Lucy R. and {Koudmani}, Sophie and {Laporte}, Nicolas and {Liu}, Boyuan and {Lyu}, Jianwei and {Mazzolari}, Giovanni and {Monty}, Stephanie and {Parlanti}, Eleonora and {P{\'e}rez-Gonz{\'a}lez}, Pablo G. and {Robertson}, Brant and {Schneider}, Raffaella and {Sijacki}, Debora and {Tacchella}, Sandro and {Trinca}, Alessandro and {Valiante}, Rosa and {Volonteri}, Marta and {Witstok}, Joris and {Zhang}, Saiyang},
        title = "{A direct black-hole mass measurement in a little red dot at high redshift}",
      journal = {\nat},
         year = 2026,
        month = may,
       volume = {653},
       number = {8116},
        pages = {1017-1021},
          doi = {10.1038/s41586-026-10579-4},
archivePrefix = {arXiv},
       eprint = {2508.21748},
 primaryClass = {astro-ph.GA},
       adsurl = {https://ui.adsabs.harvard.edu/abs/2026Natur.653.1017J}
}

@ARTICLE{Juodzbalis:2026_jades,
       author = {{Juod{\v{z}}balis}, Ignas and {Maiolino}, Roberto and {Baker}, William M. and {Lake}, Emma Curtis and {Scholtz}, Jan and {D'Eugenio}, Francesco and {Trefoloni}, Bartolomeo and {Isobe}, Yuki and {Tacchella}, Sandro and {Bunker}, Andrew J. and {Carniani}, Stefano and {Charlot}, St{\'e}phane and {Jones}, Gareth C. and {Parlanti}, Eleonora and {Perna}, Michele and {Rinaldi}, Pierluigi and {Robertson}, Brant and {{\"U}bler}, Hannah and {Venturi}, Giacomo and {Willott}, Chris},
        title = "{JADES: comprehensive census of broad-line AGN from reionization to cosmic noon revealed by JWST}",
      journal = {\mnras},
         year = 2026,
        month = mar,
       volume = {546},
       number = {3},
          eid = {stag086},
        pages = {stag086},
          doi = {10.1093/mnras/stag086},
archivePrefix = {arXiv},
       eprint = {2504.03551},
 primaryClass = {astro-ph.GA},
       adsurl = {https://ui.adsabs.harvard.edu/abs/2026MNRAS.546ag086J}
}

@ARTICLE{Carr:1974,
       author = {{Carr}, B.~J. and {Hawking}, S.~W.},
        title = "{Black holes in the early Universe}",
      journal = {\mnras},
         year = 1974,
        month = aug,
       volume = {168},
        pages = {399-416},
          doi = {10.1093/mnras/168.2.399},
       adsurl = {https://ui.adsabs.harvard.edu/abs/1974MNRAS.168..399C}
}

@ARTICLE{Carr:2024,
       author = {{Carr}, Bernard J. and {Green}, Anne M.},
        title = "{The History of Primordial Black Holes}",
      journal = {arXiv e-prints},
         year = 2024,
        month = jun,
          eid = {arXiv:2406.05736},
        pages = {arXiv:2406.05736},
          doi = {10.48550/arXiv.2406.05736},
archivePrefix = {arXiv},
       eprint = {2406.05736},
 primaryClass = {astro-ph.CO},
       adsurl = {https://ui.adsabs.harvard.edu/abs/2024arXiv240605736C}
}

@ARTICLE{CurtisLake:2026,
       author = {{Curtis-Lake}, Emma and {Cameron}, Alex J. and {Bunker}, Andrew J. and {Scholtz}, Jan and {Carniani}, Stefano and {Parlanti}, Eleonora and {D'Eugenio}, Francesco and {Jakobsen}, Peter and {Willmer}, Christopher N.~A. and {Arribas}, Santiago and {Baker}, William M. and {Charlot}, St{\'e}phane and {Chevallard}, Jacopo and {Circosta}, Chiara and {Curti}, Mirko and {Duan}, Qiao and {Eisenstein}, Daniel J. and {Hainline}, Kevin and {Ji}, Zhiyuan and {Johnson}, Benjamin D. and {Jones}, Gareth C. and {Maiolino}, Roberto and {Maseda}, Michael V. and {Perna}, Michele and {P{\'e}rez-Gonz{\'a}lez}, Pablo G. and {Rawle}, Tim and {Rieke}, Marcia and {Rinaldi}, Pierluigi and {Robertson}, Brant and {Rodr{\'\i}guez Del Pino}, Bruno and {Saxena}, Aayush and {Shivaei}, Irene and {Smit}, Renske and {Tacchella}, Sandro and {{\"U}bler}, Hannah and {Venturi}, Giacomo and {Williams}, Christina C. and {Willott}, Chris},
        title = "{JADES data release 4 ─ Paper I. Sample selection, observing strategy and redshifts of the complete spectroscopic sample}",
      journal = {\mnras},
         year = 2026,
        month = jul,
       volume = {549},
       number = {4},
          eid = {stag836},
        pages = {stag836},
          doi = {10.1093/mnras/stag836},
archivePrefix = {arXiv},
       eprint = {2510.01033},
 primaryClass = {astro-ph.GA},
       adsurl = {https://ui.adsabs.harvard.edu/abs/2026MNRAS.549ag836C}
}

@ARTICLE{Price:2021,
       author = {{Price}, S.~H. and {Shimizu}, T.~T. and {Genzel}, R. and {{\"U}bler}, H. and {F{\"o}rster Schreiber}, N.~M. and {Tacconi}, L.~J. and {Davies}, R.~I. and {Coogan}, R.~T. and {Lutz}, D. and {Wuyts}, S. and {Wisnioski}, E. and {Nestor}, A. and {Sternberg}, A. and {Burkert}, A. and {Bender}, R. and {Contursi}, A. and {Davies}, R.~L. and {Herrera-Camus}, R. and {Lee}, M.-J. and {Naab}, T. and {Neri}, R. and {Renzini}, A. and {Saglia}, R. and {Schruba}, A. and {Schuster}, K.},
        title = "{Rotation Curves in z   1-2 Star-forming Disks: Comparison of Dark Matter Fractions and Disk Properties for Different Fitting Methods}",
      journal = {\apj},
         year = 2021,
        month = dec,
       volume = {922},
       number = {2},
          eid = {143},
        pages = {143},
          doi = {10.3847/1538-4357/ac22ad},
archivePrefix = {arXiv},
       eprint = {2109.02659},
 primaryClass = {astro-ph.GA},
       adsurl = {https://ui.adsabs.harvard.edu/abs/2021ApJ...922..143P}
}

@ARTICLE{Maiolino:2025_xray,
       author = {{Maiolino}, Roberto and {Risaliti}, Guido and {Signorini}, Matilde and {Trefoloni}, Bartolomeo and {Juod{\v{z}}balis}, Ignas and {Scholtz}, Jan and {{\"U}bler}, Hannah and {D'Eugenio}, Francesco and {Carniani}, Stefano and {Fabian}, Andy and {Ji}, Xihan and {Mazzolari}, Giovanni and {Bertola}, Elena and {Brusa}, Marcella and {Bunker}, Andrew J. and {Charlot}, Stephane and {Comastri}, Andrea and {Cresci}, Giovanni and {DeCoursey}, Christa Noel and {Egami}, Eiichi and {Fiore}, Fabrizio and {Gilli}, Roberto and {Perna}, Michele and {Tacchella}, Sandro and {Venturi}, Giacomo},
        title = "{JWST meets Chandra: a large population of Compton thick, feedback-free, and intrinsically X-ray weak AGN, with a sprinkle of SNe}",
      journal = {\mnras},
         year = 2025,
        month = apr,
       volume = {538},
       number = {3},
        pages = {1921-1943},
          doi = {10.1093/mnras/staf359},
archivePrefix = {arXiv},
       eprint = {2405.00504},
 primaryClass = {astro-ph.GA},
       adsurl = {https://ui.adsabs.harvard.edu/abs/2025MNRAS.538.1921M}
}

@ARTICLE{Maiolino:2024,
       author = {{Maiolino}, Roberto and {Scholtz}, Jan and {Curtis-Lake}, Emma and {Carniani}, Stefano and {Baker}, William and {de Graaff}, Anna and {Tacchella}, Sandro and {{\"U}bler}, Hannah and {D'Eugenio}, Francesco and {Witstok}, Joris and {Curti}, Mirko and {Arribas}, Santiago and {Bunker}, Andrew J. and {Charlot}, St{\'e}phane and {Chevallard}, Jacopo and {Eisenstein}, Daniel J. and {Egami}, Eiichi and {Ji}, Zhiyuan and {Jones}, Gareth C. and {Lyu}, Jianwei and {Rawle}, Tim and {Robertson}, Brant and {Rujopakarn}, Wiphu and {Perna}, Michele and {Sun}, Fengwu and {Venturi}, Giacomo and {Williams}, Christina C. and {Willott}, Chris},
        title = "{JADES: The diverse population of infant black holes at 4 < z < 11: Merging, tiny, poor, but mighty}",
      journal = {\aap},
         year = 2024,
        month = nov,
       volume = {691},
          eid = {A145},
        pages = {A145},
          doi = {10.1051/0004-6361/202347640},
archivePrefix = {arXiv},
       eprint = {2308.01230},
 primaryClass = {astro-ph.GA},
       adsurl = {https://ui.adsabs.harvard.edu/abs/2024A&A...691A.145M}
}

@ARTICLE{Maiolino:2024_GNz11,
       author = {{Maiolino}, Roberto and {Scholtz}, Jan and {Witstok}, Joris and {Carniani}, Stefano and {D'Eugenio}, Francesco and {de Graaff}, Anna and {{\"U}bler}, Hannah and {Tacchella}, Sandro and {Curtis-Lake}, Emma and {Arribas}, Santiago and {Bunker}, Andrew and {Charlot}, St{\'e}phane and {Chevallard}, Jacopo and {Curti}, Mirko and {Looser}, Tobias J. and {Maseda}, Michael V. and {Rawle}, Timothy D. and {Rodr{\'\i}guez del Pino}, Bruno and {Willott}, Chris J. and {Egami}, Eiichi and {Eisenstein}, Daniel J. and {Hainline}, Kevin N. and {Robertson}, Brant and {Williams}, Christina C. and {Willmer}, Christopher N.~A. and {Baker}, William M. and {Boyett}, Kristan and {DeCoursey}, Christa and {Fabian}, Andrew C. and {Helton}, Jakob M. and {Ji}, Zhiyuan and {Jones}, Gareth C. and {Kumari}, Nimisha and {Laporte}, Nicolas and {Nelson}, Erica J. and {Perna}, Michele and {Sandles}, Lester and {Shivaei}, Irene and {Sun}, Fengwu},
        title = "{A small and vigorous black hole in the early Universe}",
      journal = {\nat},
         year = 2024,
        month = mar,
       volume = {627},
       number = {8002},
        pages = {59-63},
          doi = {10.1038/s41586-024-07052-5},
archivePrefix = {arXiv},
       eprint = {2305.12492},
 primaryClass = {astro-ph.GA},
       adsurl = {https://ui.adsabs.harvard.edu/abs/2024Natur.627...59M}
}

@ARTICLE{Maiolino:2026_abel,
       author = {{Maiolino}, Roberto and {{\"U}bler}, Hannah and {D'Eugenio}, Francesco and {Scholtz}, Jan and {Juod{\v{z}}balis}, Ignas and {Ji}, Xihan and {Perna}, Michele and {Bromm}, Volker and {Dayal}, Pratika and {Koudmani}, Sophie and {Liu}, Boyuan and {Schneider}, Raffaella and {Sijacki}, Debora and {Valiante}, Rosa and {Trinca}, Alessandro and {Zhang}, Saiyang and {Volonteri}, Marta and {Inayoshi}, Kohei and {Carniani}, Stefano and {Nakajima}, Kimihiko and {Isobe}, Yuki and {Witstok}, Joris and {Jones}, Gareth C. and {Tacchella}, Sandro and {Arribas}, Santiago and {Bunker}, Andrew and {Cataldi}, Elisa and {Charlot}, Stephane and {Curti}, Giovanni Cresci Mirko and {Fabian}, Andrew C. and {Katz}, Harley and {Kumari}, Nimisha and {Laporte}, Nicolas and {Mazzolari}, Giovanni and {Robertson}, Brant and {Sun}, Fengwu and {Rodriguez Del Pino}, Bruno and {Venturi}, Giacomo},
        title = "{A black hole in a near pristine galaxy 700 Myr after the big bang}",
      journal = {\mnras},
         year = 2026,
        month = may,
       volume = {548},
       number = {1},
          eid = {staf2109},
        pages = {staf2109},
          doi = {10.1093/mnras/staf2109},
archivePrefix = {arXiv},
       eprint = {2505.22567},
 primaryClass = {astro-ph.GA},
       adsurl = {https://ui.adsabs.harvard.edu/abs/2026MNRAS.548f2109M}
}

@ARTICLE{Madau:2004,
       author = {{Madau}, Piero and {Quataert}, Eliot},
        title = "{The Effect of Gravitational-Wave Recoil on the Demography of Massive Black Holes}",
      journal = {\apjl},
         year = 2004,
        month = may,
       volume = {606},
       number = {1},
        pages = {L17-L20},
          doi = {10.1086/421017},
archivePrefix = {arXiv},
       eprint = {astro-ph/0403295},
 primaryClass = {astro-ph},
       adsurl = {https://ui.adsabs.harvard.edu/abs/2004ApJ...606L..17M}
}

@ARTICLE{Madau:2026,
       author = {{Madau}, Piero and {Maiolino}, Roberto},
        title = "{Little Red Dots as Obscured Little Blue Dots: A Super-Eddington Unification Model}",
      journal = {arXiv e-prints},
         year = 2026,
        month = feb,
          eid = {arXiv:2602.22386},
        pages = {arXiv:2602.22386},
          doi = {10.48550/arXiv.2602.22386},
archivePrefix = {arXiv},
       eprint = {2602.22386},
 primaryClass = {astro-ph.GA},
       adsurl = {https://ui.adsabs.harvard.edu/abs/2026arXiv260222386M}
}

@ARTICLE{Gloudemans:2025,
       author = {{Gloudemans}, Anniek J. and {Duncan}, Kenneth J. and {Eilers}, Anna-Christina and {Farina}, Emanuele Paolo and {Harikane}, Yuichi and {Inayoshi}, Kohei and {Lambrides}, Erini and {Vardoulaki}, Eleni},
        title = "{Another Piece to the Puzzle: Radio Detection of a JWST-detected Active Galactic Nucleus Candidate}",
      journal = {\apj},
         year = 2025,
        month = jun,
       volume = {986},
       number = {2},
          eid = {130},
        pages = {130},
          doi = {10.3847/1538-4357/adddb9},
archivePrefix = {arXiv},
       eprint = {2501.04912},
 primaryClass = {astro-ph.GA},
       adsurl = {https://ui.adsabs.harvard.edu/abs/2025ApJ...986..130G}
}

@ARTICLE{Gutkin:2016,
       author = {{Gutkin}, Julia and {Charlot}, St{\'e}phane and {Bruzual}, Gustavo},
        title = "{Modelling the nebular emission from primeval to present-day star-forming galaxies}",
      journal = {\mnras},
         year = 2016,
        month = oct,
       volume = {462},
       number = {2},
        pages = {1757-1774},
          doi = {10.1093/mnras/stw1716},
archivePrefix = {arXiv},
       eprint = {1607.06086},
 primaryClass = {astro-ph.GA},
       adsurl = {https://ui.adsabs.harvard.edu/abs/2016MNRAS.462.1757G}
}

@ARTICLE{Dayal:2026,
       author = {{Dayal}, Pratika and {Maiolino}, Roberto},
        title = "{The properties of primordially-seeded black holes and their hosts in the first billion years: implications for JWST}",
      journal = {\aap},
         year = 2026,
        month = feb,
       volume = {706},
          eid = {A72},
        pages = {A72},
          doi = {10.1051/0004-6361/202555959},
archivePrefix = {arXiv},
       eprint = {2506.08116},
 primaryClass = {astro-ph.GA},
       adsurl = {https://ui.adsabs.harvard.edu/abs/2026A&A...706A..72D}
}

@ARTICLE{Jakobsen:2022,
       author = {{Jakobsen}, P. and {Ferruit}, P. and {Alves de Oliveira}, C. and {Arribas}, S. and {Bagnasco}, G. and {Barho}, R. and {Beck}, T.~L. and {Birkmann}, S. and {B{\"o}ker}, T. and {Bunker}, A.~J. and {Charlot}, S. and {de Jong}, P. and {de Marchi}, G. and {Ehrenwinkler}, R. and {Falcolini}, M. and {Fels}, R. and {Franx}, M. and {Franz}, D. and {Funke}, M. and {Giardino}, G. and {Gnata}, X. and {Holota}, W. and {Honnen}, K. and {Jensen}, P.~L. and {Jentsch}, M. and {Johnson}, T. and {Jollet}, D. and {Karl}, H. and {Kling}, G. and {K{\"o}hler}, J. and {Kolm}, M. -G. and {Kumari}, N. and {Lander}, M.~E. and {Lemke}, R. and {L{\'o}pez-Caniego}, M. and {L{\"u}tzgendorf}, N. and {Maiolino}, R. and {Manjavacas}, E. and {Marston}, A. and {Maschmann}, M. and {Maurer}, R. and {Messerschmidt}, B. and {Moseley}, S.~H. and {Mosner}, P. and {Mott}, D.~B. and {Muzerolle}, J. and {Pirzkal}, N. and {Pittet}, J. -F. and {Plitzke}, A. and {Posselt}, W. and {Rapp}, B. and {Rauscher}, B.~J. and {Rawle}, T. and {Rix}, H. -W. and {R{\"o}del}, A. and {Rumler}, P. and {Sabbi}, E. and {Salvignol}, J. -C. and {Schmid}, T. and {Sirianni}, M. and {Smith}, C. and {Strada}, P. and {te Plate}, M. and {Valenti}, J. and {Wettemann}, T. and {Wiehe}, T. and {Wiesmayer}, M. and {Willott}, C.~J. and {Wright}, R. and {Zeidler}, P. and {Zincke}, C.},
        title = "{The Near-Infrared Spectrograph (NIRSpec) on the James Webb Space Telescope. I. Overview of the instrument and its capabilities}",
      journal = {\aap},
         year = 2022,
        month = may,
       volume = {661},
          eid = {A80},
        pages = {A80},
          doi = {10.1051/0004-6361/202142663},
archivePrefix = {arXiv},
       eprint = {2202.03305},
 primaryClass = {astro-ph.IM},
       adsurl = {https://ui.adsabs.harvard.edu/abs/2022A&A...661A..80J}
}

@ARTICLE{vanderWel:2022,
       author = {{van der Wel}, Arjen and {van Houdt}, Josha and {Bezanson}, Rachel and {Franx}, Marijn and {D'Eugenio}, Francesco and {Straatman}, Caroline and {Bell}, Eric F. and {Muzzin}, Adam and {Sobral}, David and {Maseda}, Michael V. and {de Graaff}, Anna and {Holden}, Bradford P.},
        title = "{The Mass Scale of High-redshift Galaxies: Virial Mass Estimates Calibrated with Stellar Dynamical Models from LEGA-C}",
      journal = {\apj},
         year = 2022,
        month = sep,
       volume = {936},
       number = {1},
          eid = {9},
        pages = {9},
          doi = {10.3847/1538-4357/ac83c5},
archivePrefix = {arXiv},
       eprint = {2208.12605},
 primaryClass = {astro-ph.GA},
       adsurl = {https://ui.adsabs.harvard.edu/abs/2022ApJ...936....9V}
}

@article{Planck:2015,
	adsurl = {https://ui.adsabs.harvard.edu/abs/2016A&A...594A..13P},
	archiveprefix = {arXiv},
	author = {{Planck Collaboration} and {Ade}, P.~A.~R. and {Aghanim}, N. and {Arnaud}, M. and {Ashdown}, M. and {Aumont}, J. and {Baccigalupi}, C. and {Banday}, A.~J. and {Barreiro}, R.~B. and {Bartlett}, J.~G. and {Bartolo}, N. and {Battaner}, E. and {Battye}, R. and {Benabed}, K. and {Beno{\^\i}t}, A. and {Benoit-L{\'e}vy}, A. and {Bernard}, J. -P. and {Bersanelli}, M. and {Bielewicz}, P. and {Bock}, J.~J. and {Bonaldi}, A. and {Bonavera}, L. and {Bond}, J.~R. and {Borrill}, J. and {Bouchet}, F.~R. and {Boulanger}, F. and {Bucher}, M. and {Burigana}, C. and {Butler}, R.~C. and {Calabrese}, E. and {Cardoso}, J. -F. and {Catalano}, A. and {Challinor}, A. and {Chamballu}, A. and {Chary}, R. -R. and {Chiang}, H.~C. and {Chluba}, J. and {Christensen}, P.~R. and {Church}, S. and {Clements}, D.~L. and {Colombi}, S. and {Colombo}, L.~P.~L. and {Combet}, C. and {Coulais}, A. and {Crill}, B.~P. and {Curto}, A. and {Cuttaia}, F. and {Danese}, L. and {Davies}, R.~D. and {Davis}, R.~J. and {de Bernardis}, P. and {de Rosa}, A. and {de Zotti}, G. and {Delabrouille}, J. and {D{\'e}sert}, F. -X. and {Di Valentino}, E. and {Dickinson}, C. and {Diego}, J.~M. and {Dolag}, K. and {Dole}, H. and {Donzelli}, S. and {Dor{\'e}}, O. and {Douspis}, M. and {Ducout}, A. and {Dunkley}, J. and {Dupac}, X. and {Efstathiou}, G. and {Elsner}, F. and {En{\ss}lin}, T.~A. and {Eriksen}, H.~K. and {Farhang}, M. and {Fergusson}, J. and {Finelli}, F. and {Forni}, O. and {Frailis}, M. and {Fraisse}, A.~A. and {Franceschi}, E. and {Frejsel}, A. and {Galeotta}, S. and {Galli}, S. and {Ganga}, K. and {Gauthier}, C. and {Gerbino}, M. and {Ghosh}, T. and {Giard}, M. and {Giraud-H{\'e}raud}, Y. and {Giusarma}, E. and {Gjerl{\o}w}, E. and {Gonz{\'a}lez-Nuevo}, J. and {G{\'o}rski}, K.~M. and {Gratton}, S. and {Gregorio}, A. and {Gruppuso}, A. and {Gudmundsson}, J.~E. and {Hamann}, J. and {Hansen}, F.~K. and {Hanson}, D. and {Harrison}, D.~L. and {Helou}, G. and {Henrot-Versill{\'e}}, S. and {Hern{\'a}ndez-Monteagudo}, C. and {Herranz}, D. and {Hildebrandt}, S.~R. and {Hivon}, E. and {Hobson}, M. and {Holmes}, W.~A. and {Hornstrup}, A. and {Hovest}, W. and {Huang}, Z. and {Huffenberger}, K.~M. and {Hurier}, G. and {Jaffe}, A.~H. and {Jaffe}, T.~R. and {Jones}, W.~C. and {Juvela}, M. and {Keih{\"a}nen}, E. and {Keskitalo}, R. and {Kisner}, T.~S. and {Kneissl}, R. and {Knoche}, J. and {Knox}, L. and {Kunz}, M. and {Kurki-Suonio}, H. and {Lagache}, G. and {L{\"a}hteenm{\"a}ki}, A. and {Lamarre}, J. -M. and {Lasenby}, A. and {Lattanzi}, M. and {Lawrence}, C.~R. and {Leahy}, J.~P. and {Leonardi}, R. and {Lesgourgues}, J. and {Levrier}, F. and {Lewis}, A. and {Liguori}, M. and {Lilje}, P.~B. and {Linden-V{\o}rnle}, M. and {L{\'o}pez-Caniego}, M. and {Lubin}, P.~M. and {Mac{\'\i}as-P{\'e}rez}, J.~F. and {Maggio}, G. and {Maino}, D. and {Mandolesi}, N. and {Mangilli}, A. and {Marchini}, A. and {Maris}, M. and {Martin}, P.~G. and {Martinelli}, M. and {Mart{\'\i}nez-Gonz{\'a}lez}, E. and {Masi}, S. and {Matarrese}, S. and {McGehee}, P. and {Meinhold}, P.~R. and {Melchiorri}, A. and {Melin}, J. -B. and {Mendes}, L. and {Mennella}, A. and {Migliaccio}, M. and {Millea}, M. and {Mitra}, S. and {Miville-Desch{\^e}nes}, M. -A. and {Moneti}, A. and {Montier}, L. and {Morgante}, G. and {Mortlock}, D. and {Moss}, A. and {Munshi}, D. and {Murphy}, J.~A. and {Naselsky}, P. and {Nati}, F. and {Natoli}, P. and {Netterfield}, C.~B. and {N{\o}rgaard-Nielsen}, H.~U. and {Noviello}, F. and {Novikov}, D. and {Novikov}, I. and {Oxborrow}, C.~A. and {Paci}, F. and {Pagano}, L. and {Pajot}, F. and {Paladini}, R. and {Paoletti}, D. and {Partridge}, B. and {Pasian}, F. and {Patanchon}, G. and {Pearson}, T.~J. and {Perdereau}, O. and {Perotto}, L. and {Perrotta}, F. and {Pettorino}, V. and {Piacentini}, F. and {Piat}, M. and {Pierpaoli}, E. and {Pietrobon}, D. and {Plaszczynski}, S. and {Pointecouteau}, E. and {Polenta}, G. and {Popa}, L. and {Pratt}, G.~W. and {Pr{\'e}zeau}, G. and {Prunet}, S. and {Puget}, J. -L. and {Rachen}, J.~P. and {Reach}, W.~T. and {Rebolo}, R. and {Reinecke}, M. and {Remazeilles}, M. and {Renault}, C. and {Renzi}, A. and {Ristorcelli}, I. and {Rocha}, G. and {Rosset}, C. and {Rossetti}, M. and {Roudier}, G. and {Rouill{\'e} d'Orfeuil}, B. and {Rowan-Robinson}, M. and {Rubi{\~n}o-Mart{\'\i}n}, J.~A. and {Rusholme}, B. and {Said}, N. and {Salvatelli}, V. and {Salvati}, L. and {Sandri}, M. and {Santos}, D. and {Savelainen}, M. and {Savini}, G. and {Scott}, D. and {Seiffert}, M.~D. and {Serra}, P. and {Shellard}, E.~P.~S. and {Spencer}, L.~D. and {Spinelli}, M. and {Stolyarov}, V. and {Stompor}, R. and {Sudiwala}, R. and {Sunyaev}, R. and {Sutton}, D. and {Suur-Uski}, A. -S. and {Sygnet}, J. -F. and {Tauber}, J.~A. and {Terenzi}, L. and {Toffolatti}, L. and {Tomasi}, M. and {Tristram}, M. and {Trombetti}, T. and {Tucci}, M. and {Tuovinen}, J. and {T{\"u}rler}, M. and {Umana}, G. and {Valenziano}, L. and {Valiviita}, J. and {Van Tent}, F. and {Vielva}, P. and {Villa}, F. and {Wade}, L.~A. and {Wandelt}, B.~D. and {Wehus}, I.~K. and {White}, M. and {White}, S.~D.~M. and {Wilkinson}, A. and {Yvon}, D. and {Zacchei}, A. and {Zonca}, A.},
	doi = {10.1051/0004-6361/201525830},
	eid = {A13},
	eprint = {1502.01589},
	journal = {\aap},
	month = sep,
	pages = {A13},
	primaryclass = {astro-ph.CO},
	title = {{Planck 2015 results. XIII. Cosmological parameters}},
	volume = {594},
	year = 2016}

@ARTICLE{Kocevski:2023,
       author = {{Kocevski}, Dale D. and {Onoue}, Masafusa and {Inayoshi}, Kohei and {Trump}, Jonathan R. and {Arrabal Haro}, Pablo and {Grazian}, Andrea and {Dickinson}, Mark and {Finkelstein}, Steven L. and {Kartaltepe}, Jeyhan S. and {Hirschmann}, Michaela and {Aird}, James and {Holwerda}, Benne W. and {Fujimoto}, Seiji and {Juneau}, St{\'e}phanie and {Amor{\'\i}n}, Ricardo O. and {Backhaus}, Bren E. and {Bagley}, Micaela B. and {Barro}, Guillermo and {Bell}, Eric F. and {Bisigello}, Laura and {Calabr{\`o}}, Antonello and {Cleri}, Nikko J. and {Cooper}, M.~C. and {Ding}, Xuheng and {Grogin}, Norman A. and {Ho}, Luis C. and {Hutchison}, Taylor A. and {Inoue}, Akio K. and {Jiang}, Linhua and {Jones}, Brenda and {Koekemoer}, Anton M. and {Li}, Wenxiu and {Li}, Zhengrong and {McGrath}, Elizabeth J. and {Molina}, Juan and {Papovich}, Casey and {P{\'e}rez-Gonz{\'a}lez}, Pablo G. and {Pirzkal}, Nor and {Wilkins}, Stephen M. and {Yang}, Guang and {Yung}, L.~Y. Aaron},
        title = "{Hidden Little Monsters: Spectroscopic Identification of Low-mass, Broad-line AGNs at z > 5 with CEERS}",
      journal = {\apjl},
         year = 2023,
        month = sep,
       volume = {954},
       number = {1},
          eid = {L4},
        pages = {L4},
          doi = {10.3847/2041-8213/ace5a0},
archivePrefix = {arXiv},
       eprint = {2302.00012},
 primaryClass = {astro-ph.GA},
       adsurl = {https://ui.adsabs.harvard.edu/abs/2023ApJ...954L...4K}
}

@ARTICLE{Koller:2026,
       author = {{Koller}, Maria and {Maiolino}, Roberto and {{\"U}bler}, Hannah and {Duan}, Qiao and {Scholtz}, Jan and {Arribas}, Santiago and {Baker}, William M. and {Carniani}, Stefano and {Charlot}, Stephane and {Curti}, Mirko and {Graziani}, Luca and {Jones}, Gareth and {McClymont}, William and {Perna}, Michele and {Rodr{\'\i}guez Del Pino}, Bruno and {Tacchella}, Sandro and {Venditti}, Alessandra and {Venturi}, Giacomo and {Witstok}, Joris},
        title = "{Metal Mayhem at $\rm z \sim 7-10$: Diversity and Evolution of Gas-Phase Metallicity Gradients}",
      journal = {arXiv e-prints},
         year = 2026,
        month = apr,
          eid = {arXiv:2604.07076},
        pages = {arXiv:2604.07076},
          doi = {10.48550/arXiv.2604.07076},
archivePrefix = {arXiv},
       eprint = {2604.07076},
 primaryClass = {astro-ph.GA},
       adsurl = {https://ui.adsabs.harvard.edu/abs/2026arXiv260407076K}
}

\begin{appendix}

\section{Multi-band NIRCam imaging}\label{apx:nircam}
\begin{figure*}
    \centering
    \includegraphics[width=0.95\linewidth]{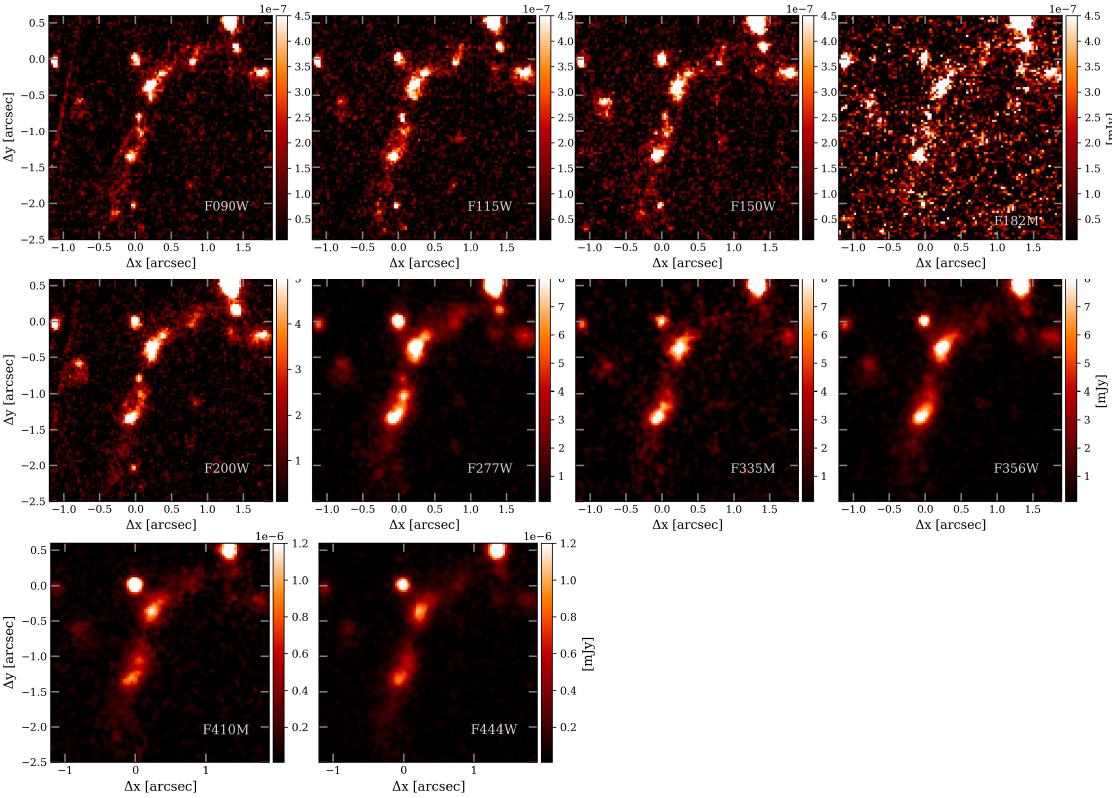}
    \caption{{\it JWST}/NIRCam 6$''$\,$\times$\,6$''$ cutouts of \target and its extended environment in units of mJy, extracted from the full mosaics released as part of the JADES DR5 \citep{Johnson:2026}. Imaging in ten bands is available for \target: five short-wavelength filters, of which one medium-band (i.e. F182M), and five long-wavelength filters, including two medium-band filters (i.e. F335M and F410M). The F182M imaging is shallower compared to the other filters, therefore, we do not include this band in our resolved SED analysis.}
    \label{fig:nircam}
\end{figure*}

{\it JWST}/NIRCam imaging in ten bands is available for \target and its close neighbours: five short-wavelength filters (F090W, F115W, F150W, F182M, F200W), and five long-wavelength filters (F277W, F335M, F356W, F410M, F444W). Figure~\ref{fig:nircam} displays 6$''$\,$\times$\,6$''$ cutouts extracted from the JADES DR5 original mosaics \citep{Johnson:2026}. At the location of \target, supplementary coverage, in addition to that from the JADES program in GOODS-N (PID: 1181; \citealt{Eisenstein:2023a}), is provided by {\it JWST} GO programs 2674 (PI: Arrabal Haro) and 6434 (PI: Egami). In particular, F182M data (top right panel) are available for \target from the former, although much shallower than the other bands. For this reason and also considering that the available higher-S/N F150W and F200W images already provide a good sampling of the SED around the observed 1.8~$\mu$m wavelength, we do not include the F182M imaging in the resolved SED modelling presented in Appendix~\ref{apx:sed_analysis}. Such analysis is carried out on 6$''$\,$\times$\,6$''$ cutouts extracted from the F444W PSF-matched mosaics \citep{Robertson:2026}.

\section{Spectroscopic confirmation of foreground galaxies}\label{apx:specz}

\begin{figure*}
    \centering
    \includegraphics[width=0.95\linewidth]{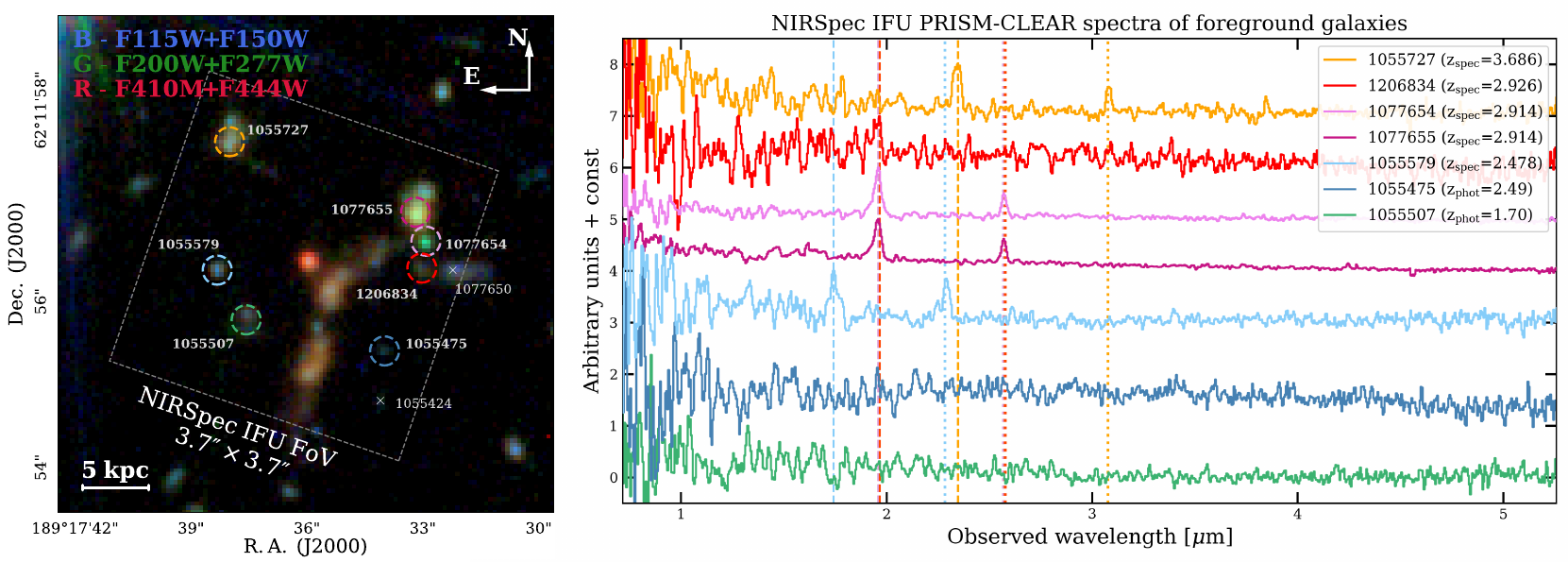}
    \caption{Spectroscopic characterisation of the field around the main source complex at $z$\,$\simeq$\,5.23. In the NIRSpec IFU PRISM-CLEAR datacube, we detect seven sources, encircled in different colours (JADES DR5 IDs in bold) in the left-hand panel; two additional objects (white crosses) lie within the NIRSpec IFU FoV, one is undetected (1055424) and the other one at the (noisy) detector edge (1077650). The right panel shows the integrated PRISM-CLEAR spectra of the seven detected sources, extracted from the 0.3$^{\prime\prime}$ coloured apertures. Five of them are spectroscopically confirmed to be foreground galaxies at $z$\,$=$\,2.48\,--\,3.69, based on the detection of \oiii\ and \ha emission lines, marked with dashed and dotted vertical lines, respectively; while no emission lines are clearly identified in 1055475 (blue) and 1055507 (green). For visual purpose, all spectra have been normalized to their emission peak at $\lambda$\,$>$\,1~$\mu$m and a constant added.}
    \label{fig:foregal}
\end{figure*}

In the NIRSpec IFU PRISM-CLEAR data cube, we also detect emission from another seven objects, all identified in the JADES DR5 photometric catalogue \citep{Robertson:2026}. In the left panel of Fig.~\ref{fig:foregal}, we highlight the seven detected sources with 0.3$''$ circular apertures in different colours and JADES DR5 IDs in bold. By means of these apertures, we then extract the PRISM-CLEAR spectra of the various sources that are shown in the right panel. In five spectra, we detect optical \oiii\ and \ha emission lines, which allow us to spectroscopically confirm these systems to be foreground galaxies at $z$\,$\sim$\,2.5\,--\,3.7. Due to its proximity in redshift ($z$\,$=$\,2.914) and space (0.4$^{\prime\prime}$ separation, corresponding to $\sim$\,3~kpc at $z$\,$=$\,2.914) to the larger galaxy 1007765, 1077654 might be either compact emission still part of 1007765, or a separate interacting system. Although marginally detected in our NIRSpec IFU data, 1055475 (blue) and 1055507 (green) do not show any clear emission line in their spectra, thus preventing us from measuring their spectroscopic redshift. Their photometric redshift from the JADES DR5 photometric catalogue is $z_{\rm phot}$\,$=$\,2.49 and $z_{\rm phot}$\,$=$\,1.70. Two additional sources identified in the JADES DR5 catalogue lie within the NIRSpec IFU FoV: 1055424, which is undetected in the PRISM-CLEAR data; and 1077650, which lies at the detector edge, where the emission is contaminated by noise and edge effects. These two sources are marked with small crosses in the left panel of Fig.~\ref{fig:foregal}. 

\section{Spectral fitting}\label{apx:specfit}

For the spectral modelling of the NIRSpec datacubes of \target, we use the fitting code presented in \citet{Marasco:2020}, designed for IFU data. The same procedure is applied to the integrated NIRSpec IFU spectra extracted in Sect.~\ref{sec:sources} and to the JADES NIRSpec MSA spectra of \target. To accurately estimate the best-fit values and errors of the various parameters, we adopt a Monte Carlo (MC) approach and repeat the spectral fit 200 times, after randomly perturbing the observed spectrum with its associated noise. From the final resulting parameter distributions, we then adopt the 50th percentile and the standard deviation as the best-fit value and associated error, respectively. The best-fit results are summarised in Tables~\ref{tab:specfit_R100} and \ref{tab:specfit_R2700}.

\subsection{NIRSpec IFU low-resolution data} \label{apx:specfit_R100}

\begin{figure*}
    \centering
    \includegraphics[width=0.95\linewidth]{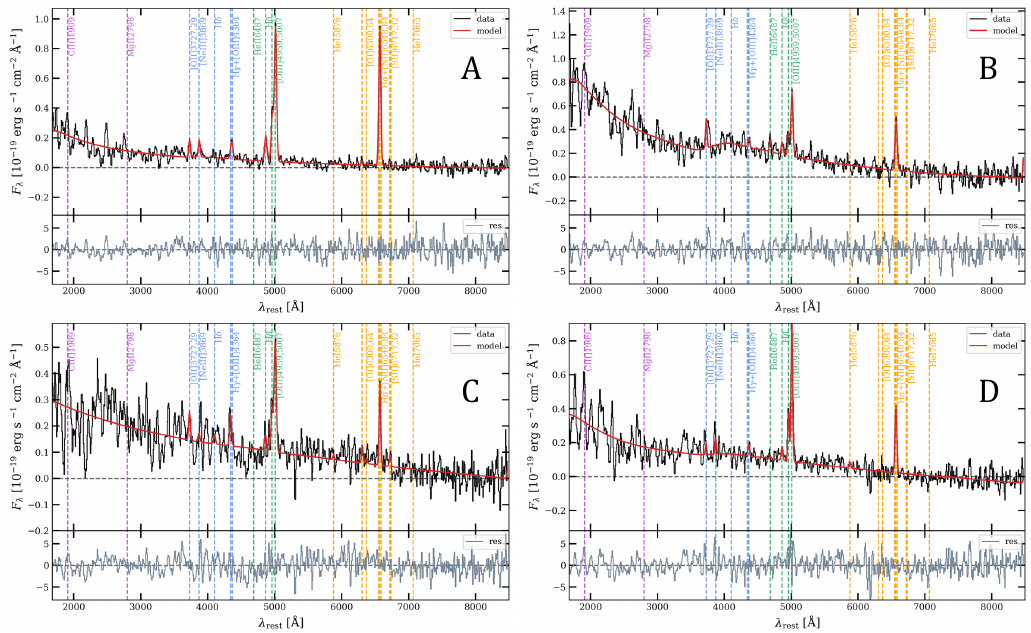}
    \caption{Best-fit models of the integrated NIRSpec IFU PRISM-CLEAR
    spectra of the various sources, extracted with the \oiii-based apertures defined in Sect.~\ref{sec:sources}. For all sources, we show the fitted rest-frame 1860\,--\,8500~\AA wavelength range, since no other emission line is detected outside this range. Data are shown in black, with the total best-fit model drawn on the top in red.}
    \label{fig:R100_bestfit}
\end{figure*}

As shown in Fig.~\ref{fig:fig2}, we detect emission lines and continuum emission over the full 0.6\,--\,5.3~$\mu$m wavelength range of NIRSpec PRISM-CLEAR data (i.e. rest-frame 0.960\,--\,8500~\AA), at variable relative intensity across the source complex. Since the NIRSpec spatial resolution varies with wavelength (from FWHM\,=\,0.13$^{\prime\prime}$ at 1~$\mu$m to FWHM\,=\,0.22$^{\prime\prime}$ at 5~$\mu$m; \citealt{DEugenio:2024}), we first create a PSF-matched version of the PRISM-CLEAR datacube by using a 2D Gaussian kernel to match the FWHM of the NIRSpec PSF at 4.1~$\mu$m (FWHM$_{4.1\rm \mu m}$\,$\sim$\,0.16$^{\prime\prime}$; \citealt{DEugenio:2024}), that is the observed wavelength of \ha, the reddest bright emission line detected.
In addition, we spatially smooth the PSF-matched data in each spectral channel by means of a Gaussian kernel of width $\sigma_{\rm k}$\,$=$\,0.05$^{\prime\prime}$ (i.e. 1 spatial pixel, corresponding to a FWHM of 0.12$^{\prime\prime}$). Since the kernel size is smaller than FWHM$_{4.1\rm \mu m}$\,$\sim$\,0.16$^{\prime\prime}$, this step enhances the visibility of faint regions, without deteriorating the spatial resolution of the data.

After these preliminary steps, we perform the spectral fitting by using the Penalized PiXel-Fitting code (\texttt{pPXF}; \citealt{Cappellari:2023}) to simultaneously model continuum and line emission in each pixel, aiming to accurately reproduce (and then remove) the former. In particular, continuum emission is modelled through a linear combination of stellar templates from the MILES SSP library \citep{Vazdekis:2010}, plus a first-degree polynomial to account for AGN emission in \target, and possibly in the other unknown sources; whereas emission lines are fitted with Gaussian components constrained to have the same kinematics (i.e. $v$ and $\sigma$). The \oiii$\lambda$5007/\oiii$\lambda$4959, [O\,{\sc i}]$\lambda$6300/[O\,{\sc i}]$\lambda$6364 and \nii$\lambda$6584/\nii$\lambda$6548 line ratios are fixed to their theoretical values of 2.98, 3.08 and 2.94 \citep{Osterbrock:2006}, respectively, based on atomic physics. After running the fit with \texttt{pPXF}, we subtract pixel by pixel the best-fit continuum model from the PRISM-CLEAR datacube, and obtain a continuum-subtracted cube that only contains emission lines. With this cube free of continuum emission, we perform a refined modelling of the emission lines with a custom \texttt{python} procedure, benefiting from more flexibility compared to \texttt{pPXF}. To deal with the NIRSpec PRISM-CLEAR spectral resolution varying with wavelength (i.e. $R$\,$\sim$\,30\,--\,300), we separately fit with independent kinematics the emission lines bluer and redder than 3.4~$\mu$m (i.e. 5500~\AA rest-frame), still adopting fixed line ratios for the \oiii, [O\,{\sc i}] and \nii\ line doublets. Best-fit model plots are shown in Fig.~\ref{fig:R100_bestfit}.

For the source C, our low-resolution data best fit likely underestimates the \hb flux (Fig.~\ref{fig:R100_bestfit}, lower left panel), as a consequence of the poor quality of the data. Therefore, for this source, we derive the low-resolution \hb flux from \oiii\ by assuming the same \hb/\oiii\ ratio as measured in the high-resolution spectrum (see Table \ref{tab:specfit_R2700}) and a 30\% error.

\subsection{NIRSpec IFU high-resolution data} \label{apx:specfit_R2700}

\begin{figure*}
    \centering
    \includegraphics[width=0.95\linewidth]{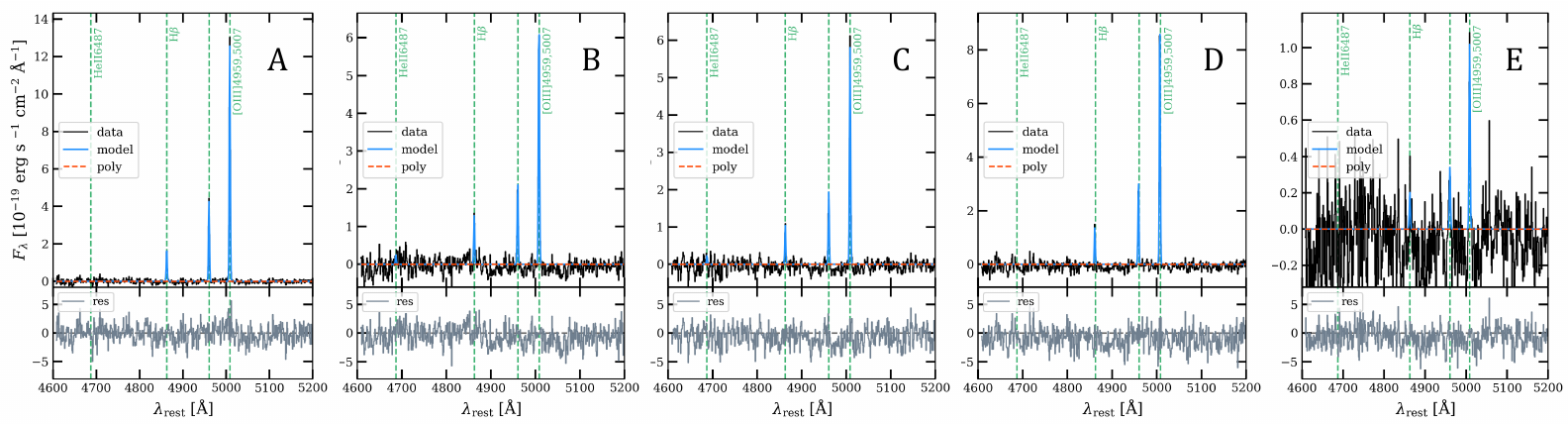}
    \caption{Best-fit models of the integrated NIRSpec IFU high-resolution spectra of the various sources, extracted with the \oiii-based apertures defined in Sect.~\ref{sec:sources}. For all sources, we show zoomed-in spectra on the rest-frame 4600\,--\,5200~\AA wavelength range, since no other emission lines are detected outside this range, except for the \hei emission line detected in \target (see Fig.~\ref{fig:R2700_bestfit_hei}). Data are shown in black, with the total model drawn on the top in light blue, and the 1st-degree polynomial as a red dashed line.}
    \label{fig:R2700_bestfit}
\end{figure*}

\begin{figure}
    \centering
    \includegraphics[width=0.8\linewidth]{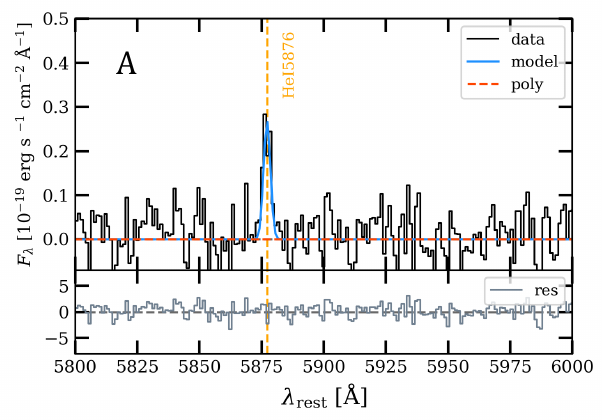}
    \caption{Zooming-in on the rest-frame 5800\,--\,6000~\AA wavelength range of the integrated NIRSpec IFU high-resolution spectrum of \target, where the the \hei emission line is detected. Data are shown in black, with the total model drawn on the top in light blue, the 1st-degree polynomial as a red dashed line.}
    \label{fig:R2700_bestfit_hei}
\end{figure}

The G395H-F290LP datacube covers the rest-frame optical wavelength range 4600\,--\,8500~\AA, and features no continuum emission because of the lower sensitivity of high-resolution data compared to the low-resolution ones. Unfortunately, we only detect the \hb and the \oiii\ emission lines (and also some \hei emission from \target; Fig.~\ref{fig:R2700_bestfit_hei}), since [O\,{\sc i}], \ha, \nii\ and \sii all fall in the NIRSpec detector gap. All emission lines feature relatively simple and symmetric profiles, which can be reproduced with a single-Gaussian model. In fact, consistently with JADES MSA data, \hb does not exhibit any clear BL component and appears as narrow as the forbidden \oiii\ line doublet. As done in the fit of low-resolution data, we impose the same kinematics to all lines, and a fixed flux ratio of 2.98 for the \oiii\ doublet components \citep{Osterbrock:2006}. In addition, we include a 1st-degree polynomial to account for any residual background emission. With all these ingredients, we fit pixel by pixel the high-resolution datacube, and the integrated high-resolution spectrum of each source. Best-fit plots zoomed-in on the detected emission lines are shown in Figs.~\ref{fig:R2700_bestfit} and \ref{fig:R2700_bestfit_hei}.

\renewcommand{\arraystretch}{1.2}
\begin{table*}[]
\caption{Best-fit line fluxes, as resulting from spectral fitting of the low- to medium-resolution NIRSpec spectra.}
\centering
\footnotesize
\label{tab:specfit_R100}
\resizebox{0.95\linewidth}{!}{
\begin{tabular}{c|c|ccccccccccc}
\hline\hline

 & Mode & [O\,{\sc ii}] & [Ne\,{\sc iii}] & H$\gamma$ $^{(\rm a)}$ & \oiii$\lambda$4363 & [He\,{\sc ii}] & \hb & \oiii$\lambda$5007 & [O\,{\sc i}] & \ha & \nii & [S\,{\sc ii}] \\
\hline\hline
A & IFU R100 & $3.9 \pm 1.0$ & $3.8 \pm 0.4$ & $5.6 \pm 0.8$ & - & $<0.919$ & $7.2 \pm 0.4$ & $38.6 \pm 0.4$ & $<1.77$ & $34.0 \pm 0.8$ & $<0.633$ & $<1.39$\\
B & IFU R100 & $7.6 \pm 1.1$ & $<3.05$ & $2.6 \pm 0.7$ & - & $<5.44$ & $3.9 \pm 0.7$ & $21.0 \pm 0.7$ & $<2.28$ & $17.6 \pm 0.8$ & $<1.66$ & $<1.05$ \\
C & IFU R100 & $<4.23$ & $<1.63$ & $3.6 \pm 0.5$ & - & $<1.31$ & $2.6 \pm 0.8$ $^{(\rm b)}$ & $15.0 \pm 0.3$ & $<2.37$ & $8.6 \pm 0.9$ & $<1.89$ & $<1.99$ \\
D & IFU R100 & $<2.44$ & $<5.55$ & $<3.21$ & - & $<1.52$ & $3.0 \pm 0.5$ & $28.9 \pm 0.5$ & $<1.62$ & $13.3 \pm 1.7$ & $<2.11$ & $<1.48$ \\
\hline
\multirow{2}{*}{A} & \multirow{2}{*}{MSA R100 $^{(\rm c)}$} & $6 \pm 5$ & $4.9 \pm 0.9$ & $4.8 \pm 1.6$ & - & $<0.814$ & $6.9\pm 0.8$ & $56.9 \pm 1.1$ & $<1.56$ & $40 \pm 5$ & $<0.845$ & $<1.50$\\
& & & & & & & & & & [$14 \pm 4$] & & \\
\multirow{2}{*}{A} & \multirow{2}{*}{MSA R1000 $^{(\rm c)}$} & $7 \pm 4$ & $10.7 \pm 1.4$ & $<9.27$ & $<3.07$ & $<0.985$ & $7.0 \pm 0.7$ & $46.2 \pm 1.1$ & $<1.27$ & $27 \pm 3$& $<0.599$ & $<0.684$ \\
& & & & & & & & & & [$23 \pm 3$] & & \\
\hline

\end{tabular}
}
\tablefoot{Observed fluxes in units of 10$^{-19}$ \ergscm of [O\,{\sc ii}]$\lambda\lambda$3727,29, [Ne\,{\sc iii}]$\lambda$3869, H$\gamma$, \oiii$\lambda$4363, \heii, \hb, \oiii$\lambda$5007, \oi, \ha, \nii$\lambda$6584 and [S\,{\sc ii}]$\lambda\lambda$6717,31 emission lines. For non-detections (S/N\,$<$\,3), we report 3$\sigma$ upper limits.\\
$^{(\rm a)}$ For low-resolution spectra (R100 in the ‘Mode' column), we provide the sum of \hg and \oiii$\lambda$6364 fluxes.\\
$^{(\rm b)}$ \hb flux derived from \oiii\ by assuming the same \hb/\oiii\ ratio as measured in the high-resolution spectrum (see Table \ref{tab:specfit_R2700}) and a 30\% error.\\
$^{(\rm c)}$ For low- and medium-resolution MSA data, we report both narrow and broad \ha fluxes, with the latter indicated in squared brackets.
}
\end{table*}

\begin{table*}[]
\caption{Best-fit line fluxes, as resulting from spectral fitting of the high-resolution NIRSpec spectra.}
\centering
\footnotesize
\label{tab:specfit_R2700}
\resizebox{0.95\linewidth}{!}{
\begin{tabular}{l|c|cccccccc|cccccc}
\hline\hline

 & Mode & \hb & \oiii$\lambda$5007 & \hei & [O\,{\sc i}] & H$\alpha_{\rm nar}$ & H$\alpha_{\rm bro}$ & \nii & [S\,{\sc ii}] & $\Delta v^{\rm [OIII]}_{\rm nar}$ & $\Delta v^{\rm H\alpha}_{\rm nar}$ & $\Delta v^{\rm H\alpha}_{\rm bro}$ & $\sigma^{\rm [OIII]}_{\rm nar}$ & $\sigma^{\rm H\alpha}_{\rm nar}$ & $\sigma^{\rm H\alpha}_{\rm bro}$ \\
 & (1) & (2) & (3) & (4) & (5) & (6) & (7) & (8) & (9) & (10) & (11) & (12) & (13) & (14) & (15)\\
\hline\hline

A & IFU & $5.1 \pm 1.6$ & $37.2 \pm 0.3$ & $0.94 \pm 0.09$ & - & - & - & - & - & $-3 \pm 1$ & - & - & $54 \pm 1$ & - & - \\
B & IFU & $3.8 \pm 0.3$ & $19.1 \pm 0.3$ & $<0.746$ & - & - & - & - & - & $-20 \pm 1$ & - & - & $54 \pm 1$ & - & - \\
C & IFU & $2.9 \pm 1.9$ & $16.5 \pm 0.2$ & $<0.866$ & - & - & - & - & - & $12 \pm 1$ & - & - & $52 \pm 1$ & - & - \\
D & IFU & $3.7 \pm 1.9$ & $25.8 \pm 0.2$ & $<0.838$ & - & - & - & - & - & $-85 \pm 1$ & - & - & $49 \pm 1$ & - & - \\
E & IFU & $<1.29$ & $3.4 \pm 0.3$ & $<1.16$ & - & - & - & - & - & $11 \pm 8$ & - & - & $68 \pm 8$ & - & - \\
\hline
A & MSA & $5.3 \pm 0.8$ & $47.8 \pm 1.2$ & - & $<2.80$ & $19.0 \pm 1.3$ & $36 \pm 2$ & $<1.52$ & $<2.16$ & $-15 \pm 2$ & $-2 \pm 4$ & $21 \pm 34$ & $53 \pm 2$ & $54 \pm 5$ & $420 \pm 40$ \\
\hline
\end{tabular}
}
\tablefoot{(1) NIRSpec mode (IFU/MSA); (2\,--\,9) emission-line fluxes in units of 10$^{-19}$ \ergscm; (10\,--\,12) velocity offset ($\Delta v$) and (13\,--\,15) intrinsic velocity dispersion ($\sigma$; i.e. corrected for instrumental resolution) of \oiii\ and \ha emission lines, in units of \kms. For non-detections (S/N\,$<$\,3), we report 3$\sigma$ upper limits. We do not provide fluxes (or upper limits) of emission lines falling either within or close to the detector gap.}
\end{table*}

\subsection{NIRSpec MSA spectra} \label{apx:specfit_msa}

\begin{figure*}
    \centering
    \includegraphics[width=0.95\linewidth]{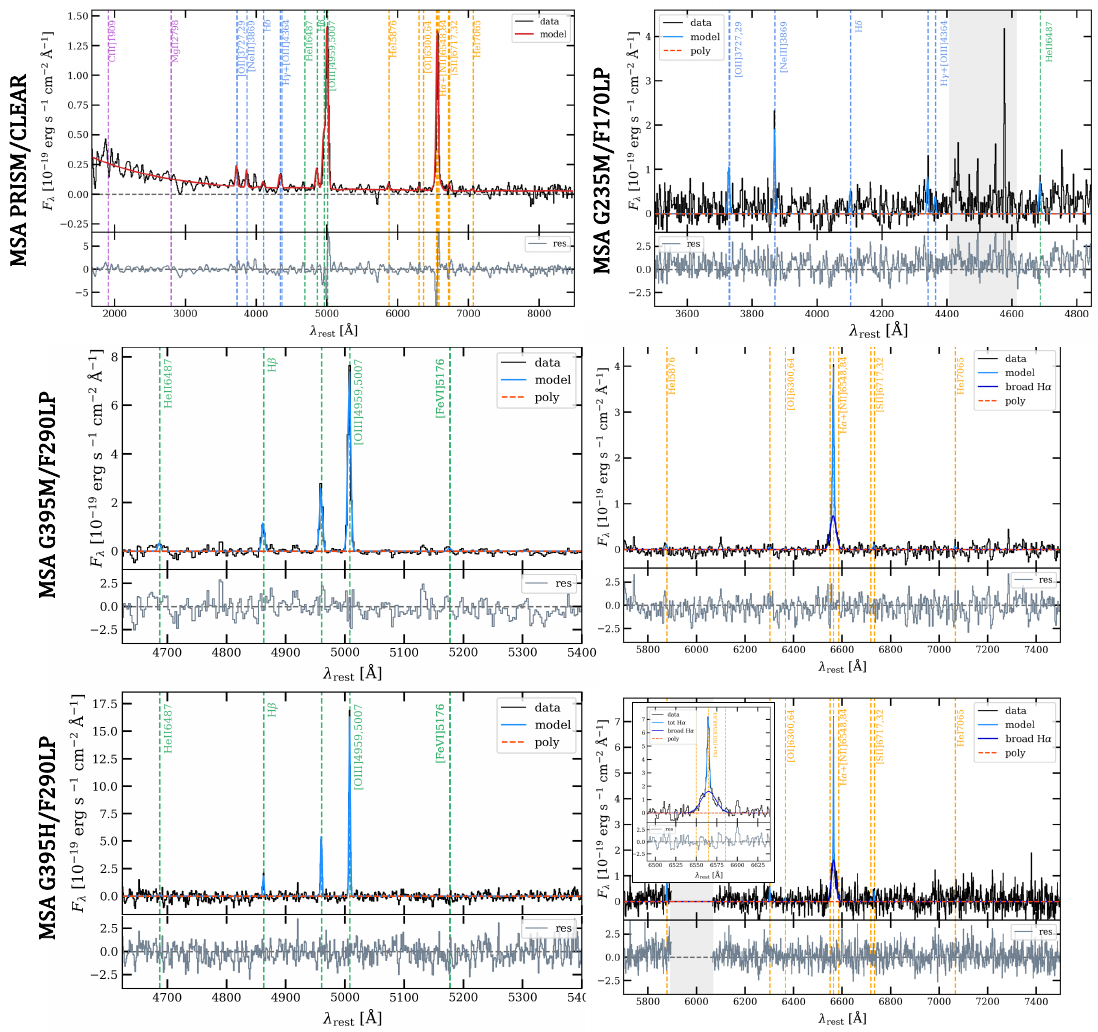}
    \caption{Best-fit models of all NIRSpec MSA spectra of \target analysed in this paper (3-pixel extraction). Data are shown in black, with the total model drawn on the top in red in the PRISM-CLEAR spectrum (top left), while in the other panels the total model is shown in light blue, the 1st-degree polynomial as a red dashed line, and the \ha BL component in blue. Grey shaded areas correspond to a particularly noisy spectral region in the G235M-F170LP spectrum (top right), and to the detector gap in the G395H-F290LP data (bottom right), which have both been masked in the spectral fitting. The bottom right panel also displays an inset zoomed in on the \ha line modelling in the high-resolution data.}
    \label{fig:msa_g395}
\end{figure*}

Following the same procedure, we also fit the JADES NIRSpec MSA spectra of \target (i.e. PRISM-CLEAR, G235M-F170LP, G395M-F290LP and G395H-F290LP), adding as many Gaussian components as needed to model all emission lines being covered. We do not fit the G140M-F070LP spectrum, since it is very noisy and does not show any clear detection of line emission. All best-fit plots are shown in Fig.~\ref{fig:msa_g395}. In contrast to the high-resolution NIRSpec IFU datacube, the \ha emission line falls outside the detector gap in the MSA G395H spectrum of \target, thus allowing us to accurately model its total line profile with two (narrow $+$ broad) Gaussian components and properly disentangle the flux of the narrow-line and BL components (bottom right panel of Fig.~\ref{fig:msa_g395}). From our \ha modelling we find an intrinsic velocity dispersion $\sigma^{\rm H\alpha}_{\rm bro}$\,$=$\,420\,$\pm$\,40~\kms, which is consistent with (yet better constrained than) the previous measurements in the medium-resolution data \citep{Maiolino:2024,Juodzbalis:2026_jades}. The resulting narrow-to-broad \ha flux ratio is 0.528, which leads to a narrow-to-total \ha ratio of 0.345.

A BL component is also needed to well reproduce the \ha line profile in the low- and medium-resolution MSA spectra of \target. As a consequence of the different spectral resolution and sensitivity, we obtain a narrow-to-broad \ha flux ratio of 2.95, with an intrinsic velocity dispersion $\sigma^{\rm H\alpha}_{\rm bro}$\,=\,1800\,$\pm$\,430~\kms for the broad component in the PRISM-CLEAR data; while a narrow-to-broad \ha flux ratio of 1.18 and $\sigma^{\rm H\alpha}_{\rm bro}$\,=\,552\,$\pm$\,97~\kms in the medium-resolution data. 
Conversely, no broad \ha component is apparently detected in the integrated IFU PRISM-CLEAR spectrum of \target. We argue that the apparent detection of such a broad \ha component in the MSA low-resolution data is extremely sensitive to inaccuracies in the best-fit continuum model, which could have produced unreal extended \ha line wings in the MSA low-resolution spectrum, resulting in a suspiciously $\times$\,3\,--\,4 broader component than in the MSA medium- and high-resolution data. The actual non-detection of a broad \ha component in low-resolution data, which instead remains blended with the narrow one, is also supported by $\sigma^{\rm H\alpha}_{\rm bro}$\,$\simeq$\,420~\kms inferred in the high-resolution MSA data, which is comparable with $\sigma^{\rm H\alpha}_{\rm nar}$\,$\simeq$\,395~\kms from low-resolution MSA data and smaller than $\sigma^{\rm H\alpha}$\,$\simeq$\,500~\kms from the single-component fit of the IFU low-resolution spectrum. Best-fit results are listed in Tables \ref{tab:specfit_R100} and \ref{tab:specfit_R2700}. Any absolute comparison for \target between line fluxes from NIRSpec IFU and MSA spectra should be avoided, since the former has been extracted from an aperture aimed to maximize the emission-line S/N (see Sect. \ref{sec:sources}), rather than to encompass the total flux.

\subsection{Emission line maps}\label{apx:line_maps}

\begin{figure*}
    \centering
    \includegraphics[width=0.98\linewidth]{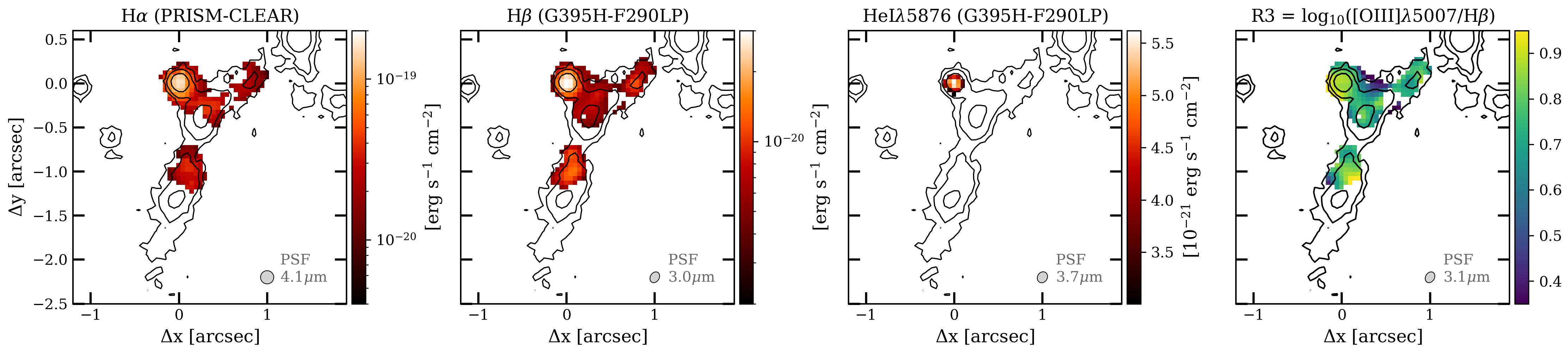}
    \caption{Observed flux maps of \ha, \hb and \hei emission lines (first to third panels), as resulting from the spectral fitting of NIRSpec datacubes, and map of the R3\,$=$\,log$_{10}$(\oiii$\lambda$5007/\hb) diagnostic (fourth panel), obtained from high-resolution emission line fluxes. The black contours trace F444W continuum emission as in Fig.~\ref{fig:oiii_map_R2700}. A S/N\,$\geq$\,2 is applied to all maps. The R3 map shows spatial variations from 0.4 to 1.0, indicating intrinsic differences in the ionisation state and metallicity across the complex.}
    \label{fig:flux_maps}
\end{figure*}

From the spectral fitting of the NIRSpec datacubes (Appendix~\ref{apx:specfit_R100} and \ref{apx:specfit_R2700}), in addition to the \oiii\ maps presented in Sect.~\ref{sec:gas_kinematics},
we obtain resolved (S/N\,$\geq$\,2) flux maps of the following emission lines (Fig.~\ref{fig:flux_maps}): \hb and \hei from the high-resolution datacube; and \ha from the low-resolution datacube (first three panels of Fig.~\ref{fig:flux_maps}). \ha and \hb feature extended emission similar to \oiii, whereas \hei is only detected in the inner compact region of \target. \oiii\ and \hb are also detected in the low resolution datacube, yet at lower S/N and with more uncertainty in the spectral modelling: the two \oiii\ doublet components are blended together at $R$\,$\sim$\,100, while \hb is only detected at S/N\,$\gtrsim$\,2 in \target.

The last panel of Figure~\ref{fig:flux_maps} displays the 2D map of the R3\,$=$\,log$_{10}$(\oiii$\lambda$5007/\hb) diagnostic, obtained from the high-resolution \oiii\/\hb ratios, which provides a relatively dust-insensitive diagnostic of (combined) chemical abundance and ionisation conditions. The R3 map reveals spatial variations across the source complex, in particular higher ratios in \target and D (R3\,$\simeq$\,0.8\,--\,1.0) than in B and C (R3\,$\simeq$\,0.6\,--\,0.8). In the outer region of B, especially close to \target, the R3 decreases down to 0.3\,--\,0.5. Although additional emission-line ratios would be needed to break the ionisation-metallicity degeneracy, the observed spatial variations in the \oiii/\hb ratios point to intrinsic differences in the metallicity content and ionisation state of the various regions.


\section{Spatially resolved SED analysis}\label{apx:sed_analysis}

We perform spatially resolved SED modeling of {\it JWST}/NIRCam imaging by means of the SED fitting code \texttt{CIGALE} \citep{Boquien:2019,Yang:2020}. For this purpose, we recover 6$''$\,$\times$\,6$''$ cutouts of \target from the F444W PSF-matched NIRCam mosaics (F444W PSF FWHM\,$\sim$\,0.16$''$; \citealt{Zhuang:2024}), released as part of the JADES DR5 on a common 0.03$''$ pixel scale \citep{Johnson:2026}, and rebin them to a 0.05$''$ pixel scale in order to match the pixel scale of the NIRSpec IFU data.
To estimate the effective background noise associated to individual pixels, accounting for pixel spatial correlations, we create for each cutout a segmentation map and measure the flux within 100 circular apertures, randomly placed in `empty’ regions and of variable diameter size $N$ (up to $N$\,$=$\,8 pixels; see also \citealt{Tacchella:2015sins}). From the resulting distribution of fluxes as a function of $N$, we derive the rms within one pixel, and associate it to all pixels of each cutout as constant background noise error. To also quantify Poisson noise, which might be non-negligible in bright regions, we recover images in units of counts, and find the Poisson noise contribution to be variable across the source complex galaxy and sensitive to the spectral band, from being negligible to exceeding the background noise by 30\%\,--\,60\%. This translates to relative Poisson uncertainties on pixel flux from a few 0.1\% to a few percent. To account for this (spatially and photometrically) variable Poisson noise, we add a conservative 5\% error on all pixel fluxes.

In order to only model reliable SEDs, we only select pixels of the $z$\,$\simeq$\,5.23 source complex with a S/N\,$\geq$\,4 on the F444W band. Such S/N threshold guarantees a continuous mapping of the extended system, and a S/N\,$>$\,3 in all other bluer filters except for the shallower F182M image. Since the available higher-S/N F150W and F200W bands still provide a good sampling of the SED around 1.8~$\mu$m, we exclude the F182M band and carry out the SED analysis with nine filters in total, which altogether cover the SED of our sources over the observed wavelength range 0.9\,--\,4.4~$\mu$m. The SED of each selected pixel is then modeled independently under the same set of assumptions. For all pixels, we fix the redshift to $z$\,=\,5.23, and assume a delayed exponential SFH, with a late starburst episode. For the main stellar population, we allow for ages since the onset of SF between 100~Myr and 1100~Myr (i.e. the age of the Universe at $z$\,=\,5.23), and $e$-folding $\tau$ times from 100~Myr to 5000~Myr, to well reproduce from more steeply declining to roughly constant SFHs; whereas for the late starburst we define a grid of ages from 10~Myr to 50~Myr, with possible $e$-folding $\tau$ times from 10~Myr to 100~Myr. We adopt stellar population synthesis models by \citet{Bruzual:2003}, with a \citet{Chabrier:2003} IMF and metallicity fixed to $Z$\,=\,0.008\footnote{Only a limited number of metallicity values are allowed in \texttt{CIGALE}. The available subsolar values are $Z$\,=\,0.0001, 0.0004, 0.004, 0.008.} (0.4\,$Z_{\odot}$), to reduce the number of free parameters and avoid degeneracy with other physical quantities. Even with a fixed metallicity $Z$\,=\,0.004 (0.2\,$Z_{\odot}$), the results do not change significantly.

\begin{figure*}
    \centering
    \includegraphics[width=0.95\linewidth]{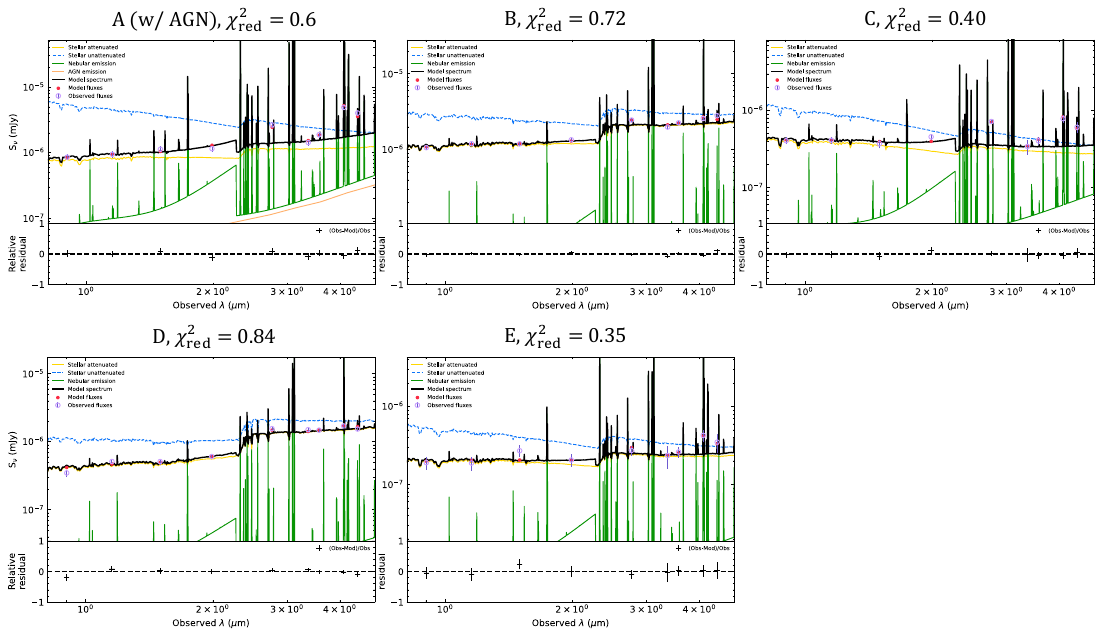}
    \caption{Best-fit SED models, as resulting from resolved modelling of the SED of individual pixels representative of the five sources. The total model spectrum is shown in black, while model spectra for the various components are shown in different colours (see legend). For the main source \target, the best-fit model includes an AGN component (orange), yet dominated by stellar emission (yellow). The SEDs of the various sources differ by spectral shape and relative intensities between the distinct spectral components, thus highlighting the need for spatially resolved SED fitting to recover the intrinsic properties across the whole source complex.}
    \label{fig:sed_plots}
\end{figure*}

In addition, we include nebular emission and assume a \citet{Calzetti:2000} attenuation law, allowing for only two possible values of electron density $n_{\rm e}$ (100 cm$^{-3}$, 1000 cm$^{-3}$), and possible values of the ionisation parameter log($U$) and stellar continuum color excess E(B\,--\,V) ranging from --\,3.0 to --\,2.0 and from 0.01 to 1.0, respectively. To account for AGN emission from \target and other possible AGN, we also the recent \texttt{skirtor2016} module \citep{Stalevski:2016}, where the rest-frame UV/optical emission from the AGN accretion disk, covered by the available NIRCam photometry, is modelled through a \citet{Schartmann:2005} spectrum. Due to the limited spectral coverage, we fix all AGN parameters but two: the AGN inclination, for which we consider two possible values ($i$\,$=$\,20$^{\circ}$, 70$^{\circ}$); and the AGN fraction (i.e. the AGN-to-galaxy luminosity ratio, computed over 0.13\,--\,5~$\mu$m), for which we define a grid of values ranging from 5\% to 90\%, to account for weak-to-dominant AGN contribution. For the polar dust component, we assume a fixed temperature of $100$~K and adopt the Small Magellanic Cloud (SMC) extinction curve, as recommended for AGN observations (e.g. \citealt{Bongiorno:2012}), with E(B-V) fixed to 0.04 for this component. Such AGN component turns out to be necessary only for those pixels within the aperture defined for \target in Sect.~\ref{sec:sed_maps}. For all other pixels, we therefore adopt results obtained without including AGN emission. With this set of assumptions, we well reproduce the variety of SED shapes of the individual pixels (see Fig.~\ref{fig:sed_plots}), accounting for intrinsic differences among the various regions across the source complex.

\section{Emission line ratios, nebular attenuation and AGN properties}\label{apx:ism}

Thanks to the emission lines detected in the NIRSpec IFU integrated spectra (\oiii-based extractions performed in Sect.~\ref{sec:sources}), we can derive a series of ISM properties for the various sources, such as nebular dust attenuation ($A_{V,{\rm neb}}$), gas phase metallicity, and ionisation mechanisms, through emission-line ratios. Due to known inaccuracies in the absolute flux calibration of NIRSpec data at different spectral resolutions \citep{DEugenio:2025}, we only make use of line ratios obtained by combining line fluxes measured at the same spectral resolution. For \target, we derive the narrow \ha flux (to use in the various line ratios) in the other data at lower spectral resolution by multiplying the the total best-fit \ha flux one by the H$\alpha_{\rm nar}$/H$\alpha_{\rm tot}$\,=\,0.345 ratio measured in MSA high-resolution data (Appendix~\ref{apx:specfit_msa}). When possible, we also check consistency between the IFU- and MSA-based line ratios. All observed line fluxes are listed in Tables~\ref{tab:specfit_R100} and \ref{tab:specfit_R2700}, with 3$\sigma$ upper limits for the non-detections. 


We derive the nebular dust attenuation from the \ha/\hb ratios measured in the integrated NIRSpec spectra of each source, except for E, where both \hb and \ha are undetected. In particular, we use: for \target, the most accurate H$\alpha_{\rm nar}$/H$\beta_{\rm nar}$ ratio measured in the MSA high-resolution spectrum, where H$\beta_{\rm nar}$ corresponds to the total \hb flux, since an \hb BL component is detected in none of the available spectra; for the sources B, C and D, the only available \ha/\hb ratio from integrated IFU low-resolution data. Assuming a \citet{Cardelli:1989} dust extinction law and a theoretical \ha/\hb ratio of 2.87 (recombination case B, low-density limit, $T$\,$=$\,10$^4$ K; \citealt{Osterbrock:2006}), we thus find the four brightest sources to feature consistent (or nearly consistent) levels of moderate dust attenuation within the uncertainties ($A_{V,{\rm neb}}$\,$\sim$\,1; see Table~\ref{tab:results} for the individual values inferred for each source). Emission line fluxes and ratios used in Sects.~\ref{sec:diag} and \ref{sec:metallicity} are corrected for $A_{V,{\rm neb}}$.

The \oiii$\lambda$4363 auroral line is only detected in the NIRSpec low-resolution data ($R$\,$\sim$\,100), where it is blended with \hg, yielding an \oiii$\lambda$4363\,$+$\,\hg detection at S/N\,$>$\,3 in \target, B and C. Yet, the poor quality of the bluer part ($<$\,5000~\AA) of the low-resolution spectrum of C makes very uncertain the fit of the lines in this spectral region (Fig.~\ref{fig:R100_bestfit}, lower left panel). Therefore, we only derive the \oiii$\lambda$4363 and \hg fluxes for \target and B. For them, we infer the de-reddened \hg flux from the de-reddened \hb one, assuming a theoretical \hg/\hb ratio of 0.466 (recombination case B, low-density limit, $T$\,$=$\,10$^4$ K; \citealt{Osterbrock:2006}), and thus attribute the remaining $A_{V, \rm neb}$-corrected \oiii$\lambda$4363\,$+$\,\hg flux to \oiii$\lambda$4363. We obtain $F_{\rm H\gamma}=7.07\times10^{-19}$~\ergscm and $F_{\rm [OIII]4363}=6.24\times10^{-19}$~\ergscm for \target; $F_{\rm H\gamma}=8.48\times10^{-19}$~\ergscm and $F_{\rm [OIII]4363}=6.72\times10^{-19}$~\ergscm for B. These fluxes are used in Sect.~\ref{sec:diagrams} to place \target and B in the \oiii$\lambda$4363 auroral diagrams (Fig.~\ref{fig:diagrams}).


To derive the properties of the BL AGN in \target, we rely on the broad \ha best fit of the high-resolution MSA NIRSpec data (Appendix~\ref{apx:specfit_msa}), after correcting the broad \ha luminosity for the nebular dust attenuation derived from the (only available) narrow \ha/\hb ratio (i.e. $A_V$\,$=$\,0.69). To estimate the BH mass, we assume the gas in the AGN BLR to be virialised and adopt the \ha-based calibration of \citet{Reines:2013}, thus finding $M_{\rm BH}$\,$=$\,$1.1^{+0.4}_{-0.3}$\,$\times$\,10$^7$~\msun. From broad \ha emission, we also estimate an AGN bolometric luminosity of $L_{\rm bol}$\,$=$\,$2.37^{+0.18}_{-0.13}$\,$\times$10$^{44}$~erg~s$^{-1}$ \citep{Stern:2012}, which, combined with the inferred $M_{\rm BH}$, leads to an Eddington ratio of $\lambda_{\rm Edd}$\,$=$\,$0.18^{+0.10}_{-0.05}$. All these properties are summarised in Table~\ref{tab:results}. We notice that our inferred BH and AGN parameters are slightly higher than, yet within the uncertainties of, the estimates recently derived by \citet{Juodzbalis:2026_jades}, based on medium-resolution broad \ha best-fit results and with no dust correction (i.e. $A_V$\,$=$\,0).

\section{Possible future evolution}\label{apx:evolution}

In Sect.~\ref{sec:discussion} we explore the possible evolution of the source complex, to see whether, as a consequence of the system coalescence and the combined host and BH growth, the compact BL AGN might become more similar to more common populations of AGN, star-forming or quiescent galaxies observed at lower redshift. In particular, we consider the possible evolution of the complex in the $M_{\rm BH}-M_\star$ and $\mu_{\rm gas}-z$ planes (Fig.~\ref{fig:evolution}), until $z$\,$=$\,4 using a simplified toy model, and until lower redshift relying on numerical simulations, as described in the following.

\subsection{Toy model until $z$\,$=$\,4}\label{apx:toy_model}
We extrapolate the evolution of the complex from $z$\,$=$\,5.23 to $z$\,$=$\,4 (i.e. over 400~Myr) using a toy model that relies on simplified assumptions, including no further gas or star accretion onto the system. All the multiple components (i.e. \target plus the sources along the filament) are considered as part of a whole single system, with $M_\star$, $M_{\rm gas}$ and SFR equal to the sum of the individual source values ($M_\star$\,$=$\,$2.7$\,$\times$\,10$^9$~\msun, $M_{\rm gas}$\,$=$\,1.2\,$\times$\,10$^{10}$~\msun, SFR10\,$=$\,$17$~\sfr). The global system is also assumed to host a central BH equal to that of \target ($M_{\rm BH}$\,$=$\,$1.1$\,$\times$\,10$^7$~\msun, $\lambda_{\rm Edd}$\,$=$\,$0.18$, derived in Appendix~\ref{apx:ism}).

We consider two possible growth modes for the host galaxy, at constant SFR\,$=$\,$17$~\sfr and at constant $t_{\rm depl}$\,$=$\,$0.7$~Myr. For the central BH, we instead assume accretion at constant $\lambda_{\rm Edd}$\,$=$\,0.18, with an AGN duty cycle of 10\%. Under these assumptions over a time frame of about 400 Myr, we obtain a BH growth of 17\% ($M_{\rm BH}$\,$=$\,$1.3$\,$\times$\,10$^7$~\msun at $z$\,$=$\,4), along with a host stellar mass growth of only a factor 2\,--\,3 assuming constant gas depletion ($M_{\star}$\,$=$\,$8.2$\,$\times$\,10$^9$~\msun) or SFR ($M_{\star}$\,$=$\,$1.0$\,$\times$\,10$^{10}$~\msun). The two evolutionary paths lead to $M_{\rm gas}$\,$=$\,$6.3$\,$\times$\,10$^{9}$~\msun and $M_{\rm gas}$\,$=$\,$4.5$\,$\times$\,10$^{9}$~\msun, corresponding to $\mu_{\rm gas}$\,$=$\,0.77 and $\mu_{\rm gas}$\,$=$\,0.45, respectively. 

\subsection{Predictions from numerical simulations}\label{apx:simulations}

For independent insights into the possible evolution of the source complex, we look at the predictions for similar systems from the numerical hydrodynamic simulation IllustrisTNG \citep{2018MNRAS.475..676S, 2018MNRAS.475..624N, 2018MNRAS.475..648P,2018MNRAS.477.1206N, 2018MNRAS.480.5113M}.  
In particular, we consider the intermediate cosmological volume with side length of 100~Mpc (TNG100). In the snapshot at $z$\,$=$\,5 (snapshot 17), we search for groups containing at least three subhalos with similar stellar masses ($0.5 \times 10^8$~\msun\,$<$\,$M_{\star}$\,$<$\,$ 20 \times 10^8$~\msun) and with galaxy-to-galaxy distances smaller than 15~kpc, similar to the observed source complex. We thus find five groups satisfying these criteria, among which: one is composed of galaxies that are close satellites of a much more massive system ($ M_{\star} > 10^{10.7}$~\msun); three 
reside at the outskirts of massive proto-clusters and are surrounded by several lower mass galaxies; 
finally, one isolated compact group consisting of four galaxies with $ M_{\star}$\,$=$\,$0.51 - 8.1 \times 10^{8}$~\msun forms a structure with less than 10~kpc radius, plus other two lower mass satellites ($M_{\star}\sim 0.4\times10^8$~\msun) at larger (10 and 31~kpc) distances. Since no other system at $z$\,$\simeq$\,5.23 is found within $R_{\rm vir}$\,$=$\,30~kpc, as expected for the halo hosting the whole complex (Sect.~\ref{sec:merger_tscale}), according to the JADES DR5 photometric redshifts \citep{Robertson:2026}, we focus on the evolution of the only isolated group of galaxies (group ID 1010 in the snapshot 17),
which more closely resembles the observed source complex.

To derive the gas masses, we only consider gas particles that are forming stars (i.e. SFR\,$>$\,0~\sfr) to best match what is measured in observations. 
We follow the evolution of the group forward in time by identifying its descendants in the the SubLink merger tree \citep[see][]{2015MNRAS.449...49R}. Already by the next simulation snapshot ($z$\,$=$\,4.6), three galaxies of the group coalesce with the central galaxy, while other two by the further next one ($z$\,$=$\,4.4), thus reaching final coalescence in less than 200~Myr. At $z$\,$=$\,3, the galaxy then undergoes a major merger with a much larger system, which dramatically affects its subsequent evolution and properties. For this reason, in Sect.~\ref{sec:evolution}, we only track the evolution of this group from $z$\,$=$\,5 to $z$\,$=$\,3.

\end{appendix}

\end{document}